\documentclass[final]{jfm}

\usepackage{graphicx}
\usepackage{epsfig}
\usepackage{epstopdf,epsfig}
\usepackage{newtxtext}
\usepackage{newtxmath}
\usepackage{natbib}
\usepackage{amsmath}
\usepackage{amsfonts}
\numberwithin{equation}{section}
\usepackage{hyperref}
\usepackage{gensymb}
\usepackage{cite}
\usepackage{cleveref}
\usepackage{upgreek}
\usepackage[title]{appendix}
\hypersetup{
	colorlinks = true,
	urlcolor   = blue,
	citecolor  = blue,
	linkcolor  = blue,
}
\usepackage{caption}
\newcommand{\RomanNumeralCaps}[1]
\linenumbers
\renewcommand{\figureautorefname}{figure}

\newcommand{\Autoref}[1]{\begingroup\renewcommand{\figureautorefname}{Figure}\autoref{#1}\endgroup}

\makeatletter
\renewcommand{\absfooterflag}{}
\renewcommand{\pagelimitfooter}{}
\makeatother

\crefformat{section}{\(\S\)~#2#1#3}
\title{Cavitation-bubble Interaction with an Initially Perturbed Free Surface}

\author{Jingyu Gu\aff{1}, Zirui Liu\aff{1,2}, A-Man Zhang\aff{1,2,3} \and Shuai Li\aff{1,2,3}}

\affiliation{\aff{1}College of Shipbuilding Engineering, Harbin Engineering University, Harbin 150001, PR China
	\aff{2}Nanhai Institute of Harbin Engineering University, Sanya 572024, PR China
	\aff{3}National Key Lab of Ship Structural Safety, Harbin Engineering University, Harbin 150001, PR China}

\begin{document}
\maketitle
\noindent
\parbox{\textwidth}{
\textbf{Corresponding authors:} Shuai Li,
\href{mailto:lishuai@hrbeu.edu.cn}{lishuai@hrbeu.edu.cn}}

\vspace{2em}

\begin{abstract}
The interaction of a spark-generated cavitation bubble with an initially perturbed free surface is investigated experimentally, numerically, and analytically. By exploiting contact-line pinning, we accurately prescribe an initial meniscus with a thin, hydrophilic-coated rod inserted into the liquid. A pronounced surface cavity, driven by the oscillating bubble, forms and penetrates downward to a scale comparable to the bubble itself. The coupled cavity--bubble system exhibits two distinct regimes --- coalescence and non-coalescence --- separated by a critical condition governed by the non-dimensional stand-off parameter \(\gamma\) and the initial meniscus height \(h_m\). In the non-coalescence regime, the cavity evolves through inception, expansion, and rebound/jetting. The maximum cavity length \(h_c\) follows a power-law scaling \(h_c\propto\gamma^{\alpha}\) with \(\alpha=-2.7\) (experiments) and \(\alpha=-2.6\) (simulations) for \(1.5\lesssim\gamma\lesssim3\), where inertia dominates. Deviations emerge for \(\gamma\lesssim1.5\) (strong nonlinearity) and \(\gamma\gtrsim3\) (surface tension and viscosity become noticeable). An analytical model based on the Rayleigh--Plesset equation combined with nonlinear Rayleigh--Taylor instability theory captures the trend and confirms that \(h_m\) plays only a secondary role relative to \(\gamma\). In the coalescence regime, atmospheric air vents into the bubble through the merged cavity, weakening the collapse intensity and reducing the associated pressure peak. We also examine air/liquid compressibility and boundary layer effects, whose significance grows as \(\gamma\) decreases. These findings are relevant to surface-jetting technologies, cavitation-erosion mitigation, and underwater-noise suppression.
\end{abstract}
		

%{\bf MSC Codes } {\it(Optional)} Please enter your MSC Codes here
\section{Introduction}
The dynamics of cavitation bubbles oscillating near the free surface have been extensively investigated over the past decades. The free surface is isobaric (aerodynamics are generally negligible in free-surface--bubble interactions), subject to zero parallel shear stresses, and can deform freely. Such characteristics can induce rich hydrodynamic behaviors. In the most common scenarios involving cavitation bubble oscillations adjacent to a plane free surface, the cavitation bubble develops a Bjerknes-type downwards re-entrant jet, and the free surface bulges into a water hump \citep{gibson1968cavitation,chahineInteractionOscillatingBubble1977b}. Further researches shows that the water hump can develop into a thin accelerating jet during the bubble collapse stage if the bubble is sufficiently close to the free surface (\(\gamma\lessapprox1\), where \(\gamma\) is the standoff parameter defined by the ratio between the bubble inception depth \(H\) and maximum bubble radius \(R_m\)) \citep{chahineInteractionOscillatingBubble1977b,blakeGrowthCollapseVapour1981}. Preliminary attempts were also made in many early studies to reproduce the free-surface--bubble interaction numerically and theoretically. One of the classical numerical works was carried out by \citet{blakeGrowthCollapseVapour1981,blakeCavitationBubblesBoundaries1987}. They successfully implemented a boundary integral (BI) method to model the oscillation of a cavitation bubble in proximity to a plane free surface, and its effectiveness in describing bubble dynamics was subsequently confirmed by numerous studies \citep{blake1987transient,wangStrongInteractionBuoyancy1996,wang2010non,li2023VerticallyNeutralCollapse}. Their further study unveiled the fundamental dynamics of free-surface--bubble interactions by adopting the Kelvin impulse theory \citep{benjamin1966collapse}, and derived a combined parameter \(\delta\gamma\) to predict the migration and jet direction of the cavitation bubble and characterize different bubble behaviors \citep{blakeCavitationBubblesBoundaries1987}. \(\delta=\sqrt{\rho g R_m/P_\infty}\) is the buoyancy parameter, where \(\rho\) and \(P_\infty\) represent the density and pressure of the undisturbed fluid, and \(g\) the gravitational acceleration. Recently, \citet{zhangTheoreticalModelCompressible2024} further established a theoretical model considering the complex dynamics of the bubble oscillation near diverse boundaries. In terms of the free surface, \citet{longuet-higginsBubblesBreakingWaves1983b} mathematically approximated the water hump on the free surface with a Dirichlet hyperboloid, and predicted the inception of the accelerating primary jet when the vertex angle of the water hump decreases to \(109.47^\circ\). 

However, challenges are posed to the traditional numerical and theoretical methods by the strong nonlinear effects of the free-surface--bubble interactions. Specifically, nonlinear surface deformation, strong discontinuities, and complex multiphase flow may occur in some particular cases, such as bubble bursting \citep{liBubbleInteractionsBursting2019}, surface re-closure \citep{tianAnalysisBreakingReclosure2018}, splashing sheet \citep{wangSplashingSealingEjecta2024}, surface jet \citep{rosselloDynamicsPulsedLaserinduced2022,zhangFreesurfaceJettingDriven2025}, Rayleigh--Taylor instability \citep{zengJettingViscousDroplets2018a,wangRayleighTaylorInstability2021}, etc. Hence, many CFD methods were adopted in previous studies. One of the most widely used methods is the volume of fluid (VOF) method, which has been proven effective in numerous studies. For instance, \citet{koukouvinisSimulationBubbleExpansion2016} conducted numerical simulations of the interaction between laser-generated bubbles and the free surface. They found that the re-entrant jet of the cavitation bubble exhibits a mushroom-shaped tip, and splits the bubble into several toroidal structures. \citet{saadeCrownFormationCavitating2021} numerically investigated the formation of a crown-shaped secondary jet surrounding the primary jet during the second expansion of the bubble. Their results indicate that this secondary jet is not generated by the shock wave emitted by the bubble when it reaches its minimum volume, but rather arises from the combined effects of a distorted pressure distribution over the free surface, which induces focusing flow and subsequent flow reversal. \citet{cerbusExperimentalNumericalStudy2022} further showed that the morphology of the crown and and the secondary jet can be well controlled by adjusting the initial position and energy of the bubble.

If the bubble is placed adjacent to a distorted interface, the nonlinear behaviors of the interface will be even intensified, including rapid surface jetting, instabilities, etc. For instance, a cavitation bubble oscillating near a meniscoid interface in a capillary tube induces a supersonic microjet, with its maximum velocity reaching 850 \(\mathrm{m\,s}^{-1}\) \citep{tagawaHighlyFocusedSupersonic2012,petersHighlyFocusedSupersonic2013}. \citet{obreschkowCavitationBubbleDynamics2006} performed the first experimental study of a cavitation bubble oscillating inside a liquid droplet in microgravity. They found that a toroidally collapsing bubble creates two opposite liquid jets escaping from the droplet, and significant secondary cavitation is induced inside the droplet due to the particular shock wave confinement. Moreover, a cavitation bubble oscillating inside a suspended spherical droplet confined by another host fluid (e.g., air or oil) can trigger even more complicated interfacial behaviors, including outward rapid jet \citep{obreschkowCavitationBubbleDynamics2006}, atomisation, and sheet formation \citep{gonzalezavilaFragmentationAcousticallyLevitating2016}. \citet{liCavitationBubbleDynamics2024} identified two distinctive fluid-mixing mechanisms by placing an oscillating cavitation bubble inside a water-in-oil (W/O) or oil-in-water (O/W) droplet. The bubble induces a rapid `needle-like' jet in the O/W droplet, while inducing overall motion and pinch-off in the W/O droplet. In some specific cases, the outward jet can become clusters of fine filaments and further shattered into finer secondary droplets, which was first recorded by \citet{robertCavitationBubbleBehavior2007} in the study of an oscillating bubble inside a cylindrical free-falling liquid jet. The speed of the filaments and droplets can exceed 100 \(\mathrm{m\,s}^{-1}\) in some particular cases. Similar phenomena have been reported by \citet{zengJettingViscousDroplets2018a} and \citet{rossello2023bubble} in a three-dimensional droplet, while a two-dimensional cavitation bubble within a cylindrical droplet could lead to three distinguished instabilities characteristics on the droplet surface, namely, splashing, ventilation, and stable state, depending on the relationship between perturbation amplitude and droplet and bubble radii \citep{wangRayleighTaylorInstability2021}. The unique characteristics of the interface in the aforementioned studies can be explained by the spherical Rayleigh--Taylor instability (RTI) caused by the low-pressure feature of the cavitation bubble, which derives from the denser fluid (liquid around the bubble) being accelerated by the lighter fluid (gas surrounding the droplet) when the cavitation bubble reaches around its maximum radius.

However, bubble bursting only occurs when the cavitation bubble is extremely close to the free surface \citep{liBubbleInteractionsBursting2019,dixitViscoelasticWorthingtonJets2025}, while the RTI of the free surface may only take place at some very specific standoff parameters, where the water film between the bubble and the atmosphere is thin enough, yet has not ruptured \citep{wangNumericalStudyFormation2022}. Ventilation between the atmosphere and bubble interior may occur even when the standoff parameter is relatively large \citep{cuiSmallchargeUnderwaterExplosion2016}. A possible explanation for this phenomenon is that the strong rarefaction wave reflected by the free surface induces cavitation across the fluid field \citep{ji2017SecondaryCavitation,rossello2023bubble}. The secondary cavitation bubbles then oscillate with the primary bubble. As a result, the instability of the free surface is triggered by the secondary cavitation bubbles, and air is inhaled into the bubble. While the dimensionless standoff parameter and initial interface geometry are recognized as key factors, the critical criteria for bubble bursting and ventilation onset have yet to be fully established, warranting further systematic investigation.

Most of existing studies focus on the interaction between the cavitation bubble and the smooth free surface, whereas in realistic scenarios, the free surface is often perturbed by minor defects, such as fibers, particles, or micro floating bubbles. Only a few experimental observations point out that under such circumstance, similar free surface instability behaviors can be stably triggered. \citet{kannanEntrapmentInteractionAir2018} reported the formation of cavity at the immersion point of a dipping copper wire on the free surface during the bubble oscillation. The cavity then exhibits a `catapult' motion in the bubble collapse stage, creating a rapid upward jet into the atmosphere. \citet{chengParticulateReshapesSurface2025} further investigated the rich dynamics of the surface jet induced by a floating particulate driven by an oscillating bubble. They categorized the surface jet into five different modes, i.e., cavity ventilation, sealed cavity, and open cavity, and studied the dependence of different jet mode on the non-dimensional immersion time of the particulate and bubble standoff parameter. 

However, the methods for deliberately seeding quantifiable surface perturbations and systematically investigating the evolution of the bubble-induced instability cavity remain largely unexplored. Hence, we still have little knowledge of the underlying mechanism of the cavity evolution, and its dependence on bubble oscillation. A more comprehensive study on the cavity dynamics is required. To fill this knowledge gap and transform surface perturbations from `uncontrolled noise' into `tunable control parameters', we systematically investigate the cavity evolution induced by the interaction between the cavitation bubble and an initially perturbed interface. During preliminary experiments, we found that the cavity behavior is related to the bubble and surface properties, and can be categorized into two major categories, separated by a critical condition. The dependence of cavity dynamics on governing parameters is investigated systematically through comprehensive approaches, including experiments, numerical simulations, and analytical modeling.

This paper is organized as follows: In \cref{section:2}, experimental and numerical methodologies are introduced. In \cref{section:3}, several typical cavity behaviors are presented and discussed. In \cref{section:4}, numerical simulations are performed to provide clearer observations of the cavity evolution. We further compare the numerical results with experimental observations to provide a qualitative understanding of the cavity dynamics. In \cref{section:5.1} and \cref{section:5.2}, we perform a theoretical modeling and scaling analysis of the cavity dynamics, using a classic Rayleigh--Plesset model. In \cref{section:5.3}, the quantitative relationship between cavity length and surface perturbations is investigated using a nonlinear Rayleigh--Taylor instability model. In \cref{section:5.4}, the influence of boundary layer are investigated qualitatively. The work is summarized in \cref{section:6}.

\section{Experimental and numerical methods}\label{section:2}

\subsection{Experimental set-up}\label{section:2.1}
\begin{figure}
	\centering
	\includegraphics[width=1\textwidth]{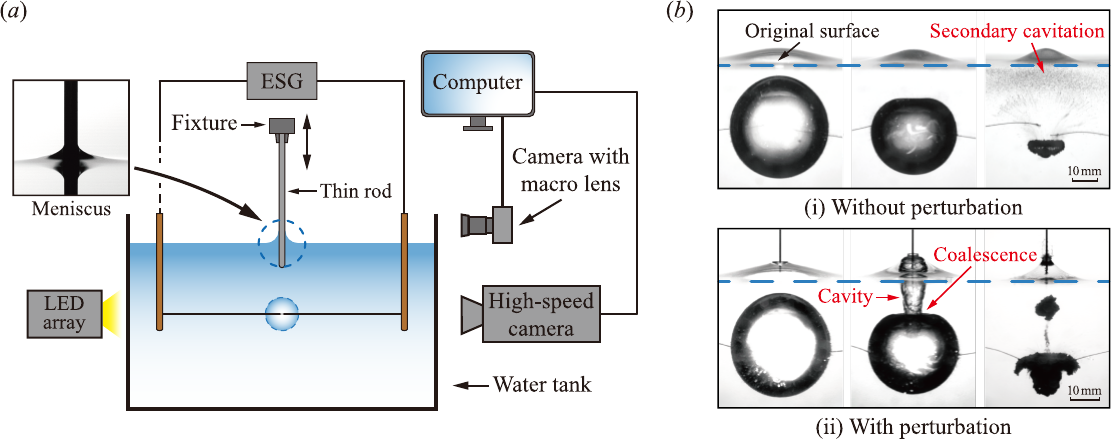}
	\setlength{\abovecaptionskip}{3mm}
	\captionsetup{width=1\textwidth}
	\caption{(\textit{a}) Schematic of experimental setup. The bubble is initiated coaxially with the thin rod covered by super-hydrophilic coating in a \(300\times300\times300\ \textrm{mm}^3\) water tank. An image of the meniscus is shown in the inset on the top left. (\textit{b}) Comparison between representative experiments of free-surface--bubble interaction (i) without and (ii) with perturbation. The standoff parameter \(\gamma\) for both experiments are 1.30. The bubble expands and collapses normally in the absence of the surface perturbation. Secondary cavitation caused by the rarefaction wave reflected by the free surface can be observed in the fluid field. In contrast, a cavity forms on the initially perturbed free surface and coalesces with the primary bubble, creating a channel that enables ventilation between the atmosphere and the bubble interior. Under such circumstances, the cavitation in the fluid field can not be observed, and the bubble exhibits a misty shape upon collapse.}
	\label{exp_schematic}
	\vspace{-5mm}
\end{figure}

As shown in \autoref{exp_schematic}, the experiment is conducted in a \(300\times300\times300\ \textrm{mm}^3\) water tank, where the cavitation bubble is generated using an Electric Spark Generator (ESG), whose reproducibility has been validated in our previous studies \citep{cuiIceBreakingCollapsing2018, hanInteractionCavitationBubbles2022a}. The tips of two copper wires are brought into contact at a certain depth below the free surface, initiating the bubble formation. Upon discharge, a current with a voltage of up to 400V is released by the ESG, creating an electric spark at the wire tips. This spark generates intense Joule heating, vaporizing the surrounding fluid and causing it to expand into a cavitation bubble with a maximum radius \(R_m\) of approximately 20 mm. A high-speed camera (Phantom V2012) operating at 30\,000--40\,000 frames per second is used to capture the transient interaction between the bubble and the free surface.

We establish an experimental technique to generate well-controlled surface perturbations by utilizing the meniscus rising mechanism \citep{clanetOnsetMenisci2002} and the contact line pinning effect \citep{schafferDynamicsContactLine1998}. A thin rod of 0.2--0.5 mm in radius is vertically mounted on a micro-controlled translation table. A stable meniscoid interface where capillary forces and gravity are balanced is created by piercing the thin rod into the liquid pool. The thin rod is positioned coaxially with the bubble initiation point. When the thin rod is vertically moved at a sufficiently low speed (less than \(10\ \textrm{mm}\,\textrm{s}^{-1}\)) fixed on a high-precision displacement platform, the three-phase contact line at the rod's surface remains pinned \citep{elliott1967dynamic, hansen1957relaxation, schafferDynamicsContactLine1998}. This leads to changes in the key parameters and geometry of the meniscus. This technique allows for the generation of menisci with high controllability and reproducibility. To extend the parameter range, the rod was coated with a super-hydrophilic material (Shangmeng Technology Wuxi Co. Ltd., SM-QS3500). As a result, the maximum meniscus height increases from 0.27 mm (without coating) to 0.95 mm (with coating). Additionally, a digital single-lens reflex (DSLR) camera (Nikon Z 6II) with a macro lens is used to capture the details of the meniscus above the free surface before the bubble initiation, with a spatial resolution of \(5.1\pm0.1\ \upmu\textrm{m}\) (one pixel of the image), which is approximately 1.4 \textit{\%} of the thin rod radius. One may easily notice that, the radius of the thin rod is much smaller than the capillary length \(l_c=\sqrt{\sigma/\rho_l g}=2.7\ \textrm{mm}\) (\(\sigma=0.073\ \textrm{N}\,\textrm{m}^{-1}\), \(\rho_l=997\ \textrm{kg}\,\textrm{m}^{-3}\), \(g=9.8\ \textrm{m}\,\textrm{s}^{-2}\)). Under such circumstances, the influence of the thin rod itself (radius and depth) can be minimized. The demonstration for the influence of the thin rod is referred to \autoref{appB}.

Two typical experiments of free-surface--bubble interactions: (i) without and (ii) with surface perturbation, are presented in \autoref{exp_schematic}(\textit{b}). As can be seen, in the case without perturbation, secondary cavitation bubbles caused by the free-surface-reflected rarefaction wave can be clearly seen. On the other hand, a cavity forms at the free surface and coalesces with the bubble in the case of a surface perturbation, while secondary cavitation is invisible. The underlying mechanisms associated with this phenomenon will be examined in \cref{section:3.3}. Since the maximum bubble radius \(R_m\) and the characteristic height of the meniscus \(h_m\) are the two crucial parameters in the experiments, the determination procedure in an experiment is given as follows. First of all, the characteristic height of the meniscoid interface \(h_m\) is directly measured from the image captured by the camera, with the axis of the macro lens aligned and parallel to the free surface. Then, the bubble oscillation is recorded with the high-speed camera. For relatively large standoff parameters, the bubble remains approximately spherical when it reaches maximum radius. Therefore, \(R_m\) can be calculated with \(R_m=D_m/2\), where \(D_m\) is the maximum bubble diameter measured directly from the experiments. As the bubble gets closer to the free surface, it will exhibit an apparent non-spherical shape. Under such circumstances, \(R_m=D_m/2\) tends to fail for insufficient accuracy. Hence, we adopt the same correction method as \citet{lishuaiBubblesphere2018}. To ensure the reliability and reproducibility of the experimental data, the experiments are repeated 3 times for each configuration. By elaborately manipulating the voltage of the ESG and the velocity and displacement of the thin rod, the reliability and reproducibility of the experiments are encouragingly good: The relative discrepancy of the maximum bubble radius \(R_m\) and meniscus height \(h_m\) can be controlled below 5.5 \textit{\%} (\(\approx 1\ \textrm{mm}\)) and 1.2 \textit{\%} (\(\approx0.01\ \textrm{mm}\)), respectively.

\subsection{Numerical model}\label{section:2.2}
\begin{figure}
	\centering
	\includegraphics[width=0.85\textwidth]{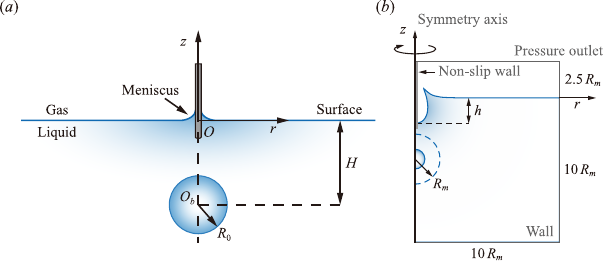}
	\setlength{\abovecaptionskip}{2mm}
	\captionsetup{width=1\textwidth}
	\caption{(\textit{a}) A schematic of the numerical model of free-surface--bubble interaction. The \(z\) axis is coaxial with the thin rod while the origin \(O\) is located on the free surface. The bubble is initiated at point \(O_b\) located beneath the free surface at depth \(H\) with a initial radius of \(R_0\). (\textit{b}) Configuration of the computational domain. The cavity length \(h\) is the distance from the cavity bottom to the initial undisturbed free surface. The radius and height of the cylindrical computational domain are set as 10\(R_m\) and 12.5\(R_m\), respectively, while the depth of the fluid is set as 10\(R_m\).}
	\label{num_model_fvm}
	\vspace{-2mm}
\end{figure}

In order to simulate the highly nonlinear interaction between the bubble and the perturbed free surface, including cavity--bubble coalescence, bubble ventilation, bursting jets, and surface splashing, a segregated two-phase flow solver based on finite volume methods (FVM) is employed in this study. The schematic of the physical problem is shown in \autoref{num_model_fvm}(a). We establish a cylindrical coordinate system \(O\textrm{--}r\theta z\) with the origin \(O\) located at the free surface and aligned with both the thin rod and the bubble initiation point \(O_b\). The bubble is initialized at a depth \(H\) with an initial radius \(R_0\).

Following the discharge, the bubble rapidly expands, during which the Mach number is estimated to be \(\textit{Ma}\sim \mathcal{\textit{O}}(10^{-3}-10^{-2})\) (peaking at 0.03 when the bubble reaches its minimum volume). Here, the Mach number is defined as \(\textit{Ma}=U_\textit{collapse}/c\), where \(U_\textit{collapse}\) represents the maximum velocity during the bubble collapse (reaching approximately 50 \(\mathrm{m\,s}^{-1}\) in our experimental measurements), and \(c\) denotes the undisturbed liquid sound speed. At such low Mach numbers, liquid compressibility effects become negligible \citep{wangMultioscillationsBubbleCompressible2014}, allowing us to safely treat the liquid as incompressible in our simulations. One can refer to a more detailed verification of the influence of the liquid compressibility in \autoref{appC}. By employing a segregated two-phase flow solver with incompressible liquid and compressible gas \citep{kwak2005computational,dupuy2020analysis}, we can capture the transient cavity--bubble interactions accurately at low Mach numbers while boosting computational efficiency and minimizing resource demands.

We also neglect mass and heat diffusion across the bubble surface, which is justified by the high Péclet numbers \citep{colombetMassHeatTransfer2013, liCavitationBubbleDynamics2024}: typically \(\textit{Pe}_\textit{m}\sim\textit{O}(10^6)\) (mass diffusion) and \(\textit{Pe}_\textit{h}\sim\textit{O}(10^5)\) (heat diffusion) for centimetre-scale bubbles in this study. Specifically, \(\textit{Pe}=R_m^2/T_\textrm{osc}D\), where \(T_\textrm{osc}\) is the bubble oscillation period, and \(D\) is either the mass diffusion coefficient or the thermal diffusivity for mass or heat transfer, respectively. The gas is treated as compressible. Viscous effects and surface tension are also retained due to the presence of the thin rod and perturbations. Based on these considerations, the compressible forms of the Navier-Stokes and continuity equations are as follows:
\begin{equation}
	\frac{\partial\left(\rho\boldsymbol{U}\right)}{\partial t}+\nabla\boldsymbol{\cdot}\left(\rho\boldsymbol{U}\otimes\boldsymbol{U}\right)=-\nabla p+\nabla\boldsymbol{\cdot}\mathsfbi{T}+\boldsymbol{I}_\sigma,
	\label{nseq}
\end{equation}
\begin{equation}
	\frac{\partial\rho}{\partial t}+\nabla\boldsymbol{\cdot}\left(\rho\boldsymbol{U}\right)=0,
	\label{continuityeq}
\end{equation}
where \(\nabla\) denotes the gradient, \(\nabla\boldsymbol{\cdot}\) the divergence and \(\otimes\) the tensorial product. \(\rho\) is the fluid density, \(\boldsymbol{U}\) the velocity field, and \(p\) the pressure field. \(\mathsfbi{T}\) is the viscous stress tensor of the Newtonian fluid which is defined as:
\begin{equation}
	\mathsfbi{T}=\mu\left[\nabla\boldsymbol{U}+\left(\nabla\boldsymbol{U}\right)^T-\frac{2}{3}\left(\nabla\boldsymbol{\cdot}\boldsymbol{U}\right)\mathsfbi{I}\right],
	\label{viscoustensor}
\end{equation}
with \(\mathsfbi{I}\) the unit tensor and \(\mu\) the viscosity. The last term \(\boldsymbol{I}_\sigma\) in (\ref{nseq}) represents the integral of the capillary force acting on the liquid/gas interface. One can refer to studies of \citet{tryggvason2001cmf} and \citet{kochNumericalModelingLaser2016a} for more details. When it comes to the incompressible liquid, the volume of fluid elements remains constant, resulting in a velocity field without divergence and a constant density, i.e., \(\nabla\boldsymbol{\cdot}\boldsymbol{U}=0\). Consequently, the divergence terms in (\ref{nseq})--(\ref{viscoustensor}) can be neglected, reducing the system to the classical incompressible Navier-Stokes equations with the continuity condition.

The two phases are immiscible and subject to a no-slip condition across the interface. To capture the sharp interface, the volume of fluid (VOF) method \citep{hirtVolumeFluidVOF1981} combined with the High-Resolution Interface Capturing (HRIC) scheme \citep{li2022breaking} is employed in this study. The pressure and velocity fields are shared by both phases, while viscosity and density are averaged using the volume fraction \(\alpha_i\) \citep{miller2013compressiblefvm}. Thus, The continuity equation (\ref{continuityeq}) for each phase reads:
\begin{equation}
	\frac{\partial\left(\alpha_i \rho_i\right)}{\partial t}+\nabla\boldsymbol{\cdot}\left(\alpha_i\rho_i\boldsymbol{U}\right)=0,\quad i=l,g,
	\label{continuityeq2phase}
\end{equation}
where \(\alpha_i\) and \(\rho_i\) denote the volume fraction and density of the liquid and gas phases. This equation applies to both compressible and incompressible fluids by retaining or neglecting the divergence term. The computational domain is discretized into cells, with the sum of volume fractions in each cell equal to unity \citep{kochNumericalModelingLaser2016a}, i.e., \(\alpha_l+\alpha_g=1\). Thus, the overall density and viscosity in (\ref{nseq})--(\ref{viscoustensor}) are given by \(\rho=\alpha_l\rho_l+\alpha_g\rho_g\) and \(\mu=\alpha_l\mu_l+\alpha_g\mu_g\). By solving (\ref{continuityeq2phase}), the two phases are distinguished: \(\alpha_l=1\) for liquid, \(\alpha_l=0\) for gas, and \(0 < \alpha_l < 1\) for the interface. Details of the scheme can be found in \citet{kochNumericalModelingLaser2016a} and \citet{miller2013compressiblefvm}. Assuming incompressibility and non-diffusivity, the liquid phase has a fixed density of 997 \(\textrm{kg}\,\textrm{m}^{-3}\), while the gas phase is treated as a compressible and adiabatic ideal gas with the equation of state (EoS) defined as:
\begin{equation}
	\rho\left(P_g\right)=\rho_0\left(\frac{P_g}{P_0}\right)^{\frac{1}{\kappa}},
	\label{eos}
\end{equation}
where \(P_g\) is the gas pressure at an arbitrary moment after the bubble inception, \(\rho\) the gas density which is the function of \(P_g\), and \(\kappa\) the ratio of the specific heats which is set as 1.4 according to the study of \citet{liCavitationBubbleDynamics2024}. The subscript `0' denotes the initial properties of the gas. The \(P_0\) and \(\rho_0\) of the atmosphere can be set as 101325 Pa and 1.2 \(\textrm{kg\,m}^{-3}\), while the initial condition for the gas of the bubble interior will be addressed later in \cref{section:2.3} due to the complexity of bubble inception. By utilizing (\ref{eos}) and the constant liquid density assumption, the system defined by governing equations (\ref{nseq})--(\ref{continuityeq2phase}) becomes closed and solvable. The validation of our numerical model with experimental data and other analytical models is presented in \autoref{appA}.

The simulation setup of the free-surface--bubble system is illustrated in \autoref{num_model_fvm}(\textit{b}). To optimize computational efficiency, an axisymmetric computational domain is established. The radius and height of the domain is set as \(10R_m\) and \(12.5R_m\). Our simulation results show that the present domain setup is sufficiently large that the boundary barely affects the bubble dynamics. Further increase of the computational domain sizes does not significantly change the bubble characteristics, with the bubble period only increasing by 0.1 and 0.3~\textit{\%} when the domain size increases to \(20R_m\) and \(30R_m\), respectively. One can refer to the details of the domain verification in \autoref{appA}. The thin rod and the boundary beneath the free surface are set as no-slip boundaries to reproduce experiments in the water tank. To accurately capture the cavity and bubble dynamics, a refined mesh with a resolution of \(5\ \upmu\textrm{m}\) is applied to the region where the cavity and bubble evolve. This local refinement approach reduces the computational demand by maintaining the total number of mesh cells below 1\,500\,000 while ensuring the accuracy of the simulation. One may also notice that, since the boundary condition of the thin rod is set as no-slip, it is certain that there will exist a boundary layer around the thin rod surface, which does reflect part of the real experimental situation and may pose an influence on the numerical simulation results depend on the mesh size. The thickness of the boundary layer can be estimated using \(\zeta=\sqrt{\nu t}\), thus evaluating its influence, where \(\nu=1.0\ \textrm{mm}^{2}\,\textrm{s}^{-1}\) is the kinematic viscosity of the water, and characteristics time \(t\approx3\ \textrm{ms}\) is taken as the first cycle of the bubble, yielding \(\zeta\approx55\ \upmu\textrm{m}\). Apparently, the mesh size we choose is much smaller than the characteristic thickness of the boundary layer, sufficiently fine to capture its detailed characteristics. To further address this issue, we perform a convergence analysis using different refinement resolutions in \autoref{appA}, with the mesh size ranging from \(2\) to \(20\upmu\textrm{m}\). Comparison between these simulation results with different mesh sizes does not exhibit significant variations in cavity behaviors (see \autoref{convergence_analysis}). A reasonable explanation given by \citet{cox2004comparison} and \citet{li2020modelling} can be presented as follows. The Reynolds number associated with the spark-generated bubble has the order of \(\textit{O}(10^5-10^6)\), and the Mach number is much smaller than one (a detailed non-dimensional analysis is referred to \cref{section:3.1}). Moreover, the scales of bubble and cavity (\(\sim10^0-10^1\ \textrm{mm}\)) are much larger than the thickness of the boundary layer (\(\sim10^{-2}\ \textrm{mm}\)). Therefore, the detailed morphology of the boundary layer and motion of the contact line do not exert a remarkable influence on the cavity dynamics. With comprehensive considerations, the mesh configuration with a total mesh number of 1\,500\,000 and a minimum mesh size of 5\(\upmu\textrm{m}\) is adopted to attain a balance between calculation efficiency and accuracy. A more detailed analysis of the boundary layer effect is performed in \cref{section:5.4}.

A modified axisymmetric boundary integral (BI) method based on our previous work concerning the interaction between an oscillating bubble and a two-phase interface \citep{hanInteractionCavitationBubbles2022a,liCavitationBubbleDynamics2024} is also adopted, where the two fluids (gas and liquid) on both sides of the interface are considered as inviscid and incompressible. Therefore, the Laplace equation is valid in both fluids, and the velocity potential \(\varphi_l\) and \(\varphi_g\) satisfy the BI equation, yielding
\begin{gather}
	\nabla^2\varphi_i=0\quad \left(i=l,g\right),\\
	\Pi\left(\boldsymbol{r}\right)\varphi_i\left(\boldsymbol{r}\right)=\iint_{S}\left[\frac{\partial\varphi_i\left(\boldsymbol{q}\right)}{\partial{n}} \frac{1}{|\boldsymbol{r}-\boldsymbol{q}|}-\varphi_i\left(\boldsymbol{q}\right)\frac{\partial}{\partial{n}}\left(\frac{1}{|\boldsymbol{r}-\boldsymbol{q}|}\right)\right]\textrm{d}S_i\left(\boldsymbol{q}\right)\quad\left(i=l,g\right),
\end{gather}
where \(\Pi\) represents the solid angle, \(\boldsymbol{r}\) and \(\boldsymbol{q}\) are the control and source points, respectively, and \(\partial/\partial {n}\) indicates the normal derivative. Here, \(S\) refers to the gas--liquid interface and the bubble surface when \(i = l\) (liquid domain below the interface), while it refers to the gas--liquid interface only when \(i = g\).

Considering the surface tension and neglecting gas viscosity, the dynamic boundary conditions on the bubble surface and at the gas--liquid interface can be written as
\begin{gather}
	\frac{\textrm{D}\varphi_l}{\textrm{D}t}=\frac{P_{\infty}}{\rho_l}-\frac{P_0}{\rho_l}\left(\frac{V_{0}}{V}\right)^{\kappa}+\frac{\sigma \mathcal{C}}{\rho_l}+\frac{1}{2}|\nabla\varphi_l|^{2}-g(z+z_b)\quad\textrm{on the bubble surface,}\\
	\frac{\textrm{D}(\varphi_l-\beta\varphi_g)}{\textrm{D}t}=\frac{1}{2}|\nabla\varphi_l|^2+\frac{\beta}{2}|\nabla\varphi_g|^2-\beta\nabla\varphi_l\boldsymbol{\cdot}\nabla\varphi_g-(1-\beta)gz+\frac{\sigma \mathcal{C}}{\rho_l}\quad\textrm{at the interface},
	\raisetag{0.7em}
\end{gather}
where \(\textrm{D}/\textrm{D}t=\partial/\partial t+\nabla\varphi_i\boldsymbol{\cdot}\nabla\) is the material derivative, \(P_\infty\) the hydrostatic pressure at \(z = 0\), \(P_L\) the liquid pressure on the bubble surface, \(P_0\) and \(V_0\) the initial pressure and volume of the bubble, \(\rho\) the density of the fluid, \(g\) the gravitational acceleration, \(\mathcal{C}\) the local curvature, and \(\beta=\rho_g/\rho_l\) the density ratio. 

On all surfaces, the kinematic boundary condition is given by:
\begin{equation}
	\frac{\textrm{D}\textit{\textbf{r}}}{\textrm{D}t}=\nabla\varphi_l.
\end{equation}
Using the equations above, one can reproduce the interaction between an oscillating bubble and a two-phase interface. The details of this method can be referred to in the studies of \citet{hanInteractionCavitationBubbles2022a}, \citet{li3DModelInertial2023}, and \citet{liCavitationBubbleDynamics2024}. One may easily notice that this method neglects the compressibility, vorticity, and other aerodynamic effects of the gas, whose influence may become prominent as the air flow becomes rapid with increasing cavity velocity, thus leading to deviations in cavity dynamics. Therefore, the purpose of implementing the BI method in this study is not to precisely reproduce the cavity dynamics, but to elucidate the factors that may have a notable influence on the cavity dynamics by comparing with the FVM simulations, such as air compressibility effects, boundary layer, etc, which can not be reproduced by BI simulations. We will further elaborate on this issue in \cref{section:4.2}, \cref{section:5.2}, and \autoref{appC}.

\begin{figure}
	\centering
	\includegraphics[width=0.85\textwidth]{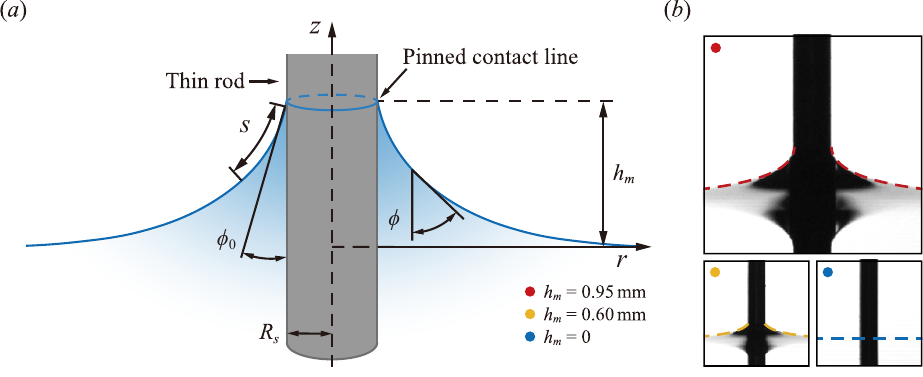}
	\setlength{\abovecaptionskip}{3mm}
	\captionsetup{width=1\textwidth}
	\caption{Schematic for the analytical model of the meniscus (\textit{a}) and comparison between experimental and analytical results of the meniscoid interface (\textit{b}). \(h_m\) is the distance from the initial undisturbed free surface to the meniscus apex. The analytical results calculated with (\ref{yleqcylindrical}) for \(h_m=0.95\ \textrm{mm}\) and 0.60 mm are plotted in red and yellow dashed lines, respectively. The control experiment for \(h_m=0\) is also given with the free surface plotted in a dashed blue line.}
	\label{num_model_meniscus}
	\vspace{-3mm}
\end{figure}

Finally, we employ an analytical model for the surface perturbation, as sketched in \autoref{num_model_meniscus}(\textit{a}). The same cylindrical coordinate as that in \autoref{num_model_fvm}(\textit{a}) is adopted. The meniscus profile can be calculated by solving the Young--Laplace equation:
\begin{equation}
	\frac{1}{R_1}+\frac{1}{R_2}=-\frac{\Delta p}{\sigma},
	\label{yleq}
\end{equation}
where \(R_1\) and \(R_2\) denote the principal radii of curvature, and \(\Delta p\) the Laplace pressure on the interface, and \(\sigma\) the surface tension. Rewriting (\ref{yleq}) in the cylindrical coordinate \citep{clanetOnsetMenisci2002} yields the following equation:
\begin{equation}
	\frac{\mathrm{d}\phi}{\mathrm{d}s}-\frac{\cos\phi}{r}=\frac{z}{l_c^2},
	\label{yleqcylindrical}
\end{equation}
where we adopt the expression for the principal radii of curvature \(1/R_1+1/R_2=-\mathrm{d}\phi/\mathrm{d}s+\cos\phi/r\), derived by \citet{bouasse1924capillarite}. For an arbitrary point on the meniscus surface, \(\phi\) represents the tangent angle of this point relative to the axis \(O\text{--}z\), while \(s\) represents the arc length measured from the meniscus tip. The coordinates \((r,z)\) depict the meniscus profile in the \(O\text{--}rz\) plane. By integrating (\ref{yleqcylindrical}) with the boundary condition \(\phi|_{r=R_s}=\phi_0\), we can obtain the meniscus profile analytical, where \(R_s\) denotes the radius of the thin rod, and \(\phi_0\) the static contact angle. One may notice that, condition \(z|_{r=R_s}=h_m\) is also needed to solve (\ref{yleqcylindrical}). Therefore, we adopt a matched asymptotic expansions method to determine \(h_m\), following the approach of \citet{jamesMeniscusOutsideSmall1974} and \citet{loMeniscusNeedleLesson1983}, whose studies suggest that \(h_m\) and \(\phi_0\) are intrinsically related. Hence we have:
\begin{equation}
	\frac{h_m}{R_s}=z_1|_{\hat{r}=1}\ln \xi+z_2|_{\hat{r}=1}+z_3|_{\hat{r}=1}\xi^2\ln^2\xi+z_4|_{\hat{r}=1}\xi^2\ln\xi+z_5|_{\hat{r}=1}\xi^2,
	\label{hmasymptotic}
\end{equation}
where \(\hat{r}=r/R_s\), \(\xi=R_s/l_c\), and \(z_i|_{\hat{r}=1}(i=1,2,\cdots,5)\) are the coefficients corresponding to different orders in the asymptotic expansion depending on the contact angle \(\phi_0\). This equation suggests that once one of \(h_m\) or \(\phi_0\) is determined, the other one is determined accordingly. The detailed expressions of the coefficients \(z_i|_{\hat{r}=1}\) are presented in \autoref{appD}. The full derivation of this method refers to the study of \citet{loMeniscusNeedleLesson1983}.

One can also notice that in \autoref{num_model_meniscus}(\textit{b}), ghosting effects at the outer rim of the meniscus due to refraction at the water-air interface can be observed. Such optical artefacts may introduce systematic uncertainties in the experimentally measured \(h_m\). To mitigate this effect, we combined experiments with analysis to determine the meniscus height \(h_m\) and its theoretical profile. First, by inserting the thin rod into the liquid pool and carefully adjusting its position, we establish a flat free surface and mark its position (blue dashed line in \autoref{num_model_meniscus}\textit{b}, \(h_m=0\)). Then the thin rod is slowly withdrawn from the liquid pool and fixed to form a meniscus with a pinned contact line. The perpendicular distance from the contact line to the free surface is the meniscus height \(h_m\). Equations (\ref{yleqcylindrical}) and (\ref{hmasymptotic}) show that, for a given solid--liquid pair, every value of \(h_m\) corresponds to a unique meniscus profile \citep{bouasse1924capillarite,jamesMeniscusOutsideSmall1974,loMeniscusNeedleLesson1983}. We therefore generated the theoretical profile corresponding to each measured \(h_m\) using (\ref{yleqcylindrical}) and (\ref{hmasymptotic}) and overlaid it on the high-resolution DSLR image. The excellent agreement (shown in \autoref{num_model_meniscus}\textit{b}, \(h_m=0.95\) and 0.60 mm) confirms that optical distortion is negligible; any residual systematic uncertainty in \(h_m\) is below 0.01 mm --- roughly two pixels at our maximum magnification --- and is insignificant relative to the rod radius and the capillary length.
	
\subsection{Non-dimensionalization and initialization}\label{section:2.3}
In this section, we present the non-dimensionalization method adopted in this study. We choose the maximum bubble radius \(R_m\), the pressure of the atmosphere on the initial free surface \(P_\infty\), and the density of the liquid (water in this study) \(\rho_l\) as the basic quantities. Based on these characteristic scales, the non-dimensional key variables are defined as follows:
\begin{equation}
	\refstepcounter{equation}
	\gamma=\frac{H}{R_m},\ \varepsilon=\frac{P_{0b}}{P_\infty},
	\tag{\theequation a,b}
	\label{nondimensionalization}
\end{equation}
where \(\gamma\) is the standoff parameter, and \(\varepsilon\) the strength parameter. Here, \(H\) is the depth of the bubble inception point, and \(P_{0b}\) is the initial pressure of the gas of the bubble interior. Another significant parameter in our study, the cavity length \(h\) and its maximum value \(h_c\), is defined as the distance from the cavity bottom to the free surface (see \autoref{num_model_fvm}\textit{b}). We use this definition due to the refraction effect: The cavity's upper part is distorted and obscured by the water dome, complicating the measurement of the total cavity length from the opening to the bottom. Apparently, this definition does not apply for the cases when the cavity fails to extend below the free surface. However, our experiments indicate that these cases happen only when \(\gamma\) is very small. In such scenarios, the cavity either coalesces with the bubble or the bubble bursts at the free surface (see \autoref{exp_gamma_below_1.36}\textit{b} and \ref{exp_gamma_below_1.36}\textit{c}), rendering the cavity length irrelevant. Similarly, \(h\), \(h_c\), and the aforementioned \(h_m\) are scaled by \(R_m\). All the variables in this study, unless otherwise annotated, are non-dimensional.

As mentioned in \cref{section:2.2}, the initial conditions of the bubble is yet to be determined, which includes the initial pressure \(P_{0b}\), density \(\rho_{0b}\), and non-dimensional initial radius \(R_0\). According to the best of the authors' knowledge, no effective models have been developed thus far to describe the initiation stage of an electric spark bubble. Due to the intense Joule heating generated by the electric spark, the surrounding liquid is vaporized, leading to complex phase changes that play a crucial role in bubble evolution, but beyond the scope of our numerical model. Hence we adopt a simplified method to initialize the bubble, i.e., the bubble is considered as an initially stationary gas bubble with high pressure \citep{klaseboer2005experimental,zeng2018wallshearstress,hanInteractionCavitationBubbles2022a} and phase change effect is ignored in this study.

Building on these discussions, we first determine the initial pressure and density of the gas inside the bubble, denoted as \(P_{0b}\) and \(\rho_{0b}\), respectively. According to previous studies of \citet{lishuaiBubblesphere2018} and \citet{hanInteractionCavitationBubbles2022a}, the maximum radius and oscillation period of the bubble exhibit relatively reasonable agreement among theoretical, numerical, and experimental results when \(\varepsilon\) is set between 100 and 200. Therefore, we set \(\varepsilon=125\) to achieve a satisfactory match, which yields \(P_{0b}=1.27\times10^7\ \text{Pa}\) and \(\rho_{0b}=37.75\ \text{kg}\,\text{m}^{-3}\) based on (\ref{eos}), where the initial atmospheric conditions are \(P_0=101325\ \textrm{Pa}\) and \(\rho_0=1.2\ \textrm{kg}\,\textrm{m}^{-3}\). Next, the non-dimensional initial bubble radius is set as \(R_0=0.1623\) to align the numerical simulation with the experimental result according to \citet{plesset1949dynamics} and \citet{klaseboer2005experimental}, using a modified Rayleigh--Plesset equation.

\section{Experimental observations of cavity--bubble-free-surface interactions}\label{section:3}
In this section, we discuss typical experimental observations of bubble interactions with a perturbed free surface. The experimental results are classified into two major categories based on cavity behavior, separated by a critical condition, and the dependence of the cavity dynamics on the standoff parameter is qualitatively illustrated.

\subsection{Non-coalescence}\label{section:3.1}
\begin{figure}
	\centering
	\includegraphics[width=1.0\textwidth]{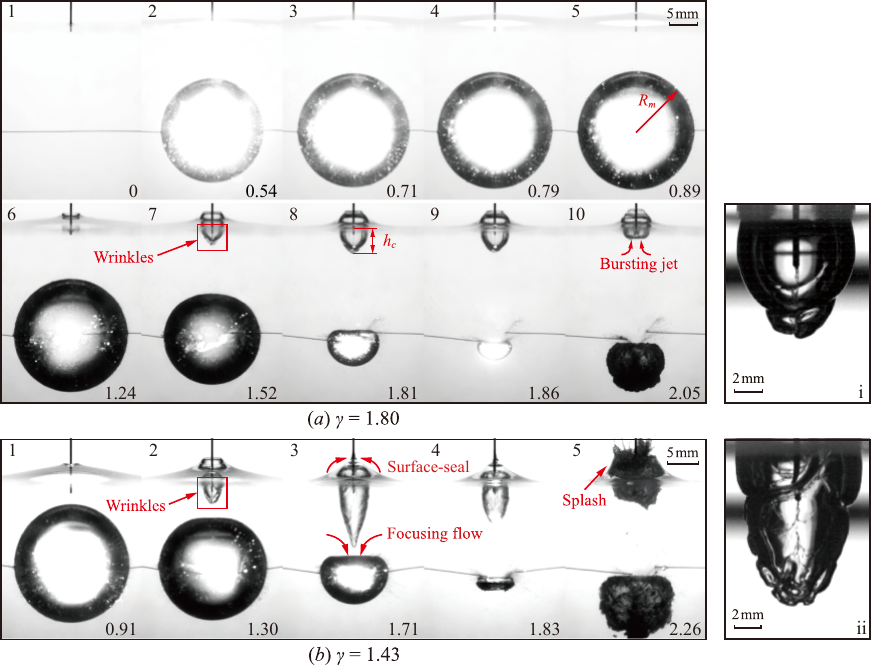}
	\setlength{\abovecaptionskip}{2mm}
	\captionsetup{width=1\textwidth}
	\caption{Two typical experiments of bubble interaction with a perturbed free surface for large standoff parameters \(\gamma\)=1.80 and 1.43. The radius of the thin rod is 0.35 mm in both experiments. (\textit{a}) The interaction between the bubble and the perturbed surface is mild; thus, the free surface elevates slightly with the bubble expansion, and the cavity maintains a smooth olive shape. A bursting jet occurs when the bubble reaches its minimum volume and rebounds (\(H=30.6\ \textrm{mm}\), \(R_m=17.0\ \textrm{mm}\), \(h_m=0.85\ \textrm{mm}\)). (\textit{b}) The cavity is elongated, and the bottom exhibits a taper shape as \(\gamma\) decreases to 1.43. A brought-forward surface-seal causes a splash when the cavity rebounds (\(H=24.6\ \textrm{mm}\), \(R_m=17.2\ \textrm{mm}\), \(h_m=0.84\ \textrm{mm}\)). The timescale of each experiment is taken as \(T_\textit{osc}=R_m\sqrt{\rho_l/P_\infty}\), which are 1.69 and 1.71 ms, respectively. Two close-up views of the cavity are also presented on the right. The parameters of these two experiments are similar with those of (\textit{a}) and (\textit{b}): (i) \(H=31.1\ \textrm{mm}\), \(R_m=18.0\ \textrm{mm}\), \(h_m=0.84\ \textrm{mm}\), \(\gamma=1.73\); (ii) \(H=27.1\ \textrm{mm}\), \(R_m=19.0\ \textrm{mm}\), \(h_m=0.84\ \textrm{mm}\), \(\gamma=1.43\).} 
	\label{exp_gamma_over_1.36}
	\vspace{-5mm}
\end{figure}
In \autoref{exp_gamma_over_1.36}, two representative experiments are presented, where the free-surface--bubble interaction remains weak due to the large standoff parameter. In the first experiment (see \autoref{exp_gamma_over_1.36}\textit{a}), the standoff parameter is \(\gamma=1.80\). As the bubble is initiated and expands, the free surface elevates mildly (frames 1--5). The fluid motion in the vicinity of the thin rod is hindered due to the no-slip boundary condition on the solid surface. This discrepancy in velocity gives rise to a small conical depression on the fluid surface around the thin rod. As the bubble radius reaches its maximum, the internal gas pressure attains its minimum, and the depression grows downward due to the pressure gradient induced by the cavitation bubble (frame 6). The free surface then descends as the bubble contracts, and the depression evolves into an olive-shaped cavity with an opening on the free surface (frame 7). Finally, the cavity reaches its maximum length \(h_c\) (frame 8), shortly before the bubble reaches its minimum volume (frame 9). At this point, the bubble's internal pressure exceeds that of the ambient fluid, thus reversing the pressure gradient. As a result, the cavity bottom rebounds and bursts upwards, exhibiting behaviors similar to a Worthington jet \citep{worthington1897}, due to the combined effects of the pressure wave generated by the bubble when it reaches its minimum volume and the reversed pressure gradient.

One may notice that we use the term `pressure wave' here to describe the pressure response in the fluid field induced by bubble oscillation, whose propagation characteristics may not be resolved experimentally or numerically in this study. However, we primarily focus on the instantaneous response of the cavity and the fluid field to the pressure wave, while neglecting its propagation hysteresis. Such simplification is feasible, since the cavity velocity (\(\sim10^{1}\ \mathrm{m\,s}^{-1}\)) is two magnitude smaller than the local sound speed \(c\), suggesting that the propagation of the pressure wave can be regarded as instantaneous --- a similar simplification adopted in the study by \citet{petersHighlyFocusedSupersonic2013}, \citet{klaseboer20064287}, and \citet{klaseboer2007interaction} and proven effective, where the effect of the pressure wave on the meniscoid surface is modeled as an instantaneous pressure pulse of characteristic amplitude and duration. Although the pressure wave cannot be resolved experimentally, its influence on the cavity dynamics and fluid field can still be partially reproduced in numerical simulations, which will be further discussed in \cref{section:4.3}.

In the second experiment (see \autoref{exp_gamma_over_1.36}\textit{b}), the standoff parameter is reduced to \(\gamma=1.43\), resulting in stronger free-surface--bubble interaction. As a result, the cavity expands more rapidly once the bubble reaches its maximum radius (frames 1 and 2). In contrast to the previous case, the entire cavity is elongated, and the bottom exhibits a tapered shape (frame 3), thus reaching a larger length. This profound difference indicates that the flow-focusing effect caused by the contraction of the bubble may contribute to different cavity dynamics. Additionally, an elongated cavity causes an increasing amount of air influx, resulting in a lower local pressure at the cavity opening. The resulting pressure imbalance triggers the surface-seal prematurely, even before the cavity reaches its maximum length, thereby isolating it from the atmosphere. Shortly after the cavity reaches its maximum length, the cavity rebounds rapidly due to the combined effects of the pressure wave emitted by the bubble and hydrostatic pressure. During this upward motion, the cavity bottom collides with the sealed surface, causing a sealed-splash event (frame 5).

Wrinkles on the cavity surface can also be observed in \autoref{exp_gamma_over_1.36}(\textit{a}), frames 7 and 8. Moreover, the wrinkles grow markedly more pronounced and asymmetric as \(\gamma\) decreases in \autoref{exp_gamma_over_1.36}(\textit{b}). The largest-amplitude wrinkle structures emerge on the lower cavity wall during the late-time evolution. Since the cavity details are difficult to observe in (\textit{a}) and (\textit{b}), we present close-up views of the cavity in another two experiments with similar \(\gamma\), i.e., (i) \(\gamma=1.73\), (ii) \(\gamma=1.43\), as shown in the images on the right. Such wrinkles may derive from the late-time secondary R--T instability evolution. Earlier work \citep{ramaprabhuRayleighTaylorInstabilityDriven2013} has demonstrated that complex acceleration histories --- especially repeated transitions between acceleration and deceleration --- promote small-scale structure formation, a behavior consistent with the cavity acceleration and rebound process discussed later in \autoref{exp_num_comparison_1}, \cref{section:4.1}. Additionally, the nonlinear late-time evolution of the primary bubble itself seeds minor bubbles and wrinkles \citep{ramaprabhu2012late}. These interpretations align with our observations: Reducing \(\gamma\) strengthens the nonlinear interactions between the free surface and the cavitation bubble, thereby amplifying the secondary wrinkles (\autoref{exp_gamma_over_1.36}i, and \ref{exp_gamma_over_1.36}ii).

In order to provide enlightenment for subsequent investigation of the governing parameters of cavity dynamics, we hereby perform a non-dimensional analysis. Since the cavity evolution and bubble collapse stage occur simultaneously, their timescales are similar, given by \(T_\textit{osc}=R_m\sqrt{\rho_l/P_\infty}\). Moreover, in \autoref{exp_gamma_over_1.36}(\textit{a}), \(h_c=7.0\ \textrm{mm}\), \(U_c=6.3\) \(\mathrm{m\,s}^{-1}\), \(R_m=17.0\) mm, suggesting that both the maximum cavity length and velocity is comparable to those of the bubble, i.e., \(h_c/R_m\sim\textit{O}(10^0)\), \(U_c/U_b\sim\textit{O}(10^0)\), where \(U_b=\sqrt{P_\infty/\rho_l}=10.1\ \textrm{m\,s}^{-1}\) denotes the characteristic velocity of the bubble oscillation, and \(U_c\) the maximum cavity velocity. Consequently, the Reynolds number, Weber number, and Froude number of the cavity evolution can be written as:
\begin{equation}
	\textit{We} = \frac{\rho_l U_c^2 h_c}{\sigma} \sim \textit{O}\left(10^3\right),\ \textit{Re}= \frac{\rho_l U_c h_c}{\mu_l} \sim \textit{O}\left(10^5\right),\ \textit{Fr}=\frac{U_c}{\sqrt{g h_c}} \sim \textit{O}\left(10^1-10^2\right),
	\label{nondimensionlessparameter}
\end{equation}
where we set liquid viscosity as \(\mu_l=10^{-3}\ \textrm{Pa}\,\textrm{s}\), density as \(\rho_l=997\ \textrm{kg}\,\textrm{m}^{-3}\), and surface tension as \(\sigma=0.073\ \textrm{N}\,\textrm{m}^{-1}\). This indicates that surface tension, viscosity, and gravity effects are all negligible while the inertia force is the primary governing factor in cavity evolution. One may also notice that, even though \(h_c/R_m\) decreases to the order of \(\textit{O}(10^{-1})\) in some cases with relatively larger \(\gamma\), the non-dimensional parameters still have the order of \(\textit{We}\sim{O}(10^2)\), \(\textit{Re}\sim{O}(10^4)\), and \(\textit{Fr}\sim{O}(10^1)\), suggesting that the aforementioned simplifications are still feasible. This leads to consistent cavity dynamics over a relatively wide range of \(\gamma\), which we will discuss in \cref{section:5.2}. Moreover, we notice that the cavity evolution is highly related to the bubble collapse stage, i.e., the small depression does not grow downwards at the beginning. Instead, it rises with the water hump as a whole due to the rapid bubble expansion stage, then the cavity evolution is initiated after the bubble reaches its maximum radius and starts to contract. This indicates that the pressure gradient induced by the bubble oscillation is the dominant factor that drives the cavity evolution. This dependence will be examined in greater detail in \cref{section:4} and \cref{section:5}.

\subsection{Critical condition}\label{section:3.2}
This section examines the critical condition of cavity--bubble interactions. As \(\gamma\) decreases into a transitional region, the system shifts from the non-coalescence state to one exhibiting increasingly complex and ambiguous interaction patterns. Despite these complex dynamics, we identify (i) a critical threshold of \(\gamma\), and (ii) a classification framework to differentiate the critical conditions from the coalescence and non-coalescence cases.

\begin{figure}
	\centering
	\includegraphics[width=0.9\textwidth]{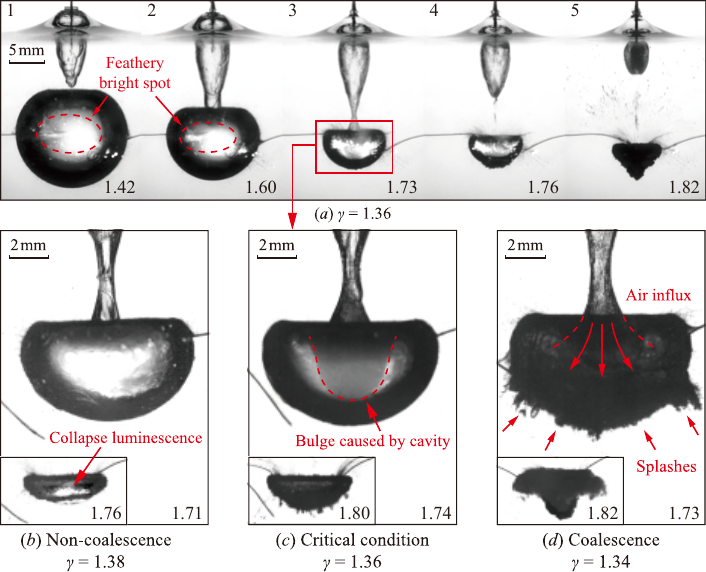}
	\setlength{\abovecaptionskip}{3mm}
	\captionsetup{width=1.0\textwidth}
	\caption{Four representative experiments of critical condition for \(\gamma\) = 1.36, 1.38, 1.36, and 1.34. The radius of the thin rod is 0.35 mm in all four experiments. (\textit{a}) The cavity is highly elongated and forms a tapered shape when \(\gamma=1.36\). The cavity bottom surpasses the bubble's upper surface and becomes invisible. After pinch-off, the cavity splits into two segments: The upper part forms an isolated bubble, while the lower part merges with the primary bubble, creating a roughened bubble surface (\(H=25.2\ \textrm{mm}\), \(R_m=18.5\ \textrm{mm}\), \(h_m=0.91\ \textrm{mm}\)). Additionally, feathery bright spots (possibly caused by continuous discharge of the wire) can be observed in frames 1 and 2. Panels (\textit{b})--(\textit{d}) show distinct cavity--bubble interaction patterns as \(\gamma\) varies slightly around 1.36, with emphasis on the dynamics just before cavity pinch-off. (\textit{b}) Though the cavity is elongated, it does not reach deep enough to coalesce with the bubble. The bubble retains a smooth surface, and strong luminescence occurs at collapses (see the inset in \autoref{exp_gamma_critical_condition}\textit{b}) (\(H=25.0\ \textrm{mm}\), \(R_m=18.1\ \textrm{mm}\), \(h_m=0.93\ \textrm{mm}\)). (\textit{c}) The cavity bottom forms a bulge inside the bubble. The primary bubble merges with the lower part of the cavity after pinch-off, forming a roughened surface (see the inset in \autoref{exp_gamma_critical_condition}\textit{c}) (\(H=25.5\ \textrm{mm}\), \(R_m=18.7\ \textrm{mm}\), \(h_m=0.90\ \textrm{mm}\)). (\textit{d}) The cavity coalesces with the bubble, creating splashes and a microbubble cloud around the lower bubble surface (see the inset in \autoref{exp_gamma_critical_condition}\textit{d}) (\(H=24.6\ \textrm{mm}\), \(R_m=18.3\ \textrm{mm}\), \(h_m=0.88\ \textrm{mm}\)). The timescales are 1.84, 1.80, 1.86, and 1.82 ms, respectively.}
	\label{exp_gamma_critical_condition}
	\vspace{-3mm}
\end{figure}

Four representative experiments are presented in \autoref{exp_gamma_critical_condition}. Since \(h_c\) varies slightly with \(h_m\), we focus on a subset of experiments where \(h_m\approx0.90\ \textrm{mm}\) and perform over 100 times of experiments, with \(\gamma\) varying from 0.49 to 3.58. It turns out that there exists a certain critical area for the cavity dynamics, and the critical condition occurs when \(\gamma=1.36\pm0.02\). In \autoref{exp_gamma_critical_condition}(\textit{a}), a typical critical case where \(\gamma=1.36\) is given. The cavity is further elongated, exhibiting the same taper shape as that in \autoref{exp_gamma_over_1.36}(\textit{b}) (frame 1). Then the cavity bottom surpasses the upper surface of the bubble (frames 2 and 3), rendering it invisible and preventing accurate cavity length measurements. After pinch-off, the cavity splits into two segments: The upper part detaches to form an isolated bubble, while the lower part merges with the primary bubble, creating a roughened bottom surface (frames 4 and 5). Unfortunately, direct visualization of the cavity through the bubble surface is not feasible with the current experimental setup, preventing clear confirmation of the cavity--bubble coalescence. Nonetheless, we can still infer the interaction pattern from post-collapse bubble morphology. The classification criteria are elaborated in \autoref{exp_gamma_critical_condition}(\textit{b})--\ref{exp_gamma_critical_condition}(\textit{d}), focusing on the cavity--bubble interactions immediately before cavity pinch-off.

In the second experiment (see \autoref{exp_gamma_critical_condition}\textit{b}), \(\gamma=1.38\), the cavity bottom surpasses the horizontal plane of the bubble's upper surface. However, there is no observable evidence that the cavity comes into contact with or merges into the bubble. Moreover, the bubble surface remains smooth until it collapses, and the bubble luminescence is still visible (see the inset in \autoref{exp_gamma_critical_condition}\textit{b}). Therefore, the cavity--bubble interaction pattern is almost the same as that in \autoref{exp_gamma_over_1.36}\textit{b}.

It is worth emphasizing that, the light emission of a spark-generated bubble is, in fact, a convoluted phenomenon: Earlier work has ascribed it to a number of concurrent mechanisms, including extreme temperature and pressure \citep{zhangLuminescencebubble,huangExperiumentalluminescence}, plasma formation \citep{stelmashukBubbleluminescence,vokurkaExperimentalbubble}, among others. Similar luminescence has been reported for non-spherical bubbles collapsing near either a free surface or a rigid boundary \citep{ohlSparkbubble,maSparkbubble}. Experimental observations also reveal feathery bright spots (as shown in \autoref{exp_gamma_over_1.36}\textit{a}, frame 7 and \autoref{exp_gamma_critical_condition}\textit{a}, frames 1--3) that emanate directly from the copper-wire tips. These filaments probably originate from a sustained micro-spark between the electrodes and could still be present when the bubble reaches minimum volume, thereby contributing to the recorded luminescence. At present, however, we do not know exactly how these structures form or how much of the total luminescence they comprise. Consequently, we cannot unambiguously attribute the observed luminescence to bubble collapse alone, to the electrode discharge, or to a combination of both. What we can state with confidence is that a pronounced luminescence appears in every non-coalescence case and is completely absent whenever coalescence occurs (\autoref{exp_gamma_over_1.36}--\ref{exp_gamma_below_1.36}). The most plausible explanation is that coalescence-driven droplets roughen the bubble wall; the resulting multiple scattering scatters most of the internal light, regardless of its physical origin. Hence, the presence or absence of luminescence provides a robust, experimentally accessible marker for distinguishing the two cavity--bubble interaction regimes.

In the third experiment (see \autoref{exp_gamma_critical_condition}\textit{c}), \(\gamma=1.36\), a bulge caused by the extending cavity inside the bubble can be observed. Though the existence of the bulge implies that the cavity is close to the bubble, there remains no definitive evidence of coalescence before the cavity neck pinch-off. After the pinch-off, the lower segment detaches from the cavity and merges with the bubble. The mixture of air and droplets collides on the bubble surface, forming some minor splashes (see the inset in \autoref{exp_gamma_critical_condition}\textit{c}). 

In the fourth experiment (see \autoref{exp_gamma_critical_condition}\textit{d}), \(\gamma=1.34\), coalescence undoubtedly occurs before the cavity pinch-off, as evidenced by the clearly visible splashes resulting from droplet collisions on the bubble surface.

Building on the discussions above, we propose a criterion for classifying these three scenarios: (1) If the cavity does not coalesce with the bubble throughout its entire evolution, and luminescence during bubble collapse is clearly visible, the case is classified as non-coalescence (\autoref{exp_gamma_critical_condition}\textit{b}); (2) If the cavity does not coalesce with the bubble before cavity neck pinch-off, but coalescence occurs afterward, the case is classified as the critical condition (\autoref{exp_gamma_critical_condition}\textit{c}); (3) If the cavity has coalesced with the bubble before cavity neck pinch-off, and luminescence is hardly visible, the case is classified as coalescence (\autoref{exp_gamma_critical_condition}\textit{d}).

As demonstrated in the experiments discussed above, the cavity--bubble interaction patterns differ significantly when \(\gamma\) varies by 0.02 around the critical value, which may indicate that \(\gamma\) is the dominant factor governing cavity dynamics. It is worth emphasizing that this drastic change between different cavity--bubble interaction patterns only occurs around the critical value, while in the rest of the \(\gamma\) region, the cavity characteristics exhibit rather good consistency. We will further investigate the critical condition through numerical simulations in \cref{section:5}.

\subsection{Coalescence}\label{section:3.3}
\begin{figure}
	\centering
	\includegraphics[width=0.9\textwidth]{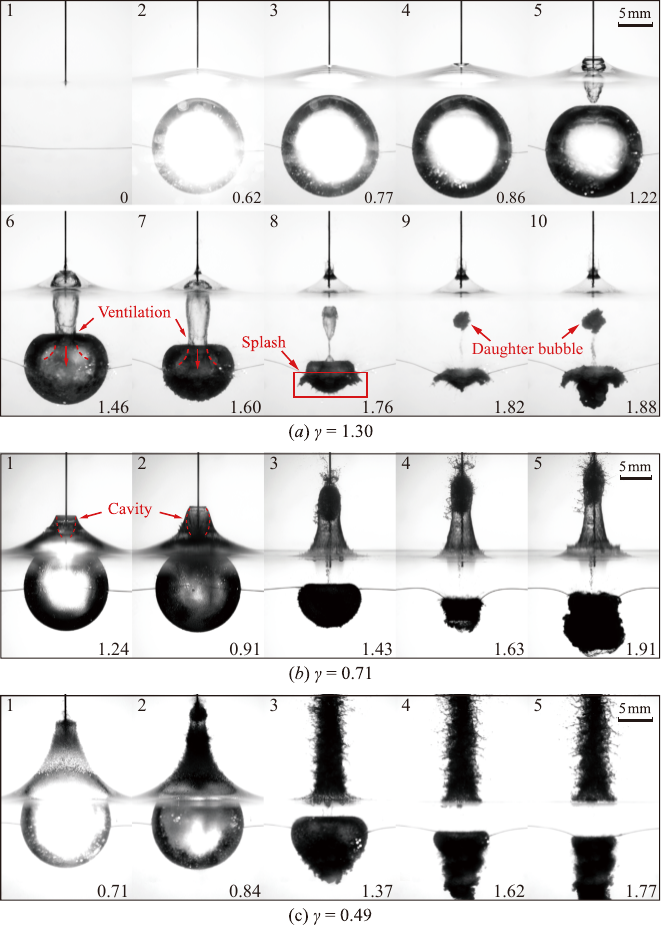}
	\setlength{\abovecaptionskip}{5mm}
	\captionsetup{width=1\textwidth}
	\caption{Three typical experiments of cavity--bubble coalescence scenario for \(\gamma\) = 1.30, 0.71, and 0.49. The radius of the thin rod is 0.35mm in all three experiments. (\textit{a}) Cavity expands rapidly and coalesces with the bubble before the surface-seal, creating a channel that connects the bubble and the atmosphere and triggering the ventilation, allowing the air to flow into the bubble (\(H=22.8\ \textrm{mm},\ R_m=17.5\ \textrm{mm},\ h_m=0.83\ \textrm{mm}\)). (\textit{b}) The decreased \(\gamma\) gives rise to an earlier coalescence and causes the water dome to elevate rapidly. As a result, the cavity neck pinches off quickly and detaches from the bubble, which ceases the ventilation. Despite the advanced pinch-off, an adequate amount of air influx balances the pressure difference between the gas pressure of the bubble interior and the ambient hydrostatic pressure (\(H=11.9\ \textrm{mm},\ R_m=16.8\ \textrm{mm},\ h_m=0.90\ \textrm{mm}\)). (\textit{c}) The bubble is so close to the free surface that the surface perturbation has little influence on the coalescence and ventilation. The bubble bursts into the atmosphere and creates an open cavity (\(H=7.9\ \textrm{mm},\ R_m=16.0\ \textrm{mm},\ h_m=0.92\ \textrm{mm}\)). The timescales are 1.74, 1.67, and 1.59 ms, respectively.}
	\label{exp_gamma_below_1.36}
\end{figure}

In this section, we present and discuss three typical experiments for cavity--bubble coalescence in \autoref{exp_gamma_below_1.36}. A coalescence case occurring at the end of the first oscillation cycle has already been discussed in \autoref{exp_gamma_critical_condition}(\textit{d}). Despite the onset of coalescence, the bubble undergoes a relatively regular collapse, with the absence of a roughened surface and splashes. Thus, the coalescence has a limited effect on the overall bubble collapse dynamics in this case. As \(\gamma\) decreases further below 1.36, the coalescence is brought forward, and the ventilation between the bubble interior and the atmosphere is initiated.

In the first experiment shown in \autoref{exp_gamma_below_1.36}(\textit{a}), \(\gamma=1.30\), the interaction between the bubble and the free surface is intensified, resulting in a faster downward growth of the cavity after the bubble reaches its maximum radius (frames 4--5). Such rapid growth, combined with the decreased standoff parameter, allows the cavity to coalesce with the bubble earlier than that in \autoref{exp_gamma_critical_condition}. Meanwhile, the cavity opening remains unsealed, allowing ambient air to flow into the bubble at high velocity through the channel. This establishes continuous ventilation between the bubble and the atmosphere above the surface until the surface-seal finally occurs (frames 6 and 7). Although the ventilation lasts for only 0.24 ms, the intense air influx is sufficient to deliver a significant amount of air into the bubble. This helps balance the pressure difference between the gas pressure of the bubble interior and the ambient hydrostatic pressure, despite the occurrence of surface-seal, until the cavity eventually pinches off (frame 8). Consequently, the bubble collapses and rebounds with a rough and rugged appearance (frames 9 and 10), in contrast to the regular-shaped bubbles observed in \autoref{exp_gamma_over_1.36}. A daughter bubble is also observed between the collapsing bubble and the free surface, which comes from the segment separated from the cavity following surface-seal and pinch-off. Notably, no evidence of secondary cavitation derived from the refraction wave reflected by the free surface is observed in the fluid field, suggesting that the amplitude of the collapse-induced pressure wave may have been attenuated due to the ventilation. Given that small cavitation bubbles might not be resolved by the high-speed camera due to large exposure time, we review the complete data set and reveals the same pattern in every non-coalescence case: The cavitation cloud persists for at least two frames, i.e., longer than the interframe time 27.02 \(\upmu\)s. This duration is an order of magnitude above the exposure time (1 \(\upmu\)s), indicating that the camera is sufficient to resolve secondary cavitation bubbles. More detailed dynamics of such microbubbles/nanobubbles refers to \citet{rossello2021demand}. Though the presence or absence of cavitation alone cannot serve as definitive evidence for pressure attenuation, this observation offers valuable insight into how the cavity influences the bubble dynamics. It also motivates further investigation into the underlying cavity behavior. The attenuation of pressure amplitude will be examined in detail in the upcoming section. 

In the second experiment shown in \autoref{exp_gamma_below_1.36}(\textit{b}), \(\gamma=0.71\), the water dome rises higher and faster due to the enhanced interaction between the bubble and the free surface (frame 1). Meanwhile, the decreased standoff parameter and enhanced cavity--bubble interaction lead to coalescence and ventilation significantly earlier than in any of the previous experiments (frame 2). However, after the onset of coalescence and ventilation, the combined effect of hydrostatic pressure, aerodynamic pressure, and the continued ascent of the water dome pinches the cavity neck, detaching the cavity from the bubble, which terminates the ventilation almost immediately. Driven by inertia, the water dome continues to rise rapidly and forms a primary jet that envelops the cavity. The bubble then starts to contract and diminishes the volume shortly after the coalescence and ventilation (frame 3). Apparently, the enormous influx of air substantially alters the internal environment of the bubble, driving the pressure across the bubble wall towards equilibrium. The weakened pressure difference slows down the contraction of the bubble, resulting in a weaker collapse and attenuated pressure wave (frame 4). The bubble and cavity both exhibit a vague and misty appearance upon and after the collapse, which is completely different from the previous experiments, where phenomena such as bursting jet (\autoref{exp_gamma_over_1.36}\textit{a}), collapse luminescence (\autoref{exp_gamma_critical_condition}\textit{b}), and splashes caused by droplets (\autoref{exp_gamma_critical_condition}\textit{c} and \ref{exp_gamma_critical_condition}\textit{d}) can still be recognized. As in the cases of \autoref{exp_gamma_critical_condition}, the majority of the cavity and the top of the bubble are invisible due to the obstruction by the water dome, so it is difficult to discuss the cavity dynamics with experimental results. Therefore, we will numerically elaborate on this case in \cref{section:4}.

In the last experiment (see \autoref{exp_gamma_below_1.36}\textit{c}), \(\gamma\) is reduced to 0.49, where the influence of perturbation on the interaction between the bubble and the free surface becomes negligible. In this case, the standoff parameter is sufficiently small (\(\gamma\lesssim0.5\)) that the bubble can burst directly into the atmosphere, leaving behind an open cavity. This may give rise to a series of rich dynamic processes, such as air entry and sealed splash \citep{wangSplashingSealingEjecta2024}. Under such circumstances, the influence of the surface perturbation becomes insignificant, without whose existence the bubble will burst and ventilate nonetheless. Previous studies such as \citet{wang2015bursting} and \citet{liBubbleInteractionsBursting2019} have thoroughly discussed this phenomenon; thus, we will not delve into this scenario in the present study.

\section{Comparison between experimental and simulation results}\label{section:4}
In this section, we compare the simulation results with three typical experiments discussed in \cref{section:3}. The FVM and VOF approaches are employed to reproduce the experimental observations. In addition, BI simulations \citep{hanInteractionCavitationBubbles2022a} are conducted for non-coalescence cases (\(\gamma\gtrsim1.36\)).

\subsection{Non-coalescence}\label{section:4.1}
\begin{figure}
	\centering
	\includegraphics[width=1\textwidth]{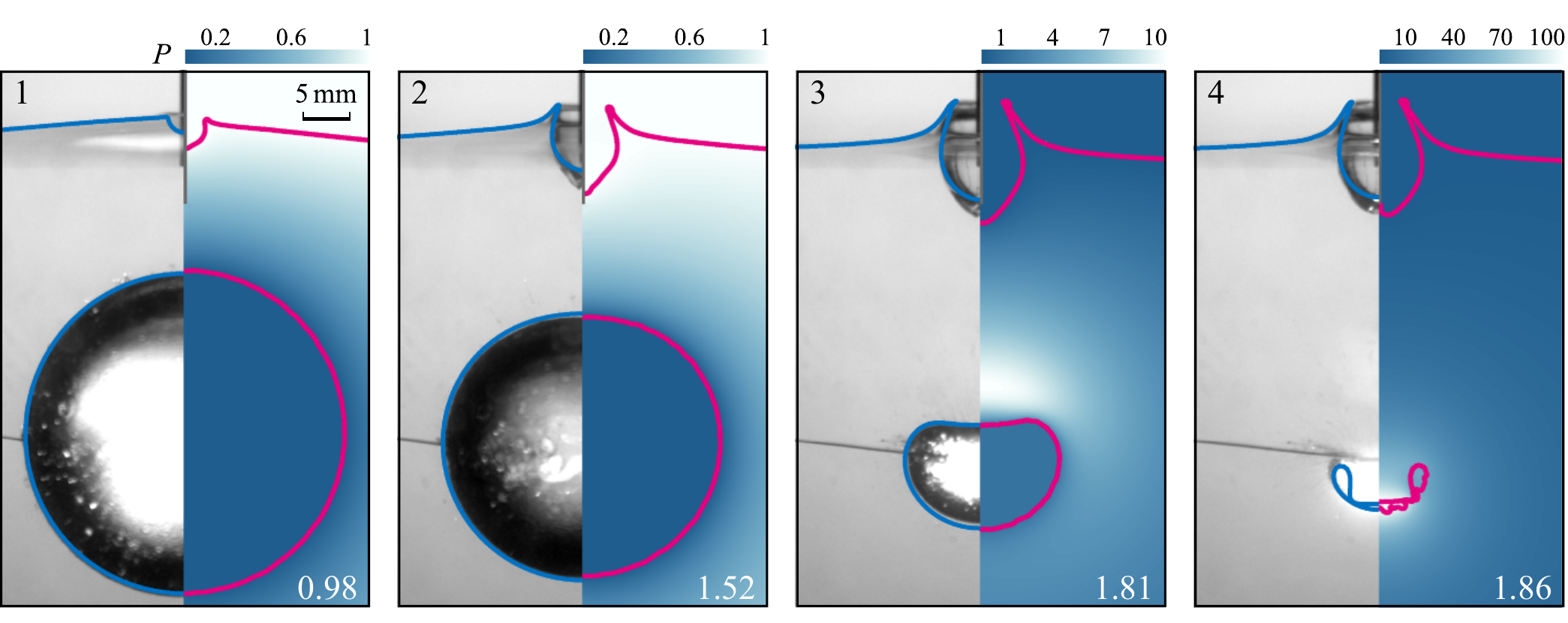}
	\setlength{\abovecaptionskip}{2mm}
	\captionsetup{width=1\textwidth}
	\caption{Comparison of cavity and bubble evolution between numerical simulation and experiment for the same case as in \autoref{exp_gamma_over_1.36}(\textit{a}). The result obtained from the FVM simulation is presented in pressure contours and magenta solid lines in the right half, while the BI simulation results is plotted with blue solid lines in the left half, along with the experimental observations. The non-dimensional parameters in this simulation are set as \(\gamma=1.80\), \(R_0=0.1623\), \(\varepsilon=125\), and \(\kappa=1.4\). All the non-dimensional parameters for the numerical simulations in \cref{section:4} are the same except for \(\gamma\). The timescale is 1.69 ms.}
	\label{exp_num_comparison_1}
	\vspace{-2mm}
\end{figure}

\Autoref{exp_num_comparison_1} presents a comparison between the experiment shown in \autoref{exp_gamma_over_1.36}(\textit{a}) and the corresponding numerical simulation results. The pressure field (colored contour faces) and interface (magenta lines) obtained from the FVM simulation and the BI simulation results (blue lines) are given. The FVM simulation results exhibit good agreement with the experimental observations, which further validates our numerical model. Frame 1 illustrates the bubble-free-surface interaction when the bubble reaches its maximum radius. A small conical depression can be observed as the water dome starts to rise. Once the bubble starts to contract, the cavity grows downward, driven by the pressure gradient (frame 2) and reaches its maximum length \(h_c\) (frame 3). Due to the large \(\gamma\), the cavity can barely exert a notable influence on the bubble. As a result, the bubble collapses with a relatively undisturbed shape, generating a pressure peak of approximately 100 in the fluid field (frame 4). Apparently, the bubble has not yet collapsed to the minimum volume when the cavity reaches its maximum length and there exists a hysteresis between these two crucial moments. In contrast, the BI simulation results exhibit a slight discrepancy with both experimental observations and numerical results in the early stage of the cavity evolution (frames 1 and 2). Nonetheless, both numerical simulations successfully reproduce the overall characteristics of the cavity.

\begin{figure}
	\centering
	\includegraphics[width=0.9\textwidth]{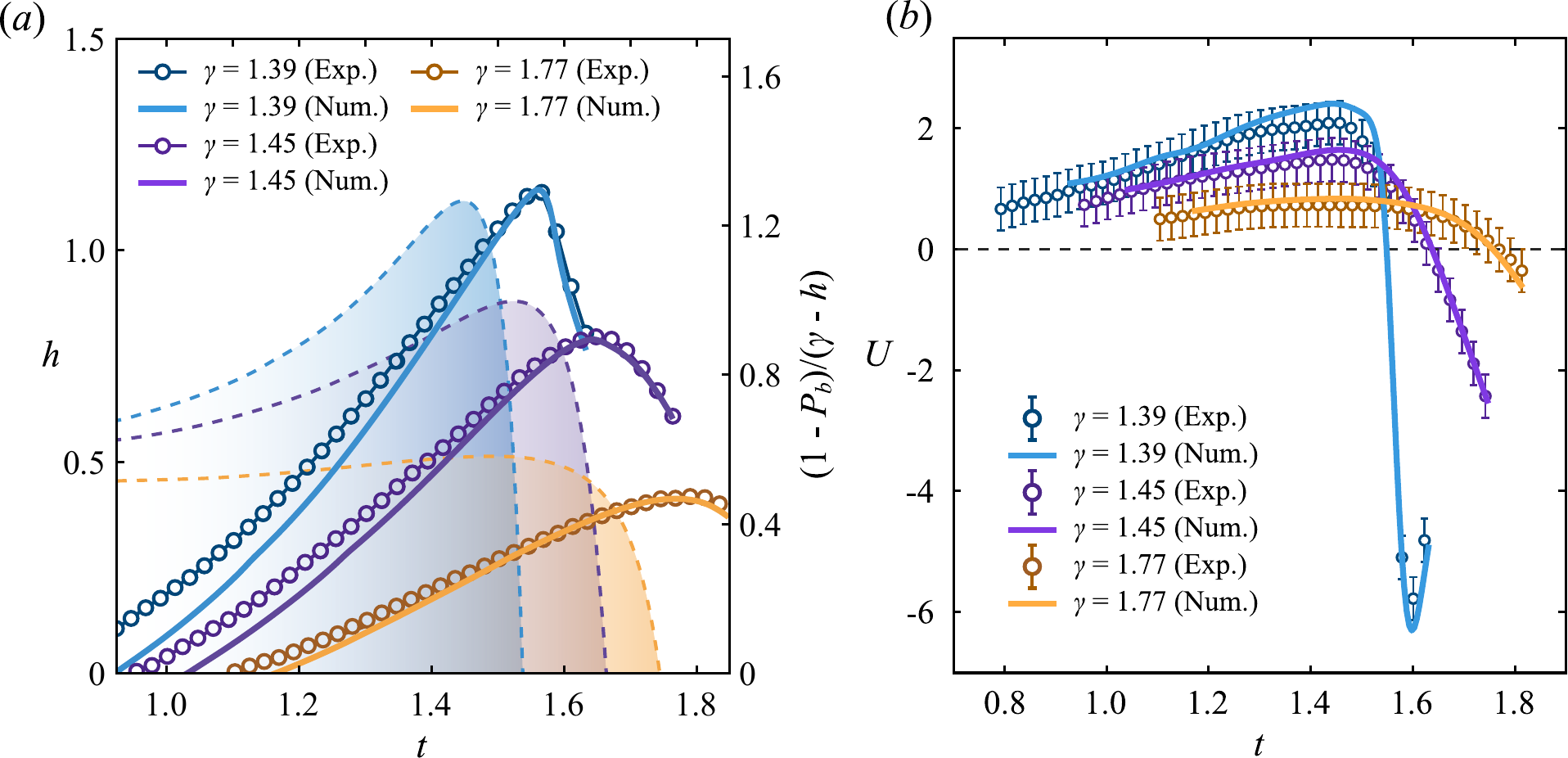}
	\setlength{\abovecaptionskip}{5mm}
	\captionsetup{width=1\textwidth}
	\caption{Comparison between simulation results and experimental data for the evolution of (\textit{a}) cavity length \(h\) and (\textit{b}) cavity velocity \(U\) when \(\gamma=1.77\), \(\gamma=1.45\), and \(\gamma=1.39\). Solid lines represent simulation results, and circular markers denote experimental data. An inflection point is observed in both the \(h\) and \(U\) curves for each case, indicating a deceleration and rebound of the cavity during the final stage of bubble collapse. In panel \textit{a}, the evolution of the pressure difference between the bubble and atmosphere is also plotted (dotted lines), scaled by \(P_\infty\). The sharp drop in pressure difference coinciding with the cavity rebound suggests that the pressure gradient is a key factor in the cavity dynamics. Cavity velocity \(U\) are scaled by \(\sqrt{P_\infty/\rho_l}\). The experimental uncertainties of \(h\) and \(U\) are approximately 0.004 and 0.36 (0.076 mm and \(3.6\ \mathrm{m\,s}^{-1}\) in dimensional form), respectively.}
	\label{exp_num_comparison_hc_vc}
	\vspace{-3mm}
\end{figure}

To further discuss this phenomenon, we compare three simulation results with corresponding experimental data for the evolution of non-dimensional cavity length \(h\) and velocity \(U\) before the bubble reaches its minimum volume, as shown in \autoref{exp_num_comparison_hc_vc}. The standoff parameter of the three cases are \(\gamma=1.77\), \(\gamma=1.45\) and \(\gamma=1.39\), respectively. The cavity velocity \(U\)\ is determined using a five-point local temporal fitting procedure: The cavity bottom position is extracted from each frame, and \(U\)\ is obtained from the local slope of a five-frame linear fit centered at each time instant. Due to the limited spatial resolution of the high-speed camera, the cavity bottom position exhibits a spatial uncertainty of one pixel (0.076~mm). Although \(R_m\) varies from different experiments (ranges from 17 to 20 mm), the non-dimensional uncertainty can be estimated to \(\approx0.004\). Consequently, the uncertainty in \(U\) is estimated to be \(3.6~\mathrm{m\,s}^{-1}\) (\(0.36\) non-dimensional). To facilitate our discussion, the cavity velocity is taken as positive when directed towards the bubble. We use the ratio between the atmosphere--bubble pressure difference and \(\gamma\), denoted by \((1-P_b)/(\gamma-h)\), as a qualitative proxy for the real pressure gradient, since the direction of the pressure gradient changes synchronously with the sign of \(1-P_b\), where \(P_b\) is scaled by \(P_\infty\). This simplified approach helps us analyze the effect of the pressure gradient on cavity dynamics qualitatively. A quantitative analysis of this dependence will be presented in \cref{section:5}. The simulation result of the cavity length (see \autoref{exp_num_comparison_hc_vc}\textit{a}) and velocity (see \autoref{exp_num_comparison_hc_vc}\textit{b}) agree well with the experimental data in the final stage of the bubble collapse and are slightly later than the experimental data in the early stage as we expected. This discrepancy in \(h\) is attributed to the continuous electric discharge discussed in \cref{section:2.3}, which causes the bubble to grow more slowly than in the simulations. As a result, the conical depression around the thin rod forms with a slightly lower altitude and velocity, triggering an earlier cavity downward growth during bubble contraction.

Upon cavity initiation, \(h\) and \(U\) accumulate gradually, while the pressure difference \((1-P_b)/(\gamma-h)\) remains relatively low (around 0.5--0.6). However, as the bubble keeps contracting, the internal pressure of the bubble increases rapidly from a minimum value to exceeding \(P_\infty\) within a transient period (approximately 0.5 ms according to numerical simulations). The increasing internal pressure lowers and finally reverses the pressure gradient between the bubble and the atmosphere. Consequently, this reversed gradient causes the cavity to decelerate, eventually bringing its velocity to zero and triggering a rebound. The deceleration and rebound process occurs almost simultaneously with the plunge of the pressure difference and takes place before the bubble reaches its minimum volume, which explains the aforementioned hysteresis in the cavity--bubble interaction. This hysteresis, along with the close synchronization between the cavity evolution and the pressure gradient changes, suggests that the pressure gradient between the atmosphere and the bubble interior governs the cavity evolution.

We also notice that both consistency and inconsistency exist in the cavity evolution shown in \autoref{exp_num_comparison_hc_vc}. On the one hand, the rebound velocity of the cavity increases as \(\gamma\) decreases, which aligns with the experimental observations in \autoref{exp_gamma_over_1.36}: When \(\gamma\) decreases, the cavity--bubble interaction becomes stronger, resulting in a larger \(h_c\) and a slender cavity shape with a tapered bottom (see \autoref{exp_gamma_over_1.36}\textit{b}, frame 4). The tapered bottom of the cavity creates a high local curvature, which leads to an earlier, faster, and thinner bursting jet than that in \autoref{exp_gamma_over_1.36}(\textit{a}) \citep{terasaki2024interaction}. On the other hand, though the rebound of the cavity occurs almost synchronously with the plunge of the pressure difference in the case of \(\gamma\) = 1.77 and 1.45, a slight temporal lag of cavity rebound can still be observed in the case of \(\gamma\) = 1.39. This deviation may be attributed to the close distance between the cavity and the bubble at a smaller \(\gamma\). In this case, the cavity evolution is no longer governed solely by the pressure gradient; flow-focusing effects become significant during the final collapse stage, which opposes the reversed pressure gradient, stretching the cavity bottom and intensifying the nonlinear interaction with the bubble. As a result, the rebound process is postponed.

\subsection{Critical condition}\label{section:4.2}
\begin{figure}
	\centering
	\includegraphics[width=1\textwidth]{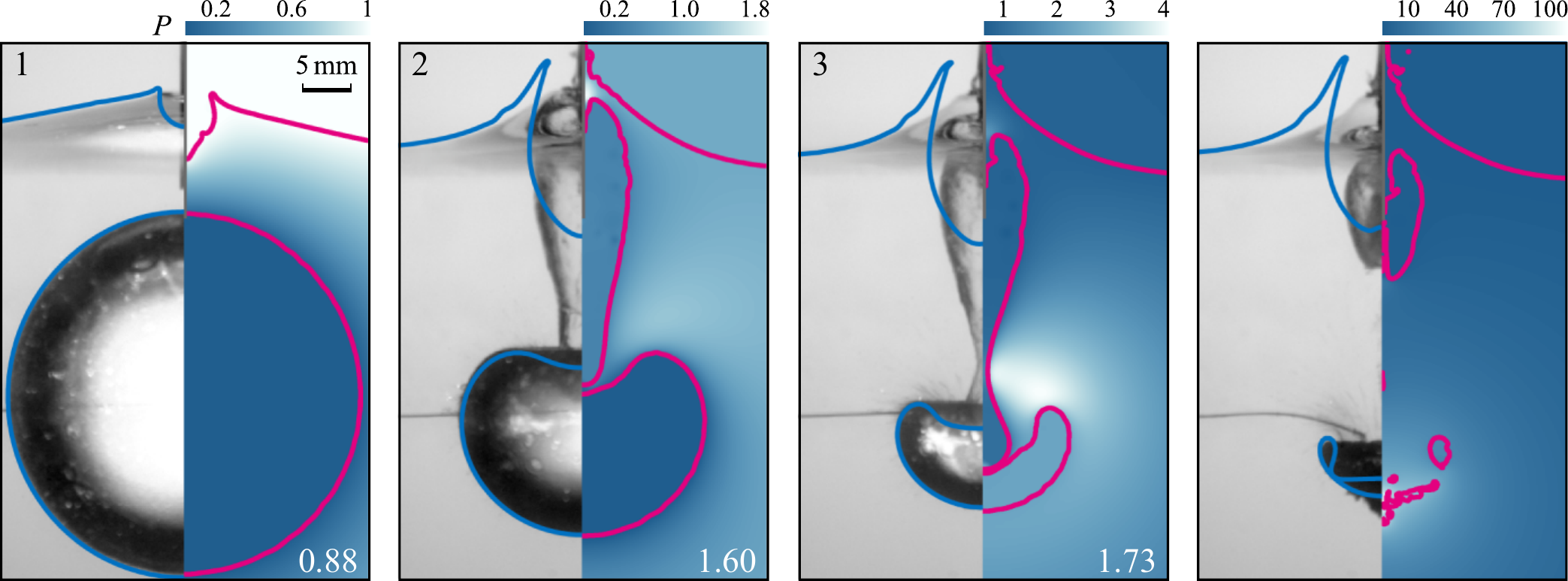}
	\setlength{\abovecaptionskip}{-1mm}
	\captionsetup{width=1\textwidth}
	\caption{Comparison of cavity and bubble evolution between numerical simulation and experiment for the same case as in \autoref{exp_gamma_critical_condition}(\textit{a}). Simulation results obtained from FVM (magenta lines) and the BI method (blue lines) are both plotted. FVM simulation results exhibit better agreement with the experimental observations than the BI method. The standoff parameter is set as \(\gamma=1.36\). The timescale is 1.84 ms.}
	\label{exp_num_comparison_2}
	\vspace{-1mm}
\end{figure}

\Autoref{exp_num_comparison_2} shows the comparison among the results obtained with FVM, BI simulation, and the experimental observations of the critical condition case in \autoref{exp_gamma_critical_condition}(\textit{a}). The simulation enables a more detailed examination of the cavity--bubble interaction characteristics, which we fail to have a clear sight in the experiments. As can be seen in \autoref{exp_num_comparison_2} (frame 2) the upper part of the bubble is depressed inward deeper due to the existence of the cavity, comparing to the BI simulation results, and forms an internal bulge. As the bubble keeps contracting, the cavity elongates downwards and the bulge keeps developing (frame 3). Different from the cases in \autoref{exp_num_comparison_1} where the cavity is far from the bubble, the cavity stretches downwards through the high-pressure zone above the bubble in the case of the critical condition. Affected by the high local pressure, the cavity starts to shrink in the middle. This explains why the cavity is squeezed and forms a thin neck near the bubble instead of elsewhere. After pinch-off, the localized high pressure near the tips initiates a rebound motion of the segments in opposite directions, producing a bursting jet reminiscent of a Worthington jet \citep{worthington1897}. The upper segment evolves into an isolated bubble that rises and bursts upwards, while the lower one merges with the primary bubble and collapses, producing a collapse pressure of \(P\approx\) 100 (frame 4). The neck pinch-off also creates rapid jets pointing oppositely inside the cavity. According to numerical simulations, the maximum jet tip velocity can exceed 300~\(\textrm{m\,s}^{-1}\). Owing to the roughened cavity surface caused by droplet impact and the mist-like appearance of the jet tip (frame 4), a direct, accurate measurement of the jet velocity after pinch-off is not feasible. As a practical alternative, we track the cavity bottom point between the frames immediately at (frame 3) and after pinch-off (frame 4), yielding an experimental cavity bottom velocity of 184.5 \(\mathrm{m\,s}^{-1}\). The same procedure applied to the numerical data gives 186.0 \(\mathrm{m\,s}^{-1}\), confirming consistency. Clearly, the actual jet tip moves faster than either value, and the procedure deliberately underestimates the true jet speed.

As discussed in \cref{section:3.2}, although coalescence between the lower segment of the cavity and the bubble occurs in the critical condition, cavitation induced by the rarefaction wave can still be observed in the fluid field (see \autoref{exp_gamma_critical_condition}\textit{a}, frame 5). Consequently, the peak pressure produced by bubble collapse remains comparable to that in \autoref{exp_num_comparison_1}, suggesting that the coalescence between a small segment of the cavity and the bubble alone does not significantly affect the collapse intensity. Therefore, stable ventilation that sustains for an sufficiently long duration of time is essential to achieve a significant attenuation in collapse pressure, which will be further examined in the next section.

Additionally, numerical results obtained with the BI simulation are also brought into comparison (plotted with blue solid lines on the left half). Within our expectations, the BI simulation results exhibit vast deviation from both experimental observations and FVM simulation results with respect to cavity characteristics in this scenario, while the bubble behavior has no major difference except for the internal bulge caused by the cavity growth. This discrepancy derives from several aspects. First, the BI model is incapable of simulating the boundary layer around the thin rod, whose influence may not be significant, but still can affect the cavity dynamics to a certain extent (the influence of the boundary layer is referred to \cref{section:5.4}). Second, which is the most significant, the air phase in a typical two-phase BI simulation is treated as an incompressible and inviscid ideal fluid. Such assumptions are feasible in the general free-surface--bubble interactions, since the air velocity above the free surface is not that high. However, when it comes to the critical condition case, the cavity grows rapidly and induces rapid air flow into the cavity, subject to the pressure difference. According to FVM simulation results, the velocity of the air flow inside the cavity can reach approximately 200--280 \(\mathrm{m\,s}^{-1}\) (\(\textit{Ma}\approx\) 0.6--0.8), and maximizes to around 300 \(\mathrm{m\,s}^{-1}\) at the cavity opening. Due to the high air velocity, the gas density and pressure inside the cavity drop by approximately 25--30 \textit{\%} compared to those of the atmosphere. This pressure plunge creates another gradient between the cavity interior and the atmosphere, thereby pumping more air into the cavity and elongating it. Such rapid air inflow may also lead to choked flow at the cavity opening, resulting in rich aerodynamics \citep{lee1997cavity,gekleSupersonic}. However, our numerical simulations show maximum gas velocities below the sonic threshold (\(\approx 280\ \mathrm{m\,s^{-1}}\), corresponding to \(\textit{Ma}\approx0.8\) at ambient conditions), as current mesh resolution is insufficient to fully capture the aerodynamics at the moment of surface seal. All the above evidence suggests that the compressibility effects of the air become prominent and cannot be neglected under such circumstances, which play a significant role in the cavity evolution \citep{petersAirFlowCollapsing2013a,kuangAirflowEffectsEvolution2025d}. However, this exceeds the capability of the BI model, thereby introducing deviations in cavity dynamics.

\subsection{Coalescence}\label{section:4.3}
\begin{figure}
	\centering
	\includegraphics[width=0.95\textwidth]{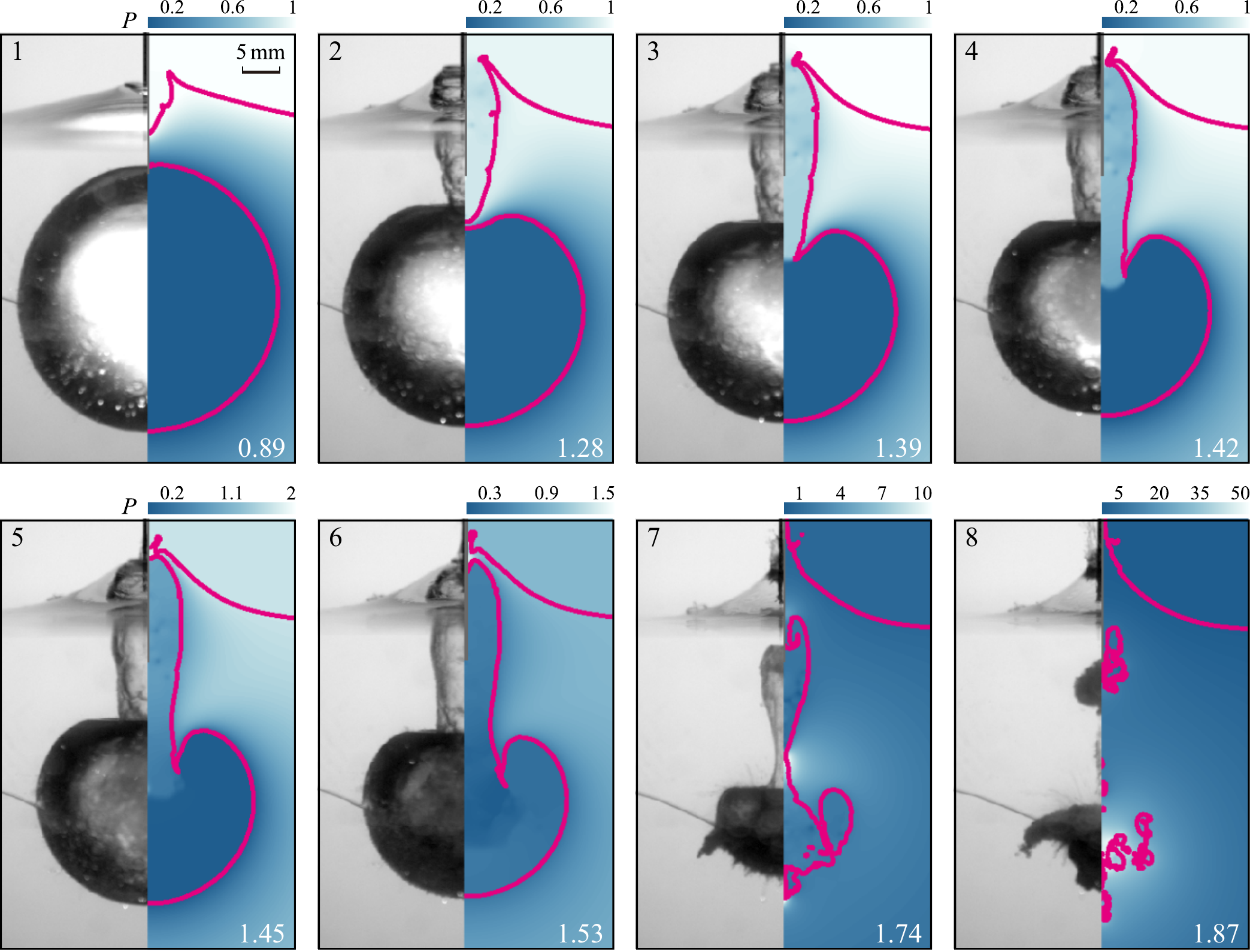}
	\setlength{\abovecaptionskip}{5mm}
	\captionsetup{width=1\textwidth}
	\caption{Comparison of cavity and bubble evolution between numerical simulation and experiment for the same case as in \autoref{exp_gamma_below_1.36}(\textit{a}). Frames 3, 5, and 7 show the three critical moments: (I) coalescence initiation, (II) surface-seal, and (III) cavity pinch-off. The standoff parameter is set as \(\gamma=1.30\). The timescale is 1.74 ms.}
	\label{exp_num_comparison_3}
	\vspace{0mm}
\end{figure}

When \(\gamma\lesssim1.36\), the coalescence and ventilation between the cavity and the bubble become inevitable, giving rise to a series of complex cavity--bubble interaction dynamics. \Autoref{exp_num_comparison_3} presents a comparison between the simulation results and the experimental observations in \autoref{exp_gamma_below_1.36}(\textit{a}) at several representative moments, to illustrate the entire coalescence-ventilation process.

The numerical simulation enables us to have a clearer vision of the cavity--bubble coalescence process. The rapidly growing cavity pushes aside the fluid between the cavity and the bubble. Then the cavity bottom coalesces with the bubble, establishing a channel between the bubble and the atmosphere. At this critical moment, the pressure inside the bubble is significantly lower than that of the atmosphere. The pressure gradient pumps air into the bubble (frames 3--4). This causes prominent air compressibility effects inside the cavity, including pressure and density plunge, as we discussed in the last section. Then the pressure difference between the cavity interior and the atmosphere triggers the surface-seal and terminates the ventilation almost immediately, shortly after the coalescence starts (frame 5). As a result, the entire ventilation process is transient and only lasts for only about 1.05 ms (frames 3--5) according to FVM simulation results. However, the transient ventilation process still leads to pressure increase inside the bubble, which can be observed from frames 2 to 5. Therefore, it is reasonable to suggest that the increased bubble pressure attenuates the collapse intensity, thereby eliminating secondary cavitation across the fluid field after the bubble reaches its minimum volume. After surface-sealing terminates the ventilation process, cavity--bubble coalescence persists (frame 6) until the cavity pinches off in the middle (frame 7), forming two cavity segments. Although the simulation results deviate slightly from the experiment observations in terms of cavity and bubble morphology after the cavity pinch-off, the overall dynamics of cavity--bubble interactions are well-captured.

\begin{figure}
	\centering
	\includegraphics[width=0.95\textwidth]{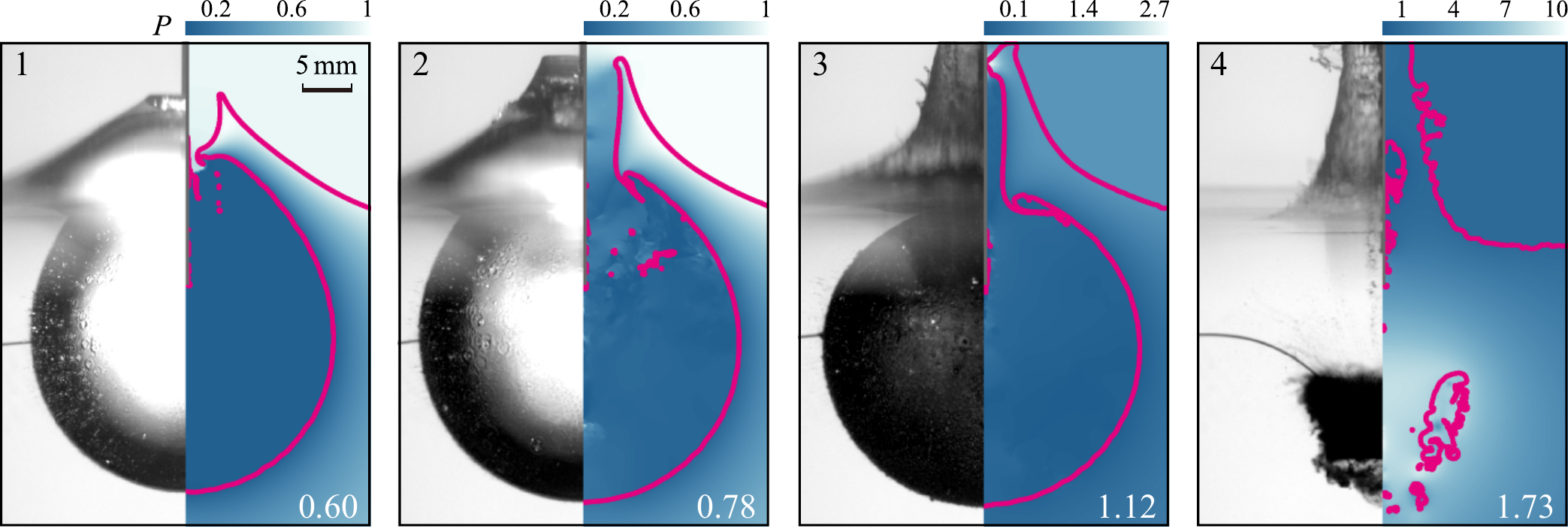}
	\setlength{\abovecaptionskip}{5mm}
	\captionsetup{width=1\textwidth}
	\caption{Comparison of cavity and bubble evolution between numerical simulation and experiment for the same case as in \autoref{exp_gamma_below_1.36}(\textit{b}). The standoff parameter is set as \(\gamma=0.72\). The timescale is 1.67 ms.}
	\label{exp_num_comparison_4}
	\vspace{-2mm}
\end{figure}
\begin{figure}
	\centering
	\includegraphics[width=1.0\textwidth]{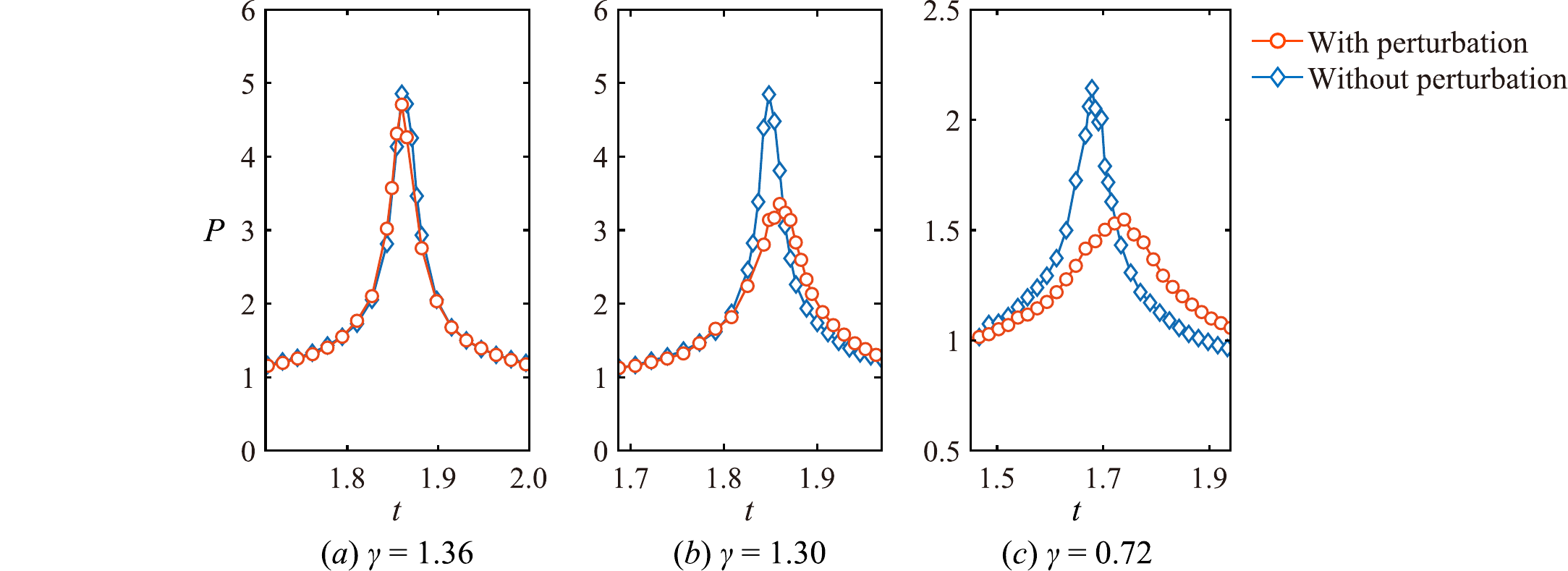}
	\setlength{\abovecaptionskip}{0mm}
	\captionsetup{width=1\textwidth}
	\caption{Comparison of the fluid field pressure between the simulations with (red circles) and without (blue diamonds) surface perturbation. A pressure monitoring point is located horizontally at a distance \(R_m\) from the bubble inception location. (\textit{a}) Critical condition in \autoref{exp_num_comparison_2}. No significant attenuation can be observed. (\textit{b}) An evident attenuation in the fluid pressure field is observed in the case of \autoref{exp_num_comparison_3}, suggesting that coalescence and ventilation can attenuate the collapse intensity. (\textit{c}) The collapse intensity is further attenuated by the coalescence and ventilation in \autoref{exp_num_comparison_4}.}
	\label{pressure_field_monitor}
	\vspace{-2mm}
\end{figure}

As \(\gamma\) keeps decreasing to 0.72, the coalescence and ventilation occur even earlier and more intensely. As shown in \autoref{exp_num_comparison_4}, the simulation results agree well with the experimental observation on the morphology of the bubble and cavity. As we expected, the coalescence occurs even earlier than the previous case due to small \(\gamma\) (frame 1), before the bubble starts to contract. As a result, a more stable channel is established between the bubble and the atmosphere (frame 2), lasting much longer than the case in \autoref{exp_num_comparison_3}. The elongated ventilation allows a large amount of air and droplets influx into the bubble, resulting in a misty bubble shape during the collapse stage (frame 4), instead of a transient collapse with strong luminescence.

One may also notice that the maximal \(P\) when the bubble reaches its minimum volume in \autoref{exp_num_comparison_3} and \autoref{exp_num_comparison_4} are attenuated to around 50 and 10, respectively, compared to \(P\approx\) 100 in \autoref{exp_num_comparison_1} and \ref{exp_num_comparison_2}. This attenuation originates from the ventilation process based on the discussions above. To justify this argument, we compare the numerical results with and without surface perturbations by examining the fluid field pressure of a monitoring point positioned horizontally to the bubble inception point at a distance of 2\(R_m\). Three groups of comparisons are presented in \autoref{pressure_field_monitor}, corresponding to the critical condition in \cref{section:4.2} and two coalescence scenarios in this section. As shown in \autoref{pressure_field_monitor}, when only the coalescence between cavity and bubble occurs, the fluid field pressure in the critical condition is barely attenuated, differing by 2.1 \textit{\%} compared to the case without surface perturbation. On the contrary, the peak pressure is attenuated to approximately 70 \textit{\%} in the two cases where ventilation between the bubble and the atmosphere takes place. These comparisons vividly demonstrate that it is the bubble-atmosphere ventilation that causes the collapse attenuation.

\section{More discussions on cavity dynamics}\label{section:5}
In this section, we further investigate the dependence of cavity dynamics on the standoff parameter as well as other factors that may influence cavity behaviors. In \cref{section:5.1}, we first present a parameter map for the dependence of different cavity behaviors on \(\gamma\) and \(h_m\). Then, a fitted power-law relationship between the maximum cavity length \(h_c\) and the standoff parameter \(\gamma\) is brought forward. A scaling analysis based on classical spherical bubble theory is then performed in \cref{section:5.2} to evaluate this dependence relationship quantitatively. Then in \cref{section:5.3}, we investigate the influence of the characteristic height of the surface perturbation \(h_m\) on cavity dynamics via numerical simulations, and interpret the simulation results through nonlinear Rayleigh--Taylor theory. Finally, \cref{section:5.4} provides a qualitative discussion on the influence of the boundary layer surrounding the thin rod.

\subsection{Overall cavity dynamics}\label{section:5.1}
\begin{figure}
	\centering
	\includegraphics[width=0.65\textwidth]{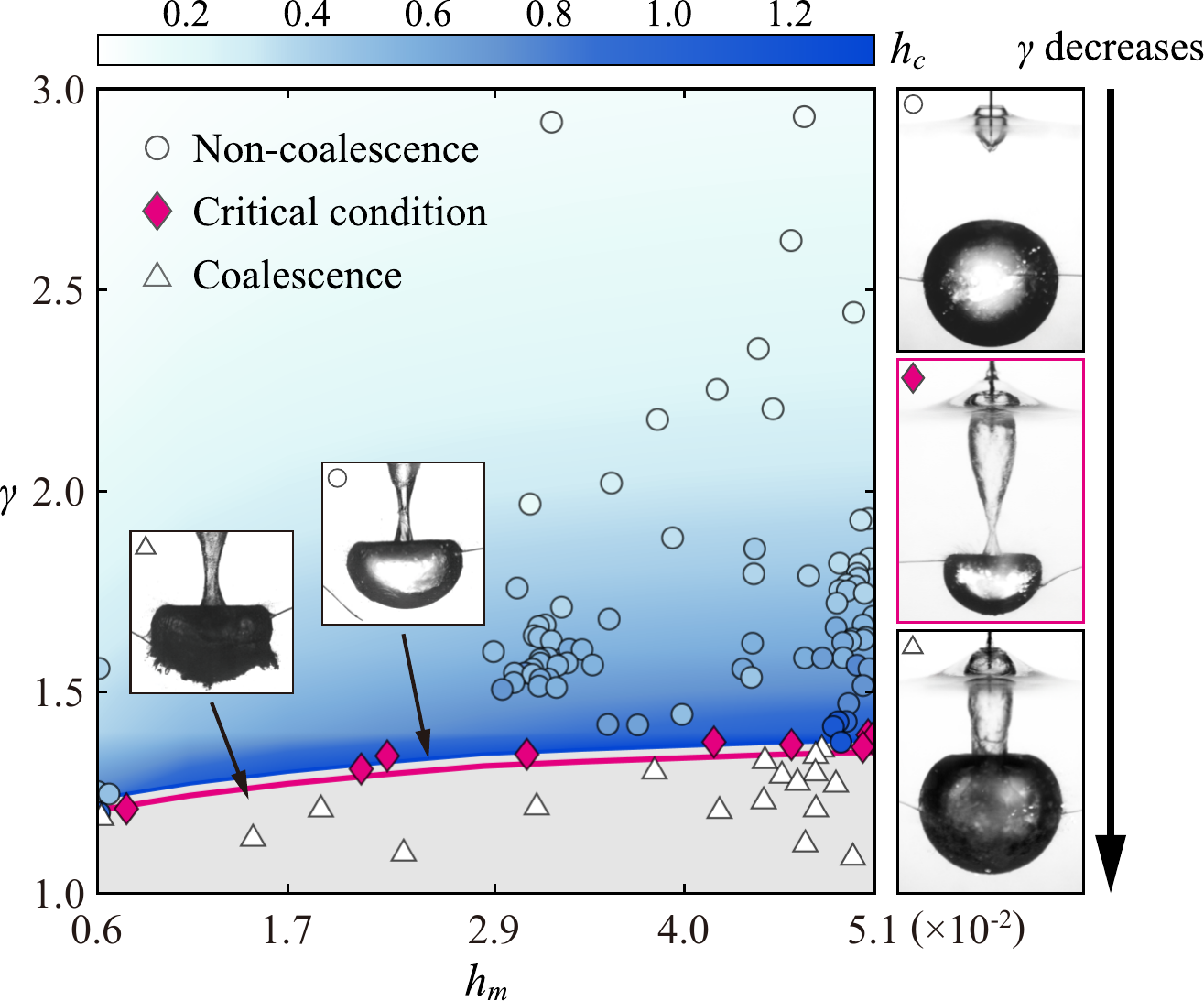}
	\setlength{\abovecaptionskip}{3mm}
	\captionsetup{width=1\textwidth}
	\caption{Dependence of cavity length on governing parameters of \(\gamma\) and \(h_m\). The cavity lengths obtained with numerical simulation are presented with the blue contours. The \(\gamma\) value of critical conditions are plotted with the magenta line. The experimental observations of different cavity behaviors are plotted with circles (non-coalescence), diamonds (critical condition), and triangles (coalescence), respectively. The experimental uncertainty of \(h_c\) is approximately 0.004 (0.076 mm in dimensional form). The corresponding experimental images for the three different cavity--bubble interaction scenarios are presented on the right.}
	\label{parameter_map_hm_gamma}
	\vspace{-3mm}
\end{figure}

\Autoref{parameter_map_hm_gamma} depicts the dependence of different cavity behaviors on \(\gamma\) and \(h_m\). The simulation results obtained with FVM is plotted in blue contours. It can be clearly seen that \(h_c\) is positively correlated with \(h_m\), while negatively with \(\gamma\). Different cavity behaviors in the experimental observations are plotted with circles, diamonds, and triangles, representing cavity--bubble interaction scenarios of non-coalescence, critical condition, and coalescence. The experimental circles share the same color gradient as the contours. It is worth emphasizing that the color distribution of the experimental results is almost the same as the simulation results, and so is the distribution of different cavity behaviors in the parameter map, suggesting that the experimental results of the coalescence cases agree well with the numerical simulation, which further demonstrates the effectiveness of our numerical model.

Moreover, the critical conditions obtained with FVM simulations are presented with a magenta line. As we expected, the critical \(\gamma\) is also correlated positively with \(h_m\). One may easily notice that there exists a gap between the non-coalescence contours and the critical line. This gap corresponds to the case in \autoref{exp_gamma_critical_condition}(\textit{b}), where the cavity stretches and surpasses the bubble top yet does not coalesce with the bubble. Under such circumstance, the cavity length cannot be measured.

\begin{figure}
	\centering
	\includegraphics[width=0.85\textwidth]{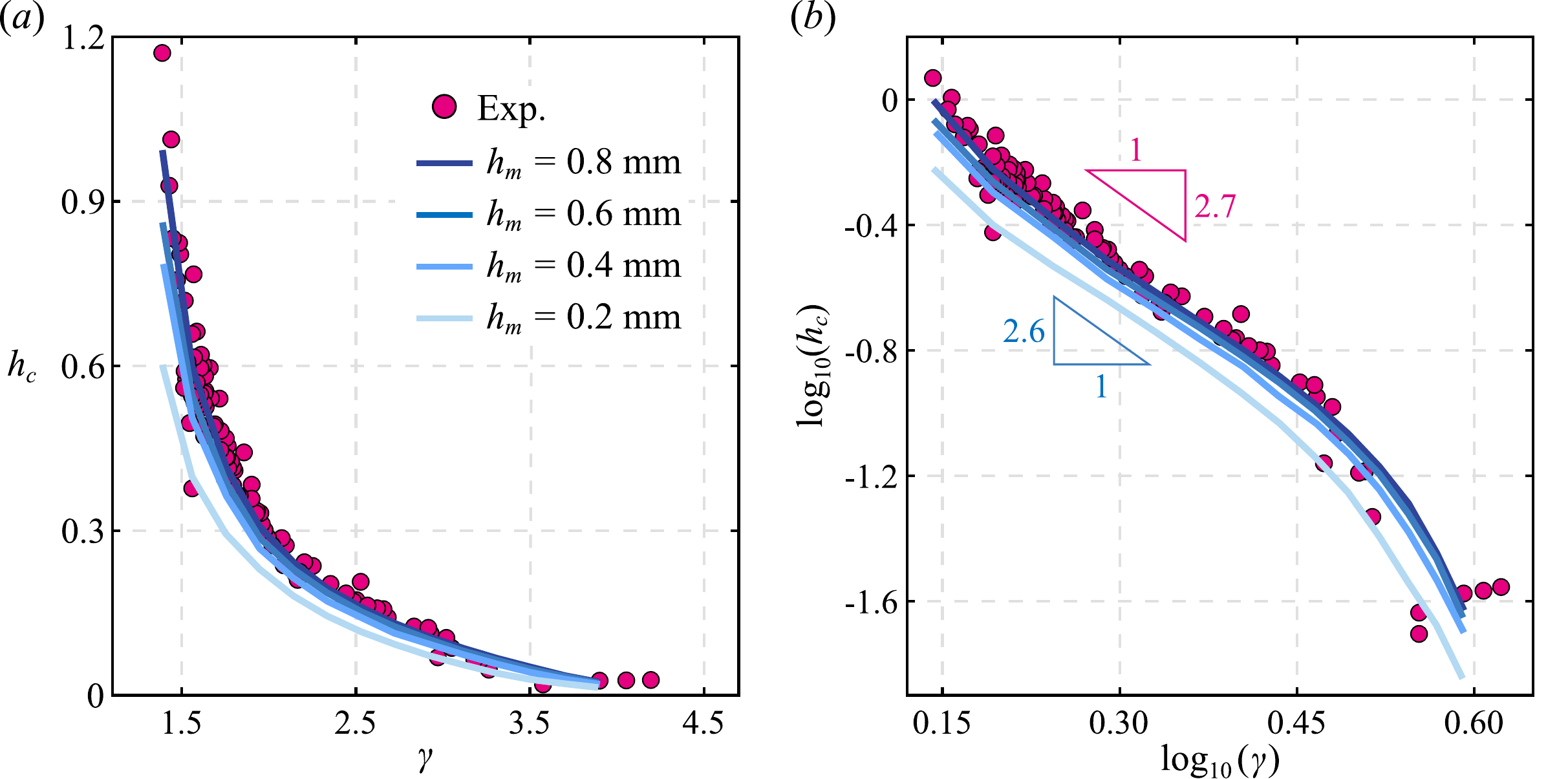}
	\setlength{\abovecaptionskip}{3mm}
	\captionsetup{width=1\textwidth}
	\caption{(\textit{a}) Maximum cavity length \(h_c\) as a function of the standoff parameter \(\gamma\). Magenta circles correspond to experimental data with \(h_m\) ranging from 0.2 to 0.8 mm, while blue solid lines represent simulation results. \(h_c\) decreases monotonically with increasing \(\gamma\). (\textit{b}) The same data in (\textit{a}) are presented in a doubly logarithmic plot. The plot within the range of \(1.5\lesssim\gamma\lesssim3\) reveals a power law of \(h_c\propto\gamma^\alpha\), with fitted exponents \(\alpha=-2.7\) (experiments) and \(-2.6\) (simulations). The experimental uncertainty of \(h_c\) is approximately 0.004 (0.076 mm in dimensional form).}
	\label{gamma_hc_diagram}
	\vspace{-3mm}
\end{figure}

\Autoref{gamma_hc_diagram} presents the variation of the maximum cavity length \(h_c\) with respect to the standoff parameter \(\gamma\). A comparison is performed between the simulation results (blue solid lines) and the experimental data (magenta circles). Overall, \(h_c\) decreases monotonically as \(\gamma\) increases, consistent with the experimental observations discussed in \cref{section:3}. Interestingly, the experimental data exhibit remarkable consistency with varying \(h_m\) (from 0.8 mm to 0.2 mm), suggesting that the size of the surface perturbation plays a secondary role in cavity dynamics. To further verify this trend, numerical results obtained from FVM simulations with \(h_m\) ranging from 0.8 mm to 0.2 mm are included for comparison. The simulation results show excellent agreement with the experiments, which demonstrates the reliability of the numerical approach and motivate further investigation. As discussed in \cref{section:3}, nonlinear cavity--bubble interactions intensify and the flow-focusing effect becomes prominent when \(\gamma\lesssim1.5\). Consequently, the cavity length \(h_c\) increases sharply, and the slope of the \(\gamma\textrm{--}h_c\) curve steepens in this regime. A doubly logarithmic plot of the same datasets shown in \autoref{gamma_hc_diagram}(\textit{a}) is presented in \autoref{gamma_hc_diagram}(\textit{b}). In the range \(1.5\lesssim\gamma\lesssim3\), the data follow a power-law relationship of the form \(h_c\propto\gamma^\alpha\), with a fitted exponent \(\alpha=-2.7\) for the experimental data (magenta circles) and \(\alpha=-2.6\) for the simulation results.

The intriguing consistency of \(h_c\) versus \(\gamma\), despite the variation in \(h_m\), motivates a further investigation into the underlying mechanism governing the cavity dynamics. As discussed in \cref{section:3}, the entire life cycle of the cavity is closely coupled with the bubble’s collapse stage. Specifically, though the small depression is initiated and develops with the bubble expansion, the cavity does not start to grow downward rapidly until the bubble reaches its maximum radius. Conventional analytical approaches based on Kelvin impulse theory \citep{benjamin1966collapse,blakeNoteImpulseDue1982}, which has been successfully adopted in several previous studies of liquid jet \citep{hanInteractionCavitationBubbles2022a,kangGravityCapillaryJetlike2019a} is not applicable in this context, as the momentum of the cavity is difficult to estimate. Moreover, the majority of the bubble momentum cannot be transferred to the cavity due to its transient life cycle. Therefore, alternative approaches are needed to analyze the cavity dynamics.

After reviewing the relation between \(h_c\) and \(\gamma\), we notice that the power law of the doubly logarithmic plot in \autoref{gamma_hc_diagram} exhibits slight deviations from the power law relationship at both ends of the parameter range of \(\gamma\). That is to say, the curve within the range of \(1.5\lesssim\gamma\lesssim3\) satisfies the fitted power law with exponent of \(\alpha=-2.6\), while the curve becomes slightly steeper when \(\gamma\lesssim1.5\) and \(\gamma\gtrsim3\). The steeper slope when \(\gamma\lesssim1.5\) can be explained by the aforementioned flow-focusing effect in \cref{section:4.1}, which becomes even more significant as \(\gamma\) decreases to the vicinity of the critical value. This results in the strong nonlinear cavity--bubble interaction, manifested as a tapered cavity bottom. On the other hand, when \(\gamma\gtrsim3\), the free-surface--bubble interactions tend to weaken. The cavity velocity and length are reduced to the order of \(\textit{O}(10^{-2}-10^{-1})\) and \(\textit{O}(10^{-1})\), respectively. As a result, \textit{We} decreases to the order of \(\textit{O}(10^0-10^1)\). Under such circumstances, the effects of surface tension are comparable to those of inertial forces. Meanwhile, the surface tension and viscosity of the fluid tend to resist the acceleration induced by the bubble oscillation and stabilize the free surface, thereby leading to smaller cavities. The above aspects jointly result in a deviated \(h_c\textrm{--}\gamma\) curve from the scaling law.

\subsection{Theoretical modeling of cavity dynamics}\label{section:5.2}
\begin{figure}
	\centering
	\includegraphics[width=1\textwidth]{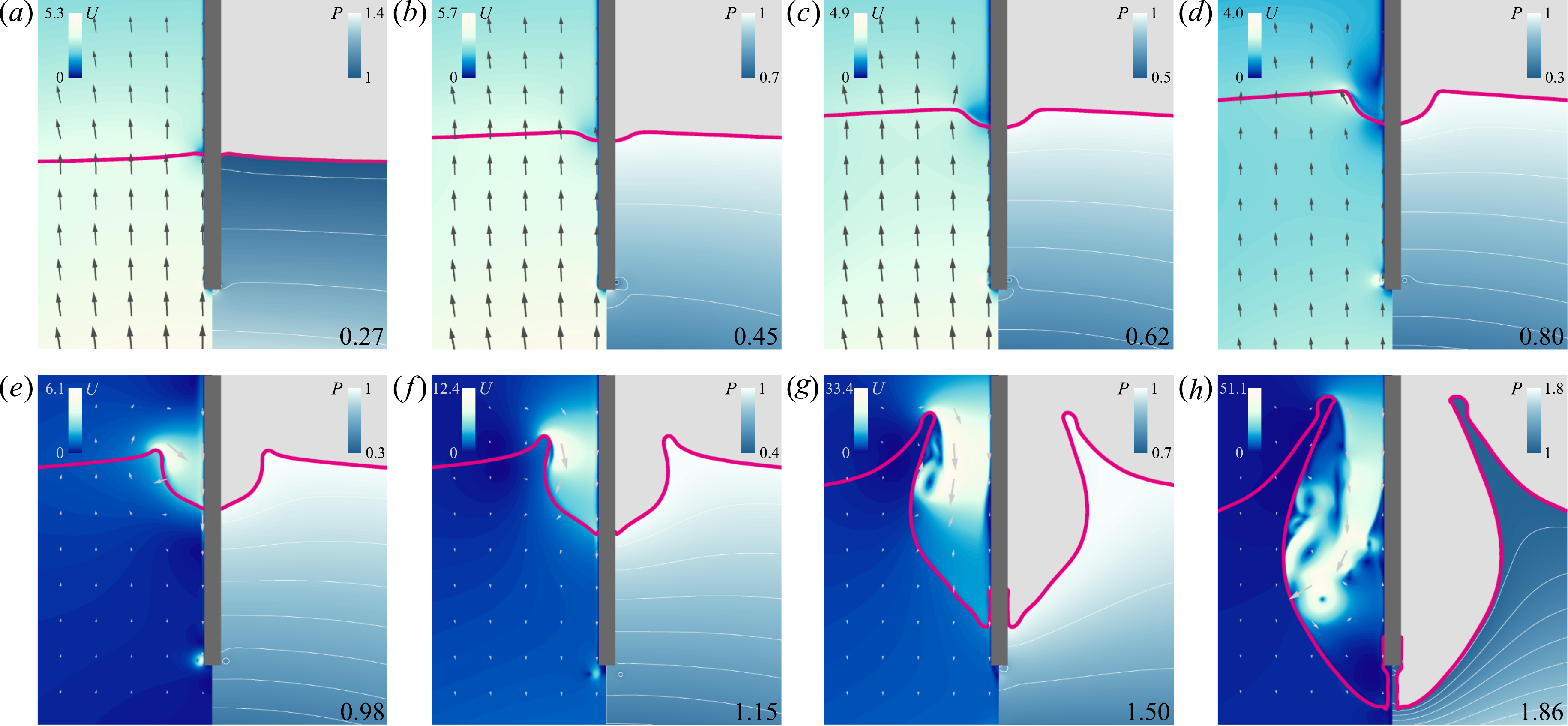}
	\setlength{\abovecaptionskip}{2mm}
	\captionsetup{width=1\textwidth}
	\caption{Flow fields in the vicinity of the cavity. Contours of velocity magnitude (left) and pressure (right) are plotted. The arrows represent the velocity direction, while the black curves represents the contour line of pressure. The free surface is plotted with magenta lines. This is the same case as in \autoref{exp_gamma_over_1.36}(\textit{a}).}
	\label{cavity_velocity_pressure_contour}
	\vspace{0mm}
\end{figure}

To advance fundamental insights into cavity dynamics, the simulation results focusing on the cavity evolution are shown in \autoref{cavity_velocity_pressure_contour}. Velocity and pressure magnitude are presented in contour faces, while the details of the fluid field are given by velocity arrows and pressure contour lines. After the bubble inception, the free surface is slightly elevated and forms a water hump due to the bubble expansion (\autoref{cavity_velocity_pressure_contour}\textit{a}--\ref{cavity_velocity_pressure_contour}\textit{d}). The no-slip boundary condition of the thin rod creates a boundary layer that hinders the fluid motion, causing a bowl-shaped small depression to form atop the water hump. Though the fluid around the thin rod is hindered, the velocity is not significantly smaller than in the outer region. Thus, the depression still moves upwards with the water hump. As a result, the depression remains relatively small during the expansion stage.

After the bubble attains its maximum radius and contracts (\autoref{cavity_velocity_pressure_contour}\textit{e}), the water hump starts to fall and triggers the cavity downward growth (\autoref{cavity_velocity_pressure_contour}\textit{f}). Slight distortion of the pressure field beneath the cavity can be observed, resulting in a higher local pressure gradient, manifested as the twisted and compressed pressure contour lines, which drag the cavity downward and trigger the accelerating process. Consequently, the growing cavity further distorts the pressure field in return. Such a positive feedback mechanism between the pressure gradient and the cavity growth drives a rapid cavity expansion, which is terminated only when the pressure gradient becomes reversed at the final stage of bubble collapse (\autoref{cavity_velocity_pressure_contour}\textit{h}).

\begin{figure}
	\centering
	\includegraphics[width=0.85\textwidth]{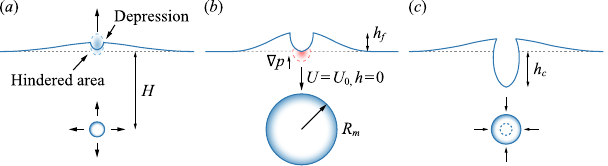}
	\setlength{\abovecaptionskip}{6mm}
	\captionsetup{width=1\textwidth}
	\caption{Schematic of cavity evolution. (\textit{a}) A small bowl-shaped depression forms due to the hindrance of the no-slip boundary condition around the thin rod during the rise of the water hump (blue dotted area). (\textit{b}) The water hump reaches its maximum height \(h_f\) when the bubble reaches its maximum radius. The cavity grows downwards due to the relatively large local pressure gradient (red dotted area) and attains its initial velocity of \(U_0\). (\textit{c}) The cavity reaches its maximum length \(h_c\).}
	\label{cavity_evolution_schematic}
	\vspace{-3mm}
\end{figure}

A schematic for the cavity evolution is shown in \autoref{cavity_evolution_schematic}. As we discussed in \cref{section:5.1} and previous sections, the cavity evolution can be divided into three stages: (I) A small depression forms concurrently with the rise of the water hump due to the bubble expansion (\autoref{cavity_evolution_schematic}\textit{a}). (II) The cavity is initiated to expand after the bubble reaches its maximum radius (\autoref{cavity_evolution_schematic}\textit{b}). (III) The cavity stretches rapidly downwards and attains its maximum length slightly before the bubble reaches its minimum volume (\autoref{cavity_evolution_schematic}\textit{c}).

According to the definition in \cref{section:2.2}, the cavity length \(h\) starts to be recorded only upon the cavity's arrival at the original horizontal surface \(z=0\). However, the growth of the cavity starts before this critical moment, which means that the cavity has an initial velocity of \(U_0\) when \(h=0\) (see \autoref{cavity_evolution_schematic}\textit{b}). Then the cavity starts to grow downward based on this initial velocity. Therefore, it will be convenient for the following analysis if we can obtain the analytical expression of \(U_0\). However, since the downward cavity growth and the rise of the water hump occur simultaneously during the bubble expansion stage, it is difficult to describe the exact cavity evolution during this period. Therefore, we adopt a simplified model to analyze \(U_0\) theoretically.
	
In order to obtain the analytical expression of \(U_0\), several assumptions are made. Firstly, we assume that the concurrent motion of the rising water hump and the expanding depression can be decoupled into two separated stages: The water hump first rises to the height of \(h_f\), then the cavity is initiated from the crest of the water hump to expand downwards and arrive at the horizontal surface. Here, \(h_f\) represents the maximum height of the water hump in the first bubble cycle. This assumption is reasonable since the cavity does not start to grow rapidly until the water hump and the bubble reach their maximum height/radius. In fact, during most of the rising stage of the water hump, the depression retains a relatively small size. Secondly, we assume that the water hump is undisturbed, with its crest being the starting position for the cavity evolution. According to \citet{yanCavityDynamicsWater2009a}, for relatively fast cavity development (\(\textit{Fr}\gtrsim1\)), the wave and splash effects can be neglected. Finally, the acceleration of the cavity \(a_c\) has the same order as the pressure gradient \(\nabla p\), yielding \(a_c\sim-\nabla p\). This assumption is feasible in the early stage of cavity growth since the pressure field has not been remarkably disturbed yet, as shown in \autoref{cavity_velocity_pressure_contour}(\textit{a})--\ref{cavity_velocity_pressure_contour}(\textit{e}).

When \(\gamma\gtrsim1.5\), the bubble can be treated as spherical during most of the first cycle due to the weak bubble-free-surface interaction \citep{blakeGrowthCollapseVapour1981,chahineInteractionOscillatingBubble1977b} and the influence of the cavity on the bubble in the non-coalescence scenarios can be neglected based on the discussion in \cref{section:4.2}. Hence, a simplified image bubble model is adopted \citep{klaseboer2005dynamics,zhangUnifiedTheoryBubble2023}. The bubble near the free surface can be approximated by a bubble pair symmetric about the free surface. The bubble pair can be described with non-dimensional RP equation, which reads:
\begin{equation}
	R_i\ddot{R_i}+\frac{3}{2}\dot{R_i}^2=\varepsilon\left(\frac{R_0}{R_i}\right)^{3\kappa}-1-\frac{2R_j\dot{R_j}^2+R_j^2\ddot{R_j}^2}{2\gamma},\ i,j=1,2,\ i\neq j.
	\label{rpeq}
\end{equation}
The suffix \(i=1,2\) represent the original and image bubble. In the case of free surface, it is required that \(R_2=R_1\), \(\dot{R}_2=-\dot{R}_1\), and \(\ddot{R}_2=-\ddot{R}_1\). The pressure of an arbitrary point \((r,z)\) in the fluid field induced by the bubble oscillation can be written as:
\begin{equation}
	p\left(r,z\right)=\ \frac{2R_1\dot{R}_1^2+R_1^2\ddot{R_1}}{\sqrt{r^2+\left(\gamma+ z\right)^2}}+\frac{2R_2\dot{R}_2^2+R_1^2\ddot{R_2}}{\sqrt{r^2+\left(\gamma-z\right)^2}}-\frac{R_1^4\dot{R}_1^2}{2\left[r^2+\left(\gamma+ z\right)^2\right]^2}-\frac{R_2^4\dot{R}_2^2}{2\left[r^2+\left(\gamma-z\right)^2\right]^2}+1.
	\label{pressurefluidfield}
\end{equation}
The relative error introduced by the image bubble model can be controlled below 5 \textit{\%} if \(\gamma>1.5\) and 1 \textit{\%} if \(\gamma>2\) according to the previous study of \citet{li2020modelling}.

Since the depression maintains a relatively small size in the bubble expansion stage and the initiation of the depression is transient, the influence that the depression imposes on the free surface can be neglected \citep{yanCavityDynamicsWater2009a}, and we can safely assume that the initiation process occurs when the water hump reaches its maximum height \(h_f\) when the bubble attains its maximum radius \(R_m\).

Building on the discussions above, theoretical solutions of \(h_f\) and \(U_0\) can be obtained henceforth. Velocity of the water hump crest \(U_f\) can be estimated by the fluid velocity at \(\left(z,r\right)=(0,0)\) induced by the real and image bubbles, which reads:
\begin{equation}
	U_f=\frac{R_1^2\dot{R}_1-R_2^2\dot{R}_2}{\gamma^2}.
	\label{waterhumpvelocity}
\end{equation}
Integrate (\ref{waterhumpvelocity}) from 0 to \(t_e\) with respect to time, where \(t_e\) is the time that the bubble expands to its maximum radius, \(h_f\) can be obtained as:
\begin{equation}
	h_f=\frac{2\left(1-R_0^3\right)}{3\gamma^2}.
	\label{hf}
\end{equation}

The acceleration of the cavity can be estimated by the \(z\) component of the pressure gradient derived from (\ref{pressurefluidfield}) on the free surface:
\begin{equation}
	a_c\sim\left. -\frac{\partial p}{\partial z}\right|_{(0,0)}=\frac{1}{\gamma^2}\sum_{i=1,2}\left(2R_i\dot{R}_i^2+R_i^2\ddot{R_i}\right).
	\label{ac}
\end{equation}
The third and fourth high order terms in (\ref{pressurefluidfield}) is neglected in case of relatively large \(\gamma\). It is worth emphasizing that this simplification will fail if the bubble is close to the free surface \citep{li2020modelling}. However, the non-coalescence cases discussed in this section only take place when \(\gamma\gtrsim1.36\), and for the specific regime where the fitted power law exists, \(\gamma\) is large enough to adopt this simplification safely. By substituting (\ref{rpeq}) into (\ref{ac}), \(a_c\) can be written as:
\begin{equation}
	a_c\sim\frac{2}{\gamma\left(2\gamma+1\right)} \sum_{i=1,2}\left[\varepsilon\left(\frac{R_0}{R_i}\right)^{3\kappa}+\frac{1}{2}\dot{R}_i^2-1\right].
	\label{acequivalent}
\end{equation}
 As we mentioned above, the initiation process occurs around the bubble reaching its maximum radius. Under such circumstance, \(R_i\approx R_m\gg R_0\), \(\dot{R_i}\ttz\). Thus, the first two terms in the square brackets can be neglected. By assuming the concavity grows in a uniformly acceleration \(a_c\), we can obtain the theoretical expression of \(U_0\), yielding:
\begin{equation}
	U_0=\sqrt{2a_ch_f}=2.31\sqrt{\frac{1}{\gamma^3\left(2\gamma+1\right)}}.
	\label{u0}
\end{equation}

\begin{figure}
	\centering
	\includegraphics[width=0.9\textwidth]{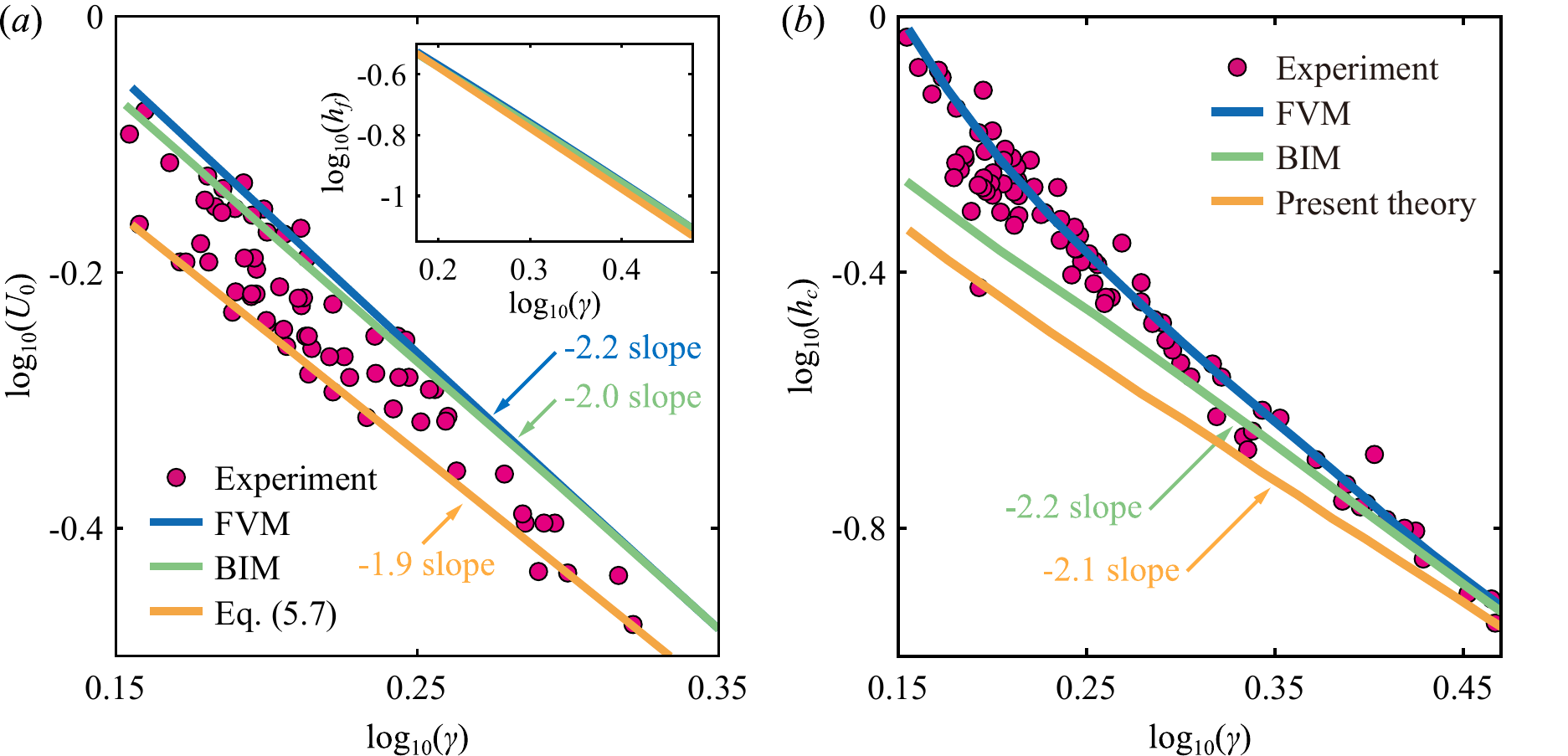}
	\setlength{\abovecaptionskip}{4mm}
	\captionsetup{width=1\textwidth}
	\caption{(\textit{a}) Doubly logarithmic plot for variation of \(U_0\) with \(\gamma\). Experimental data, numerical results obtained from the FVM and BI method, and theoretical results obtained with (\ref{u0}) are brought into comparison. The FVM simulation results are the same as those when \(h_m=0.8\ \textrm{mm}\) in \autoref{gamma_hc_diagram}. \(h_f\) obtained from (\ref{hf}), FVM simulation, and BI simulation are plotted in the inset. The experimental uncertainties of \(U_0\) and \(h_c\) are approximately 0.36 and 0.004, respectively (\(3.6\ \mathrm{m\,s}^{-1}\) and 0.076 mm in dimensional form). (\textit{b}) Comparison among experimental data, FVM and BI simulation results, and theoretical results obtained with (\ref{u0}) and (\ref{acnonuniform}). The BI simulation and theoretical results converge with both experimental data and FVM simulation results as \(\gamma\) increases, and reveal a power law with a fitted exponent of \(\alpha=-2.2\) and \(-2.1\).}
	\label{u0_hf_hc_theoretical}
	\vspace{-1mm}
\end{figure}

It is obvious that (\ref{u0}) approaches the power law of \(U_0\sim\gamma^{-2}\) asymptotically as \(\gamma\) increases. To justify the above theoretical model, we perform a comparison of \(U_0\) between experimental data, FVM simulation results, BI simulation results, and theoretical results obtained from (\ref{u0}) in \autoref{u0_hf_hc_theoretical}(\textit{a}). We choose the configuration of \(h_m=0.8\ \textrm{mm}\) in \autoref{gamma_hc_diagram}. The results obtained with FVM simulations, BI simulations, and (\ref{u0}) reveal a similar fitted power law as that of the experimental data, with the fitted exponent being \(-2.2\), \(-2\), and \(-1.9\), which are relatively close to that of the experimental data. However, the FVM and BI simulation tend to overestimate \(U_0\), while (\ref{u0}) underestimates on the contrary. This discrepancy derive from the uniform acceleration assumption in (\ref{u0}), which is actually non-uniform and continuously increasing in the early stage of the cavity evolution. Nevertheless, the similar fitted power law relationships and the relatively reasonable agreement with the experimental data justify the effectiveness of the simplified model. Additionally, results of \(h_f\) obtained with all three methods are plotted in the inset, the good agreement between the three lines justifies the accuracy of (\ref{hf}).

Once the cavity surpasses the horizontal plane of \(z=0\) and starts to accelerate downwards, the distortion of the pressure field becomes prominent, where the pressure gradient increases, and the uniform acceleration assumption is no longer applicable. However, we can still estimate \(a_c\) using the bubble-induced pressure gradient at the cavity bottom with a coordinate of \((0,-h)\), yielding:
\begin{equation}
	a_c\sim\left. -\frac{\partial p}{\partial z}\right|_{(0,-h)}\approx\frac{2R_1\dot{R}_1^2+R_1^2\ddot{R_1}}{\left(\gamma-h\right)^2}+\frac{2R_2\dot{R}_2^2+R_2^2\ddot{R_2}}{\left(\gamma+h\right)^2}.
	\label{acnonuniform}
\end{equation}

By integrating (\ref{acnonuniform}) twice over the interval of \(t_e\) to \(t_c\) with initial conditions \(h=0\) and \(U=U_0\), we obtain the theoretical results of \(h_c\), where \(t_c\) is characteristic time of the bubble collapse stage. The results are plotted in \autoref{u0_hf_hc_theoretical}(\textit{b}) alongside experimental data, FVM simulation results, and BI simulation results. The theoretical and BI simulation results exhibit a similar fitted power-law relationship, with their fitted exponent being \(\alpha=-2.1\) and \(-2.2\). Interestingly, both curves converge towards the FVM simulation results and experimental data when \(\gamma\to3\), while being distracted from them when \(\gamma\to1.5\). The causes for this discrepancy are obvious: When \(\gamma\) approaches 1.5, the nonlinear effects become significant, which are not considered in both models. Specifically, the BI method does not account for the compressibility effects of the air inflow into the cavity, viscosity and vorticity effects of the fluid, and the boundary layer on the thin rod surface. In addition to these, the theoretical model further neglects the nonlinear boundary condition of the free surface, pressure field distortion caused by cavity growth, and the non-spherical behaviors of the bubble. Despite these simplifications, both the results of the BI simulation and the theoretical model agree reasonably well with experimental data and FVM simulation results. Especially when \(\gamma\to3\), where inertia dominates while nonlinear effects are not prominent, the consistency between the four results is particularly good, which further justifies that the pressure gradient induced by bubble oscillation plays a major role in the cavity evolution.

However, one may notice that the theoretical model we present in this section does not comprise another key parameter \(h_m\), although whose influence on the cavity evolution is not as significant as \(\gamma\), but still can not be neglected. We will address this parameter in the next section.

 \subsection{Nonlinear Rayleigh--Taylor Instability Analysis}\label{section:5.3}
Based on the discussions above, we have already learned that cavity evolution is highly related to the collapse stage of the cavitation bubble. During this stage, the denser fluid (water) is accelerated by the lighter fluid (air). We also learned that the dominant factor in bubble dynamics is the standoff parameter \(\gamma\), and the pressure gradient/acceleration field in the fluid induced by the bubble oscillation is the major cause for cavity growth. Meanwhile, as we mentioned in \cref{section:5.1}, the surface perturbation can also exert a secondary, yet not negligible effect on the cavity dynamics, i.e., the cavity length increases monotonically with the increasing \(h_m\). This acceleration-driven cavity growth from the lighter fluid to the heavier one, affected by the amplitude of the initial interface perturbations, is very similar to the characteristics of the Rayleigh--Taylor instability (RTI), thus inspiring us to perform an RTI analysis in this section.

\begin{figure}
	\centering
	\includegraphics[width=1\textwidth]{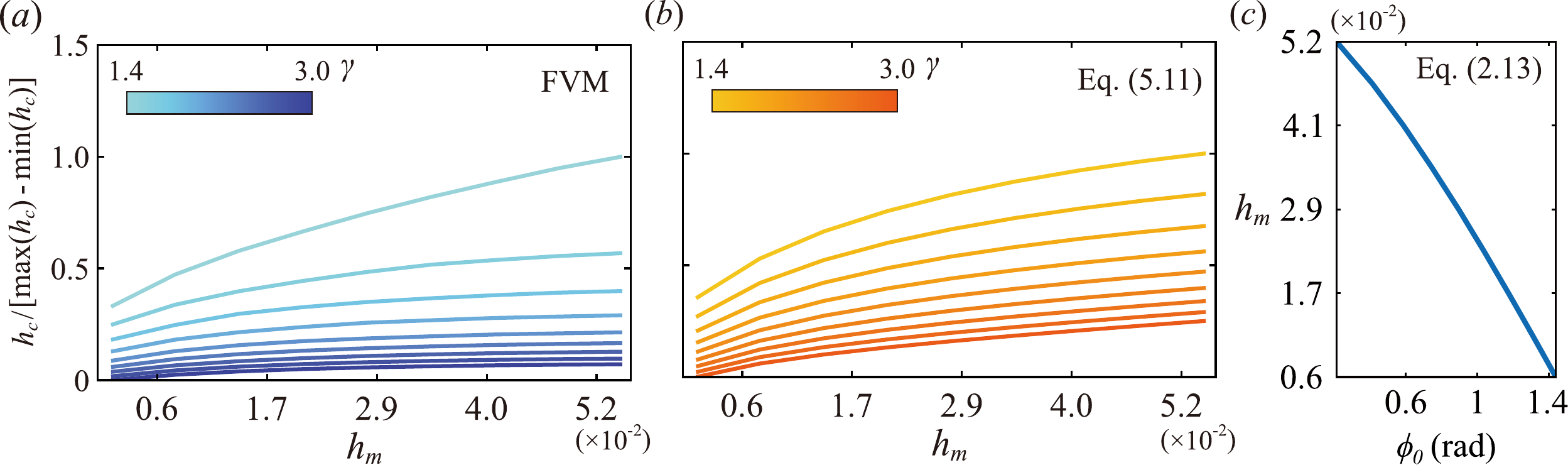}
	\setlength{\abovecaptionskip}{1mm}
	\captionsetup{width=1\textwidth}
	\caption{Comparisons between (\textit{a}) numerical results obtained from FVM (blue solid lines) and (\textit{b}) analytical results obtained from (\ref{cavitynonlinearsolution}) (orange solid lines). The dependence between \(h_m\) and \(h_c\) are plotted in (\textit{a}) and (\textit{b}). \(\gamma\) increases from 1.4 to 3 at a 0.2 interval. Both results are normalized to [0,1] for clearer comparison. The analytical calculations exhibit a similar tendency to the numerical results. The non-dimensional parameters in the FVM simulations are set to the same as \cref{section:4}. The length scale is \(R_m=17.2\ \textrm{mm}\). The relationship between \(h_m\) and \(\phi_0\) is presented in (\textit{c}).}
	\label{hm_hc}
	\vspace{-3mm}
\end{figure}

We first perform several groups of FVM simulations to have a preliminary insight into the \(h_c\textrm{--}h_m\) dependence. The numerical results are presented in \autoref{hm_hc}(\textit{a}) with blue solid lines. As we expected, the \(h_c\textrm{--}h_m\) curve exhibits a monotonically increasing relationship. If the cavity evolution is subject to the linear RTI, \(h_c\) should be proportional to \(h_m\) following the equation of \(h=h_m \cosh(\sqrt{a_c kA}t)\), where \(k=2\pi/\lambda\) is the wave number, \(\lambda\) the wave length of the initial perturbation, \(A=(\rho_l-\rho_g)/(\rho_l+\rho_g)\) the Atwood number, and \(t\) the characteristic evolution time (assuming as constant). However, the \(h_c\textrm{--}h_m\) curves exhibit decreasing slope, suggesting that RTI evolution of the cavity is in fact nonlinear, which is reasonable since the cavity grows into a scale much larger than that of the initial perturbation (\(h_c\gg h_m\)) within a very short period.

Therefore, a nonlinear RTI equation proposed by \citet{mikaelianAnalyticApproachNonlinear2010} based on the widely used axisymmetric nonlinear RTI model developed by \citep{layzerInstabilitySuperposedFluids1955} is adopted, which reads:
\begin{equation}
	\frac{\mathrm{d}^2 \hat{h}}{\mathrm{d} s^2}-\hat{k} \hat{A} \hat{h}+\frac{1}{2a_c^2}\frac{\mathrm{d}a_c}{\mathrm{d}t}\frac{\mathrm{d}\hat{h}}{\mathrm{d}t}=0,
	\label{nonlinearrtieq}
\end{equation}
where \(\hat{h}=e^{(h-h_m)k}\), \(\hat{k}=d(1+d)(1+A)k/2(1+d+dA-A)\), and \(\hat{A}=2A/(1+d+dA-A)\) are the nonlinear counterparts for \(h\), \(k\), and \(A\), while \(d\) is the generalization parameter for models of different dimensions, i.e., \(d=1\) for 3D, \(d=2\) for 2D. \(s=\int_{t_0}^{t}\sqrt{a_c}\mathrm{d}t\) is the substituting variable, where \(t_0\) and \(t\) is the initial and terminal time of instability evolution, \(a_c\) the driven acceleration in (\ref{acnonuniform}). We notice that the negative acceleration in the cavity evolution will lead to imaginary numbers in the \(\sqrt{a_c}\) terms. By introducing a modification \(\mathrm{sgn}(a_c)\sqrt{|a_c|}\), we can take the negative acceleration into account \citep{ramaprabhuRayleighTaylorInstabilityDriven2013}. Such modification is reasonable, since the cavity acceleration is positive (pointing downwards) during the majority of the cavity evolution, while the deceleration and rebound process only comprises a minor part, as we discussed in \cref{section:4.1}. Therefore, it is also safe to further neglect the \(\mathrm{d}a_c/\mathrm{d}t\) terms in (\ref{nonlinearrtieq}) without significantly changing the cavity dynamics. It is also suggested by \citet{yanCavityDynamicsWater2009a} that the flow around the cavity can be treated as two-dimensional and axisymmetric when the \textit{Fr} is relatively large, yielding \(d=2\). Given the vast density difference between the two fluids (air and water) in this study, \(A\) can be set as 1. Hence, (\ref{nonlinearrtieq}) can be simplified as:
\begin{equation}
	\frac{\mathrm{d}^2 \hat{h}}{\mathrm{d} s^2}-\hat{k}\hat{A}\hat{h}=0.
	\label{nonlinearrtieqsimplified}
\end{equation}
Solving this equation with the first order Wentzel-Kramers-Brillouin (WKB) approximation \citep{bender2013advanced} yields:
\begin{equation}
	h=h_m+\frac{1}{\hat{k}}\ln\left[\cosh\left(s\sqrt{\hat{A} \hat{k}}\right)\right].
	\label{cavitynonlinearsolution}
\end{equation}

The determination of \(\hat{k}\) is relatively challenging in the present study, since the meniscoid surface perturbation does not have a representative wavelength/wave number. Unlike the common periodic initial perturbation, where the wave number and wavelength are readily obtainable both experimentally and numerically \citep{emmonsTaylorInstabilityFinite1960,herbertPerturbationMethodsNonlinear1983,jacobsThreedimensionalRayleighTaylorInstability1988a}. Many precedent works have adopted several methods to analyze the influence of single/multi-mode initial perturbations, including spectrum analysis \citep{ramaprabhuRayleighTaylorInstabilityDriven2013,olsonExperimentalStudyRayleigh2009,cohenThreedimensionalSimulationRichtmyer2002} and Fourier expansions \citep{liSinglemodeRayleighTaylor2025,liangShockinducedHeavyfluidlayerEvolution2021,luoEffectsNonperiodicPortions2019}. In the present study, the initially perturbed interface contains both high-frequency (the capillary length scale meniscus) and low-frequency (the remaining flat free surface) compositions. Although the high-frequency composition only comprises a small part of the interface, it can exert a remarkable influence on the free-surface--bubble system according to the aforementioned discussions. Meanwhile, the influence of the remaining low-frequency compositions may not be prominent, but it still constitutes the majority of the interface. Moreover, the vast scale difference between the meniscus and the bubble suggests that the Fourier series needs to be expanded from low order to very high order, otherwise the interface may not be accurately depicted, not to mention the pinning point at the contact line may introduce a strong singularity. Therefore, efforts to approximate the interface analytically are challenging and costly in both time and calculation resources, which deviates from the original intention of adopting the analytical model.

Nonetheless, this nonlinear RTI model can still explain some issues by choosing a characteristic wavelength/wave number. \citet{mikaelianAnalyticApproachNonlinear2010} suggested \(h_m\sim1\) or \(h_mk\gtrsim1\), since many nonlinear RTI processes start with a weakly nonlinear amplitude and grow from there, and rapidly enter the nonlinear regime \(hk\gg1\), which corresponds to the transient cavity evolution in the present study. Meanwhile, \citet{zhaoEffectLongwavelengthPerturbations2023} and \citet{huang2010nonlinearrayleightayloreffects} pointed out that the compositions with larger wavelength/smaller wave number in the initial perturbations tend to grow faster within the nonlinear region and stimulate the development of the large-scale RTI structures. This argument is often related to the bubble competition and merging in the multi-mode RTI evolution, where multiple instability bubble evolve simultaneously. However, in the present study, only one primary cavity can be observed in experiments and simulations, which is actually the superposition of multiple cavities that evolve from various modes. Therefore, the dominant mode demands a smaller \(k\) according to this criterion to achieve a faster growth velocity. The paradox between these two criteria indicates that the perturbation composition with the largest wavelength/smallest wave number that satisfies the weakly nonlinear assumption tend to dominate the cavity evolution, thus leading us to a relation of \(h_mk=1\) and \(\hat{k}=1.5/h_m\).

Once \(\hat{A}\) and \(\hat{k}\) are determined, (\ref{cavitynonlinearsolution}) can be solved by iterating \(a_c\) and \(h\) throughout the cavity life cycle. The calculation results are presented in \autoref{hm_hc}(\textit{b}) with orange solid lines. To gain a clearer vision of the analytical results and compare the \(h_m\textrm{--}h_c\) tendency in both results, \(h_c\) is normalized to [0,1]. Encouragingly, the analytical results exhibit a similar dependence as the numerical results, i.e., the \(h_c\) increases monotonically as \(h_m\) with a decreasing slope. However, the analytical results still have deviations from those of the numerical simulations, as expected. Specifically, the ratio between the numerical and analytical results varies from the order of \(\textit{O}(10^1)\) to \(\textit{O}(10^0)\), as \(\gamma\) increases from 1.4 to 3. The best agreement between the two results is obtained when \(\gamma\to\) 3, while they deviate from each other when \(\gamma\to1.5\). This deviation can be explained as follows. First, the same as the analytical model that we proposed in \cref{section:5.2}, the nonlinear RTI model does not account for the influence of the boundary layer, whose influence would be increasingly prominent as \(\gamma\) decreases. Moreover, non-spherical bubble and flow-focusing effects both can cause strong nonlinear cavity--bubble interactions and increases \(h_c\). Additionally, other factors such as viscosity, vorticity, air compressibility, and surface tension are also neglected, whose influence may not be significant as the previous mentioned ones, but still can exert slight changes on the cavity dynamics. Second, since we only choose one characteristic wavelength/wave number for simplification, and neglect the remaining compositions of the initial perturbation, it is foreseeable that this deviation occurs. Moreover, the meniscus contact angle \(\phi_0\), which characterizes the meniscus geometry, is also considered. (\ref{hmasymptotic}) shows that the meniscus height \(h_m\) decreases monotonically with increasing \(\phi_0\) (see \autoref{hm_hc}\textit{c}), indicating that \(h_m\) and \(\phi_0\) are intrinsically related. Consequently, once the \(h_c\textrm{--}h_m\) relationship is established, the corresponding \(h_c\textrm{--}\phi_0\) dependence is implicitly determined, and \(h_c\) is expected to decrease with increasing \(\phi_0\). Since \(h_m\) can be measured far more accurately than \(\phi_0\) in the current experimental setup, we therefore focus on \(h_m\) as the primary parameter in the analysis.

One can also notice that the significant increase of \(h_c\) (from \(0.6\times10^{-2}\) to \(5.2\times10^{-2}\)) does not result in equivalent changes in the magnitude of \(h_m\). This observation is consistent with the discussions in \cref{section:5.1} associated with \autoref{gamma_hc_diagram}, where the scaling law \(h_c\propto\gamma^\alpha\) remains relatively identical as \(h_m\) varies, thus demonstrating the secondary role of \(h_m\) in the cavity dynamics. 

Nonetheless, the dependence of \(h_m\textrm{--}h_c\) revealed by the nonlinear RTI model is similar with the numerical results, suggesting that the cavity evolution on the initially perturbed interface driven by an oscillating cavitation bubble is actually a nonuniform acceleration driven nonlinear Rayleigh--Taylor instability phenomenon, where the acceleration field induced by the bubble oscillation governs the cavity dynamics, while the amplitude of the initial perturbations plays a secondary role by varying the cavity maximum length without significantly changing the fitted power law.

\subsection{Influence of boundary layer}\label{section:5.4}
Apparently, the analytical model in \cref{section:5.3} will fail if \(h_m=0\). However, during the experiments, we notice that if we use the contact line pinning effect by carefully moving the thin rod to eradicate the meniscoid and obtain a flat surface, i.e., \(h_m\approx0\), the cavity still occurs, but with a significantly smaller \(h_c\). If the \(\gamma\) is not very large at the same time, the cavity may be pierced by the downward jet and break up into pieces. This suggests that the boundary layer around the thin rod may have a minor influence on the cavity dynamics.

\begin{figure}
	\centering
	\includegraphics[width=0.95\textwidth]{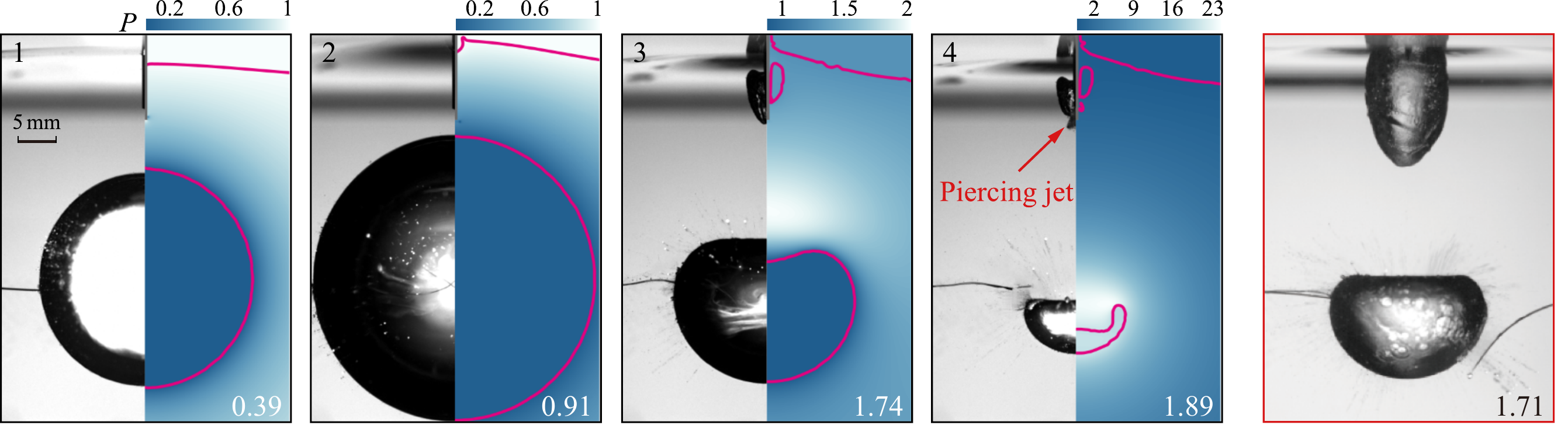}
	\setlength{\abovecaptionskip}{3mm}
	\captionsetup{width=1\textwidth}
	\caption{Comparison of cavity and bubble evolution between numerical simulation and experimental observation in a representative case without meniscus perturbation (\(H=27.5\ \textrm{mm}\), \(R_m=18.5\ \textrm{mm}\), timescale is 1.84 ms). Under such circumstance, the cavity is significantly smaller than that of the regular cases (red frame on the right, \(H=28.2\ \textrm{mm}\), \(R_m=18.6\ \textrm{mm}\), \(h_m=0.94\ \textrm{mm}\), timescale is 1.85 ms), and the cavity becomes an isolated bubble due to the much earlier surface-seal. The sealing surface creates a downwards jet piercing the bubble rapidly. Although slight deviation exists in the size and position of the isolated cavity between the two results, the numerical simulation successfully reproduces the main features.}
	\label{exp_num_no_meniscus}
	\vspace{-3mm}
\end{figure}

A representative experiment of the free-surface--bubble interaction without the meniscus perturbation is shown in \autoref{exp_num_no_meniscus}. It clearly shows that the cavity is significantly smaller than that in the regular meniscus-perturbed experiments presented in the red box. And due to much earlier surface-seal, the cavity becomes an isolated small bubble and moves downward under the influence of the pressure gradient. At the same time, the sealed surface creates a downward jet and pierces the bubble before the cavity attains its maximum length.

We perform a numerical simulation of the surface without meniscus perturbation to study the difference between the two experiments. By setting the free surface as flat, we can obtain a non-meniscus-perturbed interface \(h_m=0\). The comparison between the numerical simulation results and experimental observation is presented in \autoref{exp_num_no_meniscus}. The numerical simulation successfully reproduces the main characteristics of the isolated cavity and the piercing jet. The simulation results intuitively explain the significant difference between the cavity dynamics with and without meniscus perturbation: In the absence of the meniscus perturbation, only the boundary layer in the close proximity of the thin rod is hindered by the no-slip boundary condition. As a result, the depression on the top of the water hump is much smaller than that in the regular perturbed case when the bubble attains its maximum radius (frame 2). The smaller depression leads to a much smaller radial and longitudinal size of the cavity when the bubble starts to contract (frame 3). Moreover, the influx of air into the cavity causes the small cavity opening to seal much earlier, creating an isolated bubble and a downward jet piercing the cavity bottom (frame 4). All the aforementioned factors conjointly cease the cavity evolution, resulting in a much smaller cavity.

However, a slight deviation still exists in the size and position of the cavity between simulation results and experimental observations. This deviation derives from the disturbance on the `flat' free surface. Since the meniscus and the free surface is highly sensitive to external disturbances (\cref{section:5.3}), minor distractions such as slight vibrations and breezes could still introduce deformation on the initially flat free surface, even if the macro camera did not capture any evidence of shape changes (see the meniscus image with blue marker in \autoref{num_model_meniscus}\textit{b}). Therefore, the `flat' free surface in the experiments may not be strictly flat, and the small meniscus may result in a slightly increased cavity length \(h_c\).

\begin{figure}
	\centering
	\includegraphics[width=1.0\textwidth]{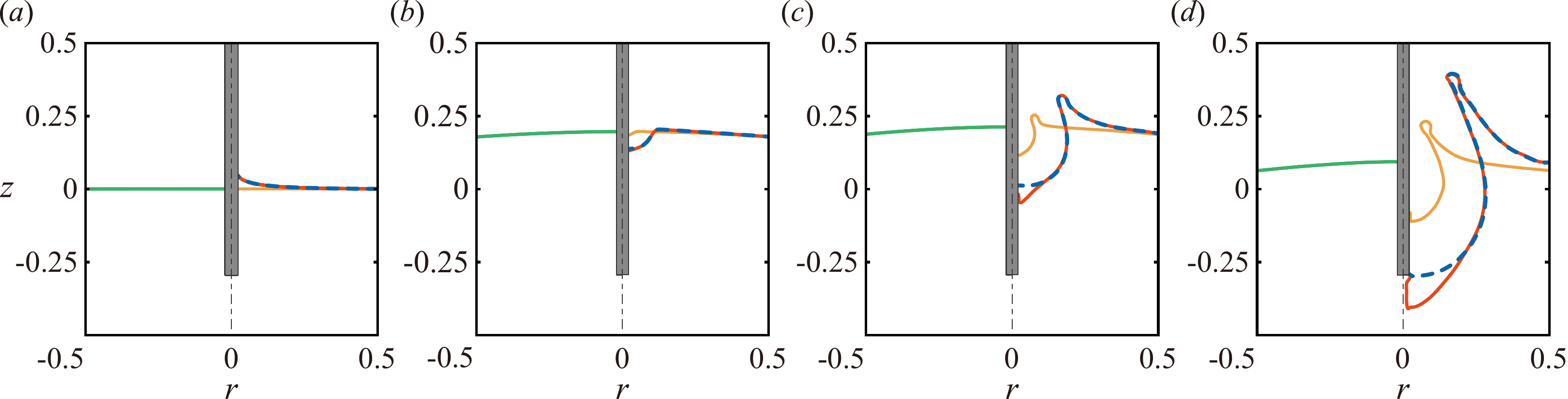}
	\setlength{\abovecaptionskip}{2mm}
	\captionsetup{width=1\textwidth}
	\caption{Comparison of cavity evolution across four distinct configurations: (i) free surface without perturbation (green solid lines), (ii) flat surface with no-slip BC (yellow solid lines), (iii) meniscus perturbed surface with no-slip BC (red solid lines), and (iv) meniscus perturbed surface with slip BC (blue dashed lines). Profiles of the interface at four different times are presented: (\textit{a}) 0, (\textit{b}) 0.62, (\textit{c}) 1.17, (\textit{d}) 1.86. The timescale is 1.69 ms, and the length scale is \(R_m=17.0\ \textrm{mm}\). The non-dimensional parameter of the bubble is the same as that in \autoref{exp_gamma_over_1.36}(\textit{a}).}
	\label{boundary_layer_effect}
	\vspace{-3mm}
\end{figure}

Given our model's limitations in quantitatively analyzing the boundary layer, we qualitatively assess its effects by toggling the rod surface's no-slip boundary condition (BC). \autoref{boundary_layer_effect} compares four configurations: (i) Unperturbed free surface (green solid lines), (ii) flat surface with no-slip BC (yellow solid lines), (iii) meniscus perturbed interface with no-slip BC (red solid lines), (iv) and meniscus perturbed interface with slip BC (blue dashed 1lines). The unperturbed free surface serves as a benchmark. Comparisons between each one of these results suggest the following arguments: (1) The slight discrepancy between configuration (iii) and (iv) indicates that turning off the no-slip BC does not significantly vary the cavity characteristics, suggesting that the boundary layer does not exert a remarkable influence on the cavity dynamics when \(\gamma\) is relatively large. (2) The vast difference between configuration (ii) and (iii) indicates that the boundary layer alone is not enough for the sufficient development of the cavity. If only the meniscoid perturbation exists, the main characteristics of the cavity occur. (3) The comparison among configurations (ii), (iii), and (iv) suggests that the meniscoid perturbation plays a major role in the cavity dynamics, while the existence of the boundary layer facilitates the cavity evolution. Changing the boundary condition varies the partial cavity details, e.g., the tapered or semi-spherical cavity bottom morphology in configurations (iii) and (iv). In conclusion, the overall radial and longitudinal scale of the cavity is still determined by the geometrical characteristics of the meniscoid perturbation. The detailed discussions between the effects of air compressibility and boundary layer are presented in \autoref{appC}.

\section{Summary and conclusions}\label{section:6}

The interaction between a spark-generated cavitation bubble and the cavity formed on an initially perturbed free surface is systematically investigated through a combined experimental, numerical, and analytical approaches. In the experiments, well-controlled millimetre-scale perturbations are introduced by inserting a thin hydrophilic-coated rod into the liquid pool and adjusting the meniscus height via the contact line pinning effect. A spark-generated cavitation bubble is then initiated coaxially with the thin rod to interact with the initially perturbed free surface. The FVM and BI framework enables a systematic, one-to-one experimental validation of how initial free surface perturbations influence macroscopic bubble dynamics. The cavity dynamics is further interpreted through analytical models based on Rayleigh--Plesset (R--P) equation and nonlinear Rayleigh--Taylor instability theory. The key findings of this study are summarized as follows.

Our experiments reveal a previously uncharacterized physical pathway: Sub-millimeter surface perturbations trigger a multiscale cascade. Specifically, the meniscoid surface perturbation evolves into a small depression on top of the rising water dome during the expansion phase of the cavitation bubble. Once the bubble begins to contract, the depression rapidly grows downward into a cavity whose size soon becomes comparable to the bubble itself. The cavity--bubble interaction is highly sensitive to the non-dimensional stand-off parameter \(\gamma\). When \(\gamma\gtrsim1.36\), the cavity rebounds after reaching its maximum depth \(h_c\) and produces an upward-bursting jet slightly before the bubble reaches its minimum volume. When \(\gamma\lesssim1.36\), the cavity coalesces with the bubble, forming an open channel that vents the bubble interior to the atmosphere and markedly attenuates the collapse intensity. The critical value of \(\gamma\) increases weakly with the initial meniscus height \(h_m\).

A parameter map of \(h_c(\gamma, h_m)\) summarizes the cavity--bubble dynamics. Experiments and FVM simulations agree over the entire tested range and show that \(h_c\) is controlled primarily by \(\gamma\), whereas \(h_m\) exerts only a minor influence. The map distinguishes non-coalescence and coalescence regimes by a single critical curve. Scaling analysis yields a power law \(h_c\propto\gamma^\alpha\), with \(\alpha=-2.7\) (experiment) and \(\alpha=-2.6\) (FVM); the relation is valid for \(1.5\lesssim\gamma\lesssim 3\), where nonlinear effects are modest and inertia dominates surface-tension and viscous influences. A combined Rayleigh--Plesset and image bubble model gives \(\alpha=-2.1\). As \( \gamma\to3 \), the theoretical curve approaches the measured values, indicating that free-surface nonlinearity, though weak, is still considerable at the lower end of the interval and becomes negligible near the upper bound. The theoretical results are justified by the BI method (\(\alpha=-2.2\)), whose agreement with experiments and FVM results is better, while still remarkably deviating from the experiments and FVM results as \(\gamma\to1.5\). This deviation originates from the increasingly prominent nonlinear free-surface--bubble interactions as the cavity growth becomes faster.

A nonlinear RT instability model is adopted to account for the influence of surface perturbations. Assuming the cavity undergoes a nonlinear RTI evolution, and determining the wave number \(k\) with \(h_mk=1\) by satisfying the weakly nonlinear initial perturbation condition and dominant wavelength condition, the cavity evolution equation (\ref{cavitynonlinearsolution}) can be solved, thus obtaining the dependence of \(h_c\) on \(h_m\) and \(\gamma\), as shown in \autoref{hm_hc}. The analytical results deviate quite remarkably from the numerical results and experimental data when \(\gamma\) is small, while still exhibiting a relatively similar dependence in terms of the \(h_m\textrm{--}h_c\) relationships, and tend to converge with both results when \(\gamma\) increases. Specifically, the \(h_m\textrm{--}h_c\) curves obtained from the analytical model share the same monotonically increasing tendency with a decreasing slope as the FVM numerical results.

We also investigate the influence of the boundary layer on the thin rod surface and air compressibility inside the cavity qualitatively. The results indicate that the influence of the boundary layer and air compressibility on cavity dynamics becomes more pronounced as the standoff parameter decreases. When \(\gamma\) is relatively large, they only affect detailed features of the cavity morphology. Only when \(\gamma\) is reduced toward the critical regime do they begin to exert a significant impact on the overall cavity morphology. Nevertheless, Variations in these two factors do not result in an order-of-magnitude variation in the maximum cavity length. The overall cavity evolution is still inertia-dominated.

The findings provide practical guidance for surface-jetting applications such as laser-induced forward transfer (LIFT) \citep{turkoz2018impulsively,lohse2022fundamental}: Eliminating pinned menisci or free-surface defect is critical for producing repeatable jets and droplets. In contrast, deliberately introducing surface perturbations can promote cavity--bubble coalescence and vent the collapse, reducing collapse intensity and peak pressure. This insight transforms surface perturbations from `uncontrolled noise' into `tunable control parameters' for engineered cavitation mitigation. The same mechanisms are expected to persist at smaller scales: Cavitation bubbles oscillating within cylindrical or spherical droplets can induce Rayleigh--Taylor instability at the interface even without deliberately imposed perturbations, producing splashing, ventilation, and fine rapid jets \citep{wangRayleighTaylorInstability2021,zengJettingViscousDroplets2018a}. At these reduced scales, viscous and surface tension effects remain negligible for water (\(\textit{We},\ \textit{Re}\gg1\)). Whether the same holds for highly viscous liquids remains an open and intriguing question for future research.

%\backsection[Supplementary data]{\label{SupMat}}

\backsection[Acknowledgements]{The authors thank J. Wang and S. Pei from HEU for the preparation of the experiments.}

\backsection[Funding]{This work is supported by the National Natural Science Foundation of China (nos 12372239, 52525102) and the Key R\&D Program Project of Heilongjiang Province (JD24A002).}

\backsection[Declaration of interests]{The authors report no conflict of interest.}

%\backsection[Data availability statement]{}

\backsection[Author ORCIDs]{J. Gu, https://orcid.org/0009-0002-1902-9746; Z. Liu, https://orcid.org/0009-0009-7471-1052; A.-M. Zhang, https://orcid.org/0000-0003-1299-3049; S. Li, https://orcid.org/0000-0002-3043-5617.}

%\backsection[Author contributions]{}

\appendix
\section{Numerical model validation, domain verification, and convergence analysis}\label{appA}
\begin{figure}
	\centering
	\includegraphics[width=0.9\textwidth]{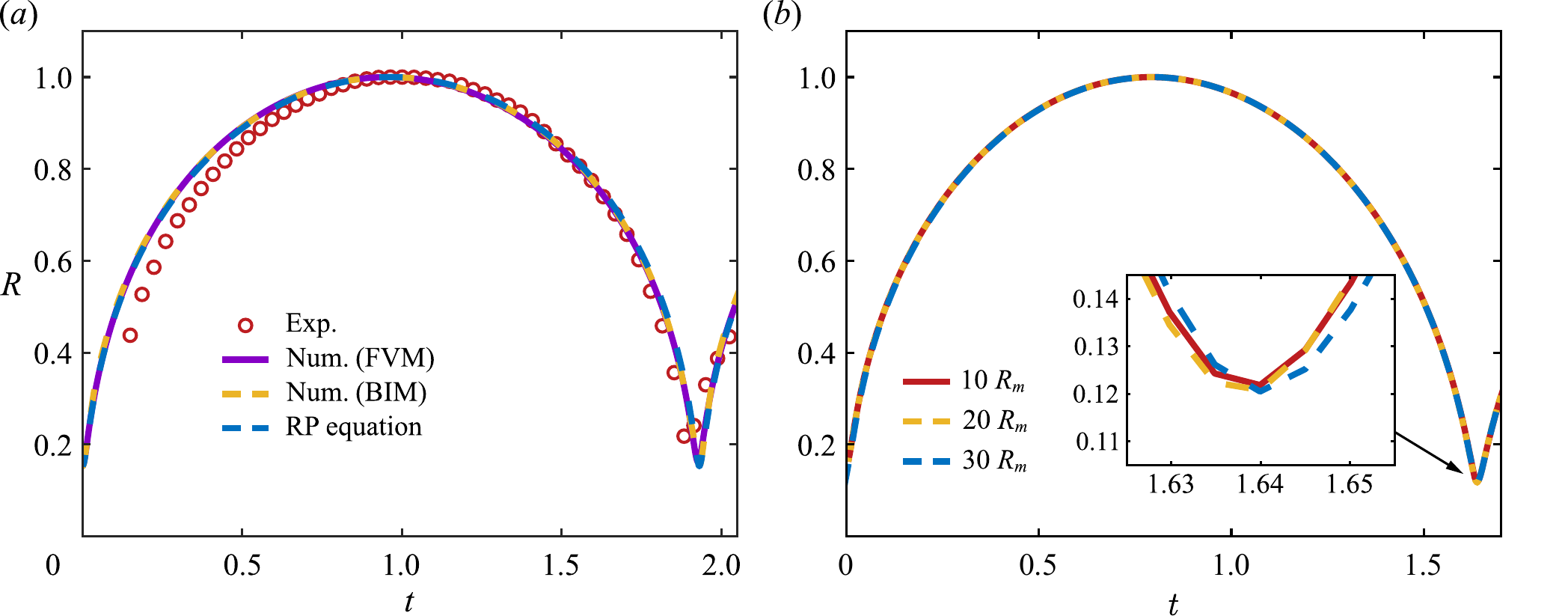}
	\setlength{\abovecaptionskip}{4mm}
	\captionsetup{width=1\textwidth}
	\caption{(\textit{a}) Comparison of non-dimensional bubble radius evolution in a free field among experimental data, theoretical predictions (derived from the RP equation), and numerical results (obtained from BI \citep{klaseboerBoundaryIntegralEquations2004} and FVM simulation). All calculations were conducted with identical \(\varepsilon\) and \(R_0\) values of 125 and 0.1623, respectively. (\textit{b}) Computational domain verification of three different scales. We conduct simulations in the domain with radius of \(10\ R_m\), \(20\ R_m\), and \(30\ R_m\). The standoff parameter \(\gamma\) of all three cases are set as 1.36. The radius and time are scaled by \(R_m\) and \(R_m\sqrt{\rho_l/P_\infty}\), respectively.}
	\label{num_model_verification}
	\vspace{5mm}
\end{figure}

Validations for the numerical model and the computational domain are performed in this section. First of all, we validate the numerical model with the aforementioned initial conditions. A comparison among experimental data, numerical results (obtained from FVM simulation and BI simulation), and theoretical results (obtained from RP equation) is presented in \autoref{num_model_verification}(\textit{a}). The bubble radius evolution reveals slight discrepancies between the experimental and numerical results during the expansion phase. These discrepancies stem from the continuous discharge process observed experimentally, which leads to gradual --- rather than instantaneous --- energy increase of the bubble. Remarkably, all datasets converge after the bubble reaches its maximum radius, demonstrating the reliability of the numerical model established in this study, and again highlighting the limited influence of the fluid compressibility on the bubble dynamics.

The verification for the computational domain is presented in \autoref{num_model_verification}(\textit{b}). According to the simulation setup in \cref{section:2.2}, the radius of the computational domain is set as \(10\ R_m\) to reproduce the experimental conditions. To assess the influence of the domain on the bubble, we conduct additional simulations with larger domains, specifically with radii of 20 \(R_m\) and 30 \(R_m\). \Autoref{num_model_verification}(\textit{b}) shows that the discrepancy between the three simulation results are small enough to be neglected, with the bubble cycle in 20 \(R_m\) and 30 \(R_m\) domain only increasing by 0.1 and 0.3 \textit{\%}, compared with that in \(10\ R_m\) (see the inset of \autoref{num_model_verification}\textit{b}). This indicates that the fluid field is sufficiently large that the solid wall boundary can hardly impose any notable influence on the bubble, not to mention the cavity.

\begin{figure}
	\centering
	\includegraphics[width=0.5\textwidth]{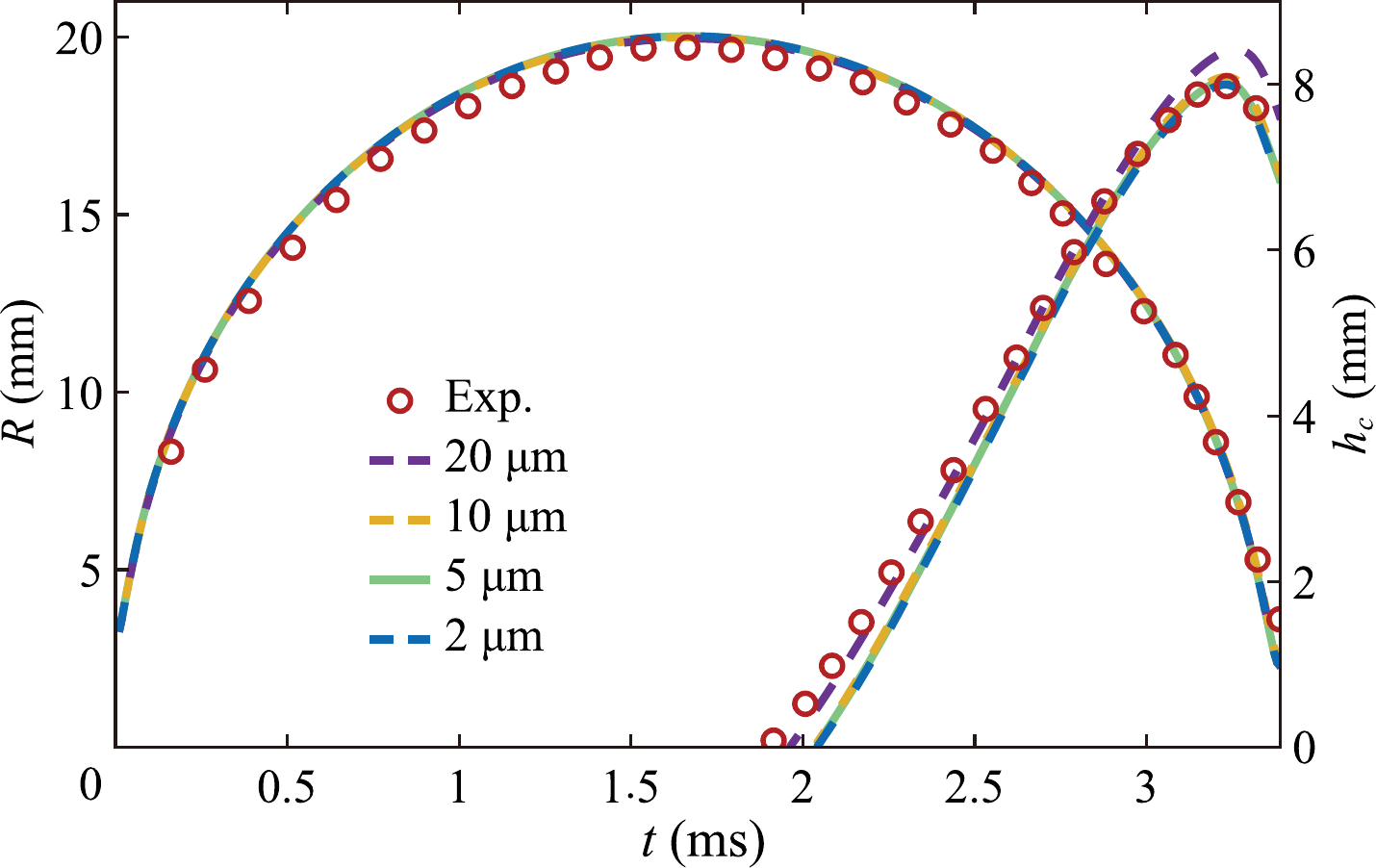}
	\setlength{\abovecaptionskip}{4mm}
	\captionsetup{width=1\textwidth}
	\caption{Convergence analysis of the mesh cell size. Simulations with four different minimum mesh cell sizes are performed and plotted with dashed and solid lines, which are 20, 10, 5, and \(2.5\ \upmu\textrm{m}\). Corresponding numbers of meshes are approximately 500\,000, 700\,000, 1\,500\,000, and 2\,000\,000, respectively. The simulations are performed in the first cycle of the bubble and \(\gamma\) is set as 1.77. Experimental results are plotted with red circles. All the results are presented in dimensional form.}
	\label{convergence_analysis}
	\vspace{0mm}
\end{figure}

We also perform a mesh convergence analysis. As shown in \autoref{convergence_analysis}, four simulations with progressively refined meshes (\(20\), \(10\), \(5\), and \(2.5\ \upmu\textrm{m}\)) correspond to the total mesh numbers of 500\,000, 700\,000, 1\,500\,000, and 2\,000\,000, respectively. All results are presented in dimensional form. The maximum bubble radius varies by less than 0.3 \textit{\%} across four simulations, confirming sufficient for capturing the bubble dynamics. For the cavity, coarser grids (\(20\), and \(10\ \upmu\textrm{m}\)) introduce larger discrepancy in cavity length, while finer grids exhibit sharp convergence with experimental data. Notably, the \(5\) and \(2.5\ \upmu\textrm{m}\) meshes differ by only 0.5 \textit{\%} in maximum cavity length. Based on these findings, we adopt \(5\ \upmu\textrm{m}\) resolution in our simulations.

\section{Influence of the thin rod}\label{appB}
\begin{figure}
	\centering
	\includegraphics[width=0.9\textwidth]{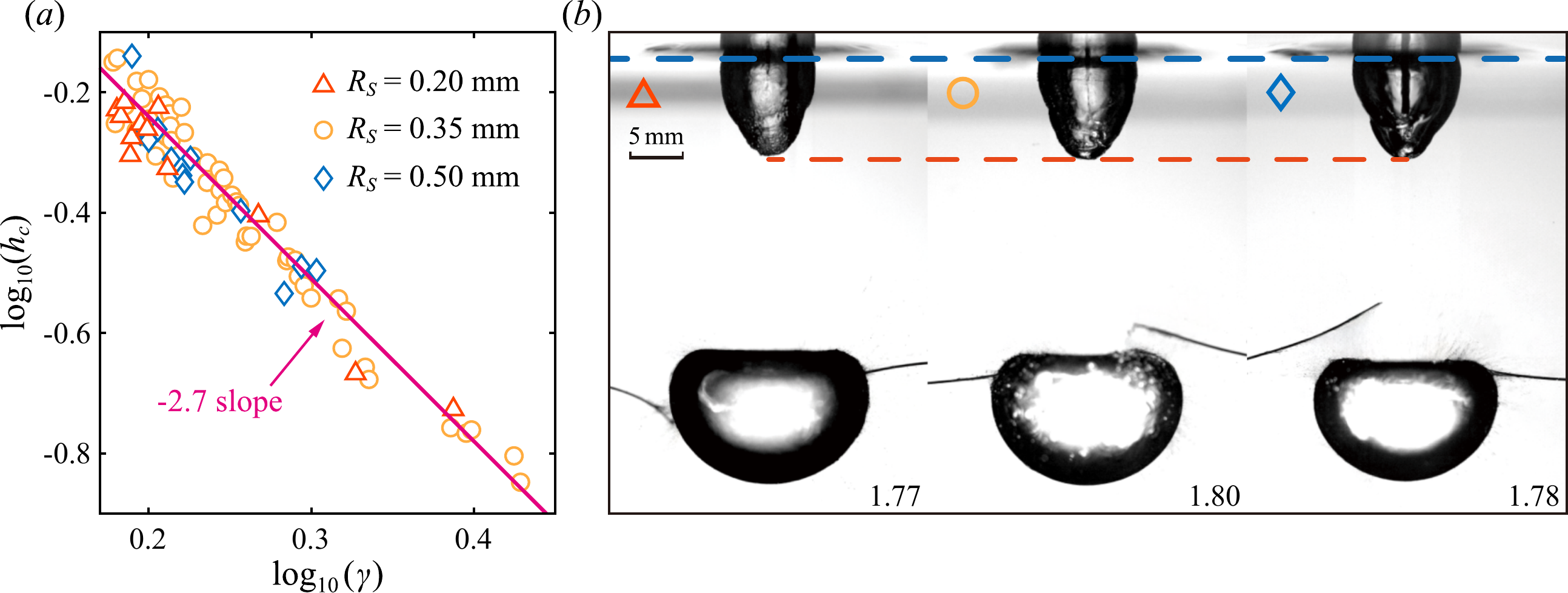}
	\setlength{\abovecaptionskip}{3mm}
	\captionsetup{width=1\textwidth}
	\caption{(\textit{a}) Doubly logarithmic plot for the variation of maximum cavity length \(h_c\) versus the standoff parameter \(\gamma\) for different radii of the thin rod. The radii of the thin rod are 0.2 mm (triangles), 0.35 mm (circles), and 0.5mm (diamonds). The experimental results exhibit the same fitted power-law relationship of \(-2.7\) exponent. (\textit{b}) Three typical experiments performed with the three thin rod radii. The maximum cavity lengths of the three cases are 9.54, 9.76, and 9.77 mm, and the timescales are 1.82, 1.84, and 1.86 ms, respectively.}
	\label{thin_rod_size}
	\vspace{0mm}
\end{figure}
\begin{figure}
	\centering
	\includegraphics[width=1\textwidth]{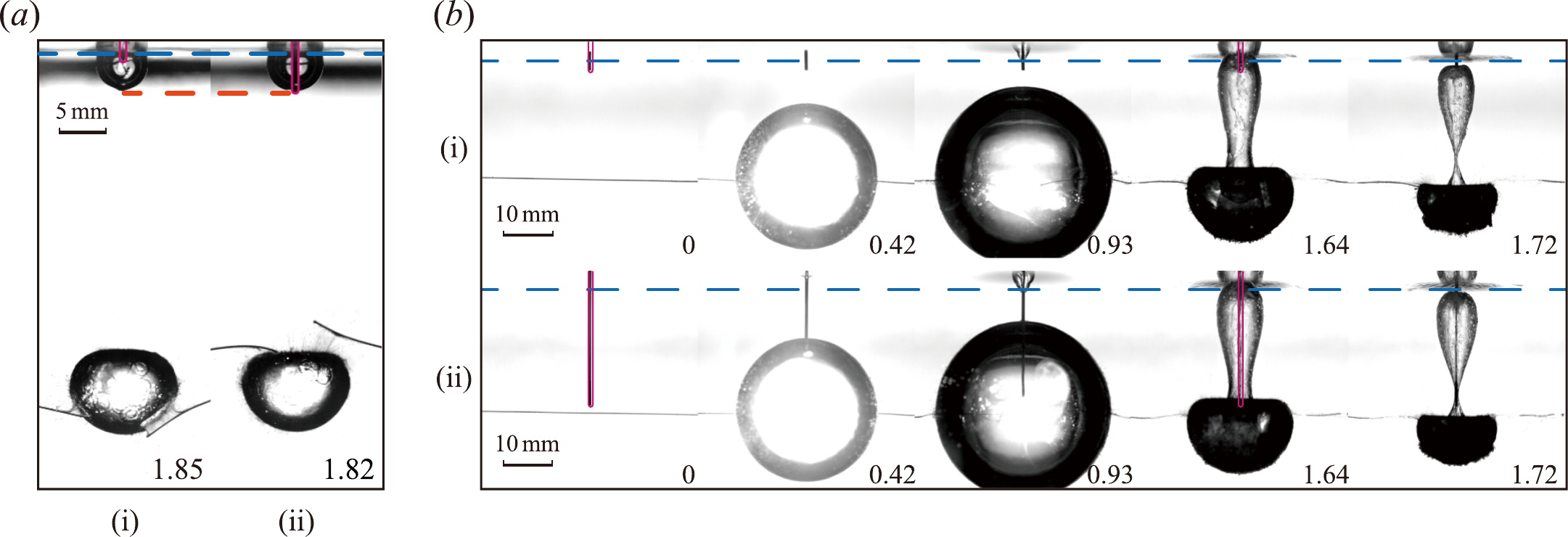}
	\setlength{\abovecaptionskip}{3mm}
	\captionsetup{width=1\textwidth}
	\caption{Comparisons of experiments with different depths of the thin rod. The thin rod is highlighted with magenta frames. From left to right, the parameters of each frame are: (\textit{a}) (i) \(\gamma=2.16\), \(h_m=0.79\ \textrm{mm}\), \(R_m=15.7\ \textrm{mm}\), \(h_c=4.1\ \textrm{mm}\); (ii) \(\gamma = 2.17\), \(h_m=0.80\ \textrm{mm}\), \(R_m=16.1\ \textrm{mm}\), \(h_c=4.1\ \textrm{mm}\). The timescales are 1.56 and 1.60 ms. The thin rod depths are 0.4 and 4.2 mm, and the time scales are 1.56 and 1.60 ms, respectively. (\textit{b}) (i) \(\gamma=1.35\), \(h_m=0.80\ \textrm{mm}\), \(R_m=18.2\ \textrm{mm}\); (ii) \(\gamma = 1.35\), \(h_m=0.81\ \textrm{mm}\), \(R_m=18.2\ \textrm{mm}\). The thin rod depths are 2.0 and 23.8 mm, respectively. The timescale is 1.81 ms. Cavity behaviors and bubble morphology do not exhibit much differences between each one of the both two groups of experiments.}
	\label{thin_rod_depth}
	\vspace{0mm}
\end{figure}

The thin rod is treated as a fixed rigid boundary in our numerical simulations. Since the radius of the thin rod is small compared to the sizes of the cavity and bubble, the influence of the radius and position of the thin rod is negligible. Doubly logarithmic plot for the variation of \(h_c\) versus \(\gamma\) for different \(R_s\) are presented in \autoref{thin_rod_size}(\textit{a}). We choose three different thin rod radii for comparison, namely, 0.2 mm (red triangles), 0.35 mm (yellow circles), and 0.5 mm (blue diamonds). The experimental data exhibits the same fitted power-law relationship as \autoref{gamma_hc_diagram}, with the fitted power exponent being \(\alpha=-2.7\). \Autoref{thin_rod_size}(\textit{b}) shows comparison of three typical experimental observations for different thin rod radii. The morphology of the cavities does not change significantly as \(R_s\) ranges from 0.2 to 0.5 mm, as well as the cavity length (distance between the blue and red dashed lines), suggesting that the changes in the radii of the thin rod do not remarkably affect the cavity dynamics.

As we discussed in \cref{section:5.4}, the boundary layer around the thin rod has a minor influence on the cavity evolution. Based on this discussion, one may notice that the depth of the thin rod piercing beneath the free surface may affect the cavity dynamics, since the deeper the thin rod pierces beneath the free surface, the longer the boundary layer could affect the cavity evolution. By adjusting the position of the thin rod with respect to the free surface, we look into the influence of the thin rod depth on the cavity dynamics. Other parameters, such as standoff parameters \(\gamma\) and perturbation sizes \(h_m\), are set as almost identical. Similarly, the thin rod depth does not substantially affect the cavity--bubble interaction characteristics across a wide range of \(\gamma\), as shown in \autoref{thin_rod_depth}. Slight discrepancies can be observed in the wrinkle patterns on the cavity surface, which can be explained by the secondary Rayleigh--Taylor instability we discussed in \cref{section:3.1}. Although such secondary instability is inevitable with current experiment conditions, its influence on the primary cavity is negligible.

\section{Effects of liquid/air compressibility and boundary layer on cavity--bubble interactions}\label{appC}

\begin{figure}
	\centering
	\includegraphics[width=0.75\textwidth]{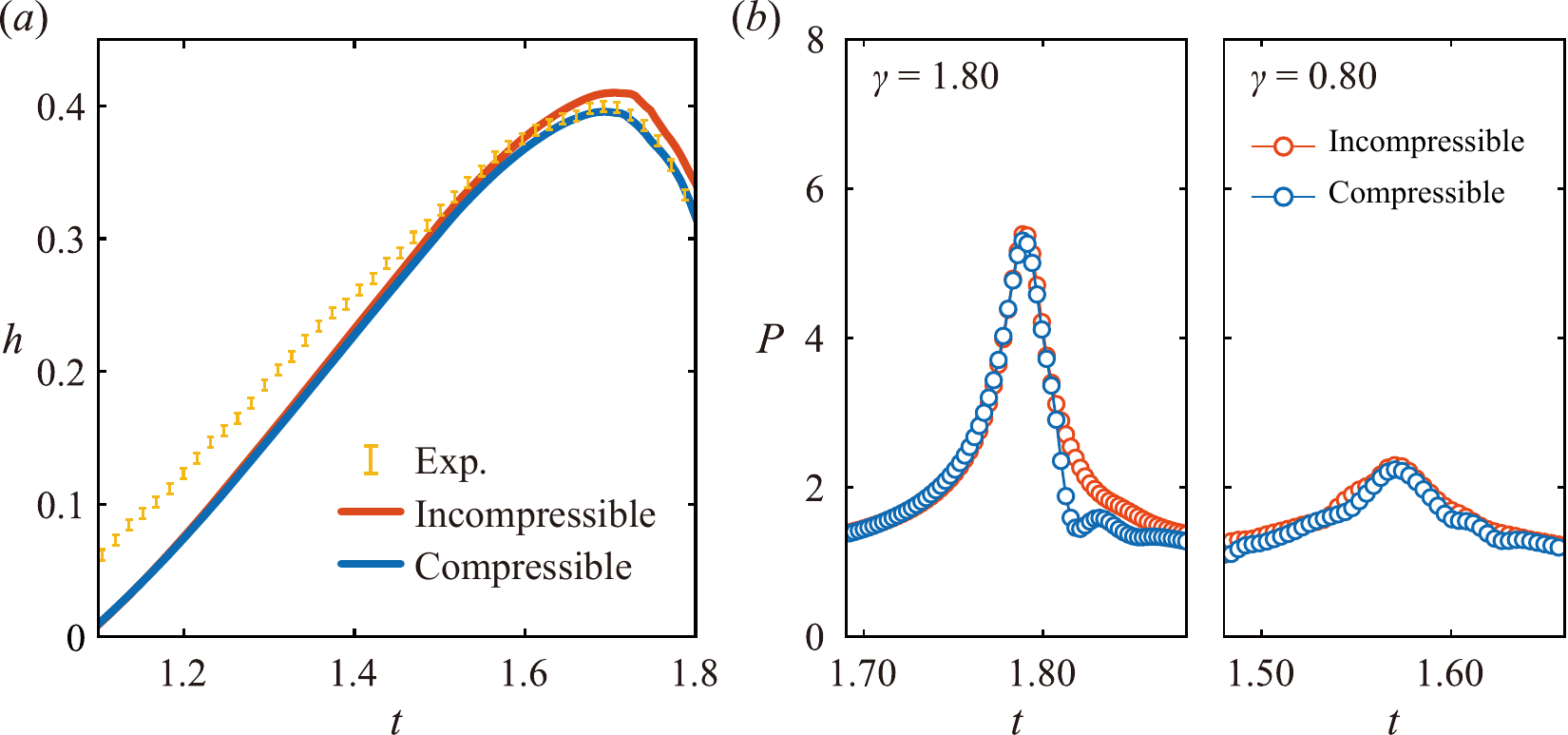}
	\setlength{\abovecaptionskip}{2mm}
	\captionsetup{width=1\textwidth}
	\caption{Comparisons between compressible and incompressible FVM models for the (\textit{a}) cavity evolution and (\textit{b}) fluid field pressure. The length and time scales are 17.0 mm and 1.70 ms, respectively. The experimental uncertainty of \(h\) is approximately 0.004 (0.076 mm in dimensional form).}
	\label{liquid_compressibility_verification}
	\vspace{0mm}
\end{figure}

Verification of the liquid/air compressibility effect on cavity--bubbles interactions is presented in this section. Numerical results obtained with compressible and incompressible FVM models are presented in \autoref{liquid_compressibility_verification}. The numerical results obtained from the compressible model exhibit slight discrepancies from the incompressible counterpart with respect to the maximum cavity length and peak pressure when the bubble reaches its minimum volume. Such discrepancies can be explained by the energy dissipation of the bubble oscillation introduced by the liquid compressibility. The discrepancies in \autoref{liquid_compressibility_verification}(\textit{a}) and \ref{liquid_compressibility_verification}(\textit{b}) are 2.8 and 2 \textit{\%}, respectively, which are relatively small and can be neglected safely. Moreover, no distinctively new cavity behaviors are introduced by the compressible model, which further suggests that the compressibility can be reasonably neglected in this study.
	
\begin{figure}
	\centering
	\includegraphics[width=1.0\textwidth]{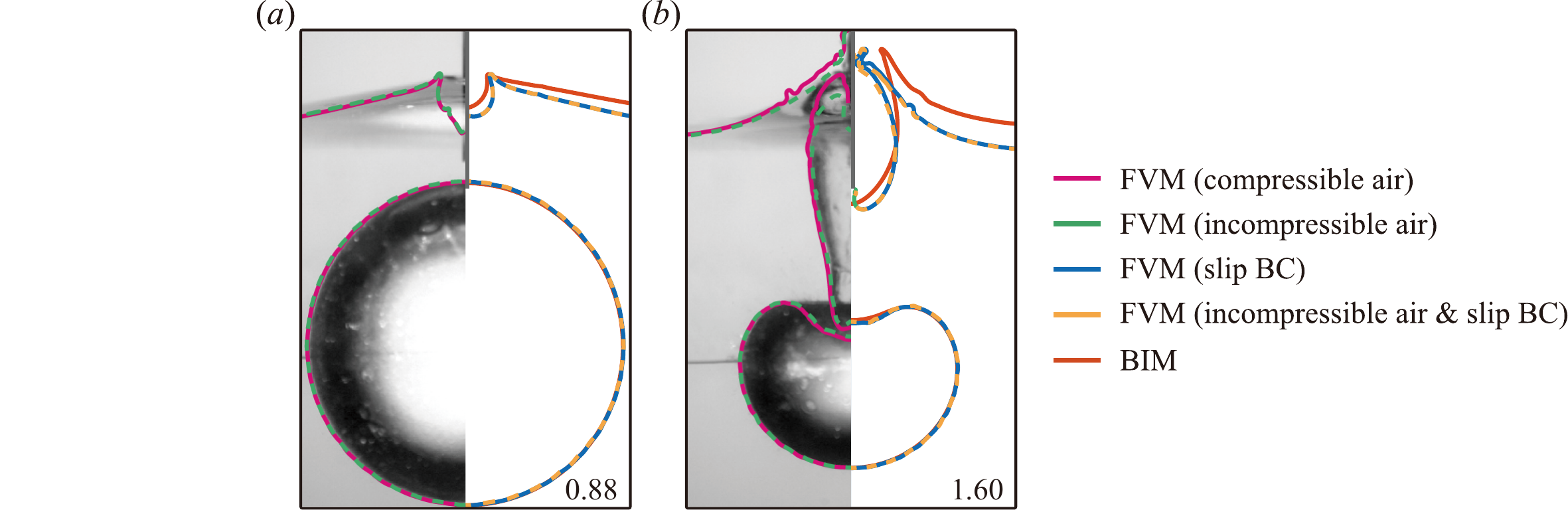}
	\setlength{\abovecaptionskip}{1mm}
	\captionsetup{width=1\textwidth}
	\caption{Comparisons between experimental observations and numerical predictions obtained using different FVM configurations and the BI model. Results with compressible (magenta solid lines) and incompressible air (green dashed lines) show good agreement with the experiments, although the incompressible air model predicts a slightly smaller cavity. In contrast, simulation results with slip BC (blue solid lines), incompressible air with slip BC (yellow dashed lines), and the BI model (red solid lines) deviate markedly from the experiments, while remain relatively consistent with one another with respect to cavity length. The parameters of the bubble and surface perturbations are identical with those in \autoref{exp_gamma_critical_condition}(\textit{a}).}
	\label{air_compressibility_verification}
	\vspace{-2mm}
\end{figure}

Moreover, the effects of air compressibility and boundary layers are examined in \autoref{air_compressibility_verification}. As can be seen, the FVM results obtained with compressible (magenta solid lines) and incompressible air (green dashed lines) agree well with the experimental observations, although the incompressible air model predicts a smaller cavity in the late-time evolution (\autoref{air_compressibility_verification}\textit{b}). These results confirm our statement in \cref{section:4.2} and \cref{section:4.3}, indicating that air compressibility becomes increasingly prominent as gas velocity rises (decreasing \(\gamma\)), especially in the late-time cavity evolution. In contrast, the results obtained with slip BC (blue solid lines), incompressible air with slip BC (yellow dashed lines), and the BI model (red solid lines) show noticeable deviations from the experiments and no-slip BC results, while remaining in close agreement with one another in terms of cavity length. The good consistency suggests that the air compressibility effect is negligible if the cavity and gas velocity are not sufficiently large, whereas the deviation between slip and no-slip BC results indicates that boundary layer effects become more pronounced as \(\gamma\) decreases, compared to those in \autoref{boundary_layer_effect}. Nevertheless, the radial and axial scales of the cavity remain comparable between the two groups of predictions, implying that the influence of the boundary layer is not fundamental in cavity evolution, but rather acts in conjunction with air compressibility to affect the cavity development. We also notice that the no-slip BC results exhibit slight discrepancies in cavity morphology from the BI simulation results, particularly near the cavity opening and bottom. These differences can be attributed to the vortex formed inside the cavity due to the rapid airflow, which can significantly influence the cavity shape and the surface-sealing process \citep{du2022cavitydynamics,wang2022Cavityairflow}.

\section{Supplementary materials of methodology}\label{appD}
Regarding the meniscus calculation, the full deduction of the asymptotic solution of \(h_m\) is complicated and lengthy. One can refer to the study of \citet{loMeniscusNeedleLesson1983} for more detailed information. We only present the necessary coefficients for calculating with (\ref{hmasymptotic}) here:
\allowdisplaybreaks
\begin{align}
	\allowdisplaybreaks
	z_1|_{\hat{r}=1}=&-\cos\phi_0,\notag\\
	z_2|_{\hat{r}=1}=&\cos\phi_0\left[\ln4-\ln(1+\sin\phi_0)-E\right],\notag\\
	z_3|_{\hat{r}=1}=&\frac{1}{2}\cos\phi_0\left(1-\cos^2\!\phi_0\right),\notag\\
	z_4|_{\hat{r}=1}=&\cos\phi_0\left\{\sin^2\!\phi_0\left[E+\frac{1}{4}+\ln\frac{1}{4}\left(1+\sin\phi_0\right)\right]+\left[\frac{1}{4}-\sin\phi_0\right]\right\},\notag\\
	z_5|_{\hat{r}=1}=&\frac{1}{4}\sin\phi_0\cos\phi_0\left(\ln4-E\right)-\frac{A}{\sin\phi_0}+\frac{1}{4}\cos\phi_0+A\left(1-\ln4-E\right)\notag\\
	&+\frac{1}{4}\cos^3\!\phi_0\left(\frac{1}{2}-E^2-\ln^22+E\ln4-\ln2\right)\notag\\
	&+\left[\ln\left(1+\sin\phi_0\right)\right]\left[\frac{1}{4}\cos^3\!\phi_0\left(\ln4-E+\frac{1}{\sin\phi_0}\right)+A-\frac{1}{4}\cos\phi_0\sin\phi_0\right]\notag\\
	&-\left[\ln^2\left(1+\sin\phi_0\right)\right]\frac{1}{4}\cos^3\!\phi_0,\notag\\
	&A=\frac{1}{4}B\left(2-B^2\right)\ln\left[1+\left(1-B^2\right)^{\frac{1}{2}}\right]-\left(1-B^2\right)\frac{1}{2}B\left(\ln4-E\right)-\frac{1}{4}B\left(1-B\right)^{\frac{1}{2}},\notag\\
	&B=\cos\phi_0,
\end{align}
where \(\phi_0\) is the meniscus contact angle, \(E\) the Euler--Mascheroni constant.

\bibliographystyle{jfm}

\bibliography{reference}

\end{document}